\documentclass[aps,prd,superscriptaddress,nofootinbib,11pt]{revtex4}
\usepackage[english]{babel}
\usepackage[utf8]{inputenc}
\usepackage{graphicx}   % need for figures
\usepackage{slashed}
\usepackage{epstopdf}
\usepackage{verbatim}   % useful for program listings
\usepackage{color}      % use if color is used in text
\usepackage{subfigure}  % use for side-by-side figures
\usepackage{multirow}
\usepackage{hyperref}   % use for hypertext links, including those to external documents and URLs
\usepackage{float}
\usepackage{epsfig,rotating}
\usepackage{amsmath,amssymb}
\usepackage{dsfont}
\restylefloat{table}
\numberwithin{equation}{section}

\newcommand{\vx}{\vec{x}}

\newcommand{\vq}{\vec{q}}
\newcommand{\vk}{\vec{k}}

\newcommand{\be}{\begin{equation}}
\newcommand{\ee}{\end{equation}}
\newcommand{\bea}{\begin{eqnarray}}
\newcommand{\eea}{\end{eqnarray}}

\newcommand{\ket}[1]{|#1\rangle}
\newcommand{\bra}[1]{\langle#1|}

\newcommand{\X}{\mathcal X}

\begin{document}
\title{What spectators do during inflation.}
\author{Daniel  Boyanovsky}
\email{boyan@pitt.edu} \affiliation{Department
of Physics, University of Pittsburgh, Pittsburgh,
Pennsylvania 15260, USA}
\date{\today}
\begin{abstract}
The inflaton equation of motion including one loop radiative corrections from spectator fields is obtained.
  We consider a massless scalar conformally coupled to gravity and  a massless fermion Yukawa coupled to the inflaton as models for spectators that \emph{do not feature} gravitational particle production,   their production during slow roll is solely  a consequence of their coupling to the inflaton. The one-loop self energy and  the fully renormalized equation of motion of the inflaton   are obtained   and solved explicitly for an   inflaton potential $m^2\varphi^2/2$.  The solution   features Sudakov-type logarithmic secular terms, which are resumed via the dynamical renormalization group and compared to the solutions with a phenomenological friction term. During $N_e$ e-folds of slow roll inflation the inflaton evolves as $\varphi^{(0)}_{Isr}(t)\,e^{\frac{m^2\Gamma}{9H^3}\,N_e(t)}$ for the phenomenological friction term  $\Gamma$ and $\varphi^{(0)}_{Isr}(t)\,e^{\Upsilon N^2_e}$ with $\Upsilon =  -\frac{\lambda^2}{24\pi^2 H^2} ; \frac{y^2_R}{12\pi^2}$ for the radiative  corrections from bosonic and fermionic spectators respectively where $\varphi^{(0)}_{Isr}(t)$ is the slow roll solution in absence of interactions,   showing that a phenomenological  friction term is not reliable. A generalization of  the optical theorem
 to a finite time domain and cosmological expansion is introduced   to obtain the distribution function $f(k,t)$ and total number of spectators produced \emph{during slow roll}.  $f(k,t)$ is peaked at superhorizon scales and the total number of particles grows $\propto e^{3N_e}$. A non-perturbative mean field theory is introduced to describe the self-consistent evolution of the inflaton coupled to spectators, its linearized version reproduces the self-energy, the inflaton equation of motion and the results on particle production.

\end{abstract}

\maketitle

\section{Introduction}\label{sec:intro}

An early period of accelerated cosmological expansion, namely an inflationary stage,   provides a framework which successfully solves many of the problems of the standard  Big Bang cosmology. Precise observations of the cosmic microwave background (CMB) anisotropies by the WMAP\cite{wmap} and PLANCK\cite{planck} missions support one of the main predictions of inflationary cosmology,   a  nearly Gaussian and scale invariant power spectrum of adiabatic perturbations which seed the anisotropies in the  (CMB) radiation. A main paradigm of inflationary cosmology posits that the inflationary stage is dominated by a   scalar field slowly rolling down a potential landscape leading to a nearly de Sitter inflationary stage. Cosmological perturbations   are generated by  quantum fluctuations of this field that are amplified when their wavelengths become larger than the Hubble radius during inflation. Within this  theoretical framework of inflation one scalar field dominates the cosmological dynamics, although many scalar fields may be introduced at the expense of generating isocurvature perturbations which are severely constrained by observations of (CMB) temperature anisotropies. Even without extensions beyond the Standard Model of particle physics, there is a plethora of other fields, including a scalar Higgs whose degrees of freedom do not simply ``vanish'' during inflation, but their contribution to the energy momentum tensor is assumed to be subleading, thereby  not playing a substantial role in the cosmological dynamics beyond being simple ``spectators''. Extensions beyond the Standard Model, perhaps necessary to explain Dark Matter add to the number of spectator fields. These spectator fields eventually must be excited so as to merge the inflationary stage with the Standard Big Bang radiation dominated era, this process is generically referred to as pre or reheating and has been and continues to be the focus of  much attention\cite{rehe}-\cite{ferpre5}.

Some consequences  of spectator fields have been previously studied, from the generation of entropy from spectator loops during inflation\cite{proko}, as a source of non-Gaussian correlations\cite{nongau}, gravitational waves\cite{gw1,gw2,gw3,gw4},  primordial black holes\cite{kaiser,verner},  stochasting inflation\cite{vennin,hardwick} or light moduli fields from string theories\cite{string1,string2,string3}. Most paradigmatic models of reheating (or preheating) post inflation invoke a coupling of spectator fields to the inflaton\cite{rehe}-\cite{ferpre5}, with some exceptions\cite{khlopov} where the coupling to the inflaton is indirect, via its gravitational effects. The process of reheating is important as an energy transfer mechanism from the inflaton, oscillating at the bottom of a potential, to the degrees of freedom
that populate the radiation dominated era post inflation.

The influence of spectator fields coupled to the inflaton \emph{during} inflation is often described phenomenologically via a local friction term in the equation of motion of the inflaton condensate\cite{kolb,berera} usually associated with the S-matrix decay rate \emph{in Minkowski space-time}. This phenomenological approach of replacing the effects of radiative corrections on the equation of motion of a scalar field by  a  local friction term has been recently scrutinized in a radiation dominated cosmology in ref.\cite{cao} with the conclusion that such simplification \emph{does not reliably describe} the effects of radiative corrections to the equations of motion of a scalar condensate in an expanding cosmology.

\vspace{1mm}

\textbf{Main objectives:} Our main objectives are the following \textbf{i:)} to study the effects of spectator fields \emph{coupled to the inflaton}  upon the dynamics of the scalar inflaton condensate by obtaining and solving
its {  linearized} equation of motion   including self-energy radiative corrections from these fields up to one loop. In particular, to compare the solution to the
equations of motion with one loop radiative corrections to the phenomenological description with a local friction term. We also obtain and solve these equations with self-energy corrections in Minkowski space time in order to establish unambiguously the differences with the cosmological case. \textbf{ii:)} Dissipative effects from radiative corrections, namely radiation reaction, are related to the production of particles in the self-energy loops, we seek to study the production of particles coupled to the inflaton \emph{during} the slow roll inflationary stage, in distinction to particle production during post-inflationary reheating. Considering the cases of inflaton coupling to massless scalar conformally coupled to gravity, and to massless fermions Yukawa coupled to the inflaton we can effectively focus on particle production as a consequence of the inflaton-spectator coupling without   gravitational production of spectator fields\cite{parker1,ford,parker,BD,fulling,long,agullo}. \textbf{iii:)} We seek to study particle production of spectator fields via their coupling to the inflaton condensate by generalizing and extending the optical theorem to the cosmological realm in a finite time domain, obtaining the distribution function of the produced particles and comparing to Minkowski space-time to explicitly exhibit the consequences of cosmological expansion.

\vspace{1mm}

\textbf{Summary of Results:}
\begin{itemize}

\item{We consider the inflaton coupled to scalar and fermion spectator fields, the latter via a Yukawa coupling and derive the {  linearized} equations of motion of the homogeneous inflaton condensate including radiative corrections  up to one loop, namely  leading order in the couplings, implementing the Schwinger-Keldysh formulation of non-equilibrium quantum field theory\cite{schwinger,keldysh,maha,jordan,beilok}.  The one loop self energy  is obtained analytically for the cases when the scalar field is massless and conformally coupled to gravity and the fermion field is massless. These cases are relevant to understand particle production solely via inflaton-spectator coupling.  The equations of motion are renormalized, including field renormalization in the case of fermion spectators.   Alternative derivations of the { linearized} equation of motion are obtained from linear response theory, and from a non-perturbative mean field approach in the weak coupling linearized approximation reproducing the results and confirming independently its validity.  }

 \item{The renormalized equations of motion with the non-local self-energy kernel are solved in the case of a   quadratic inflaton potential $V(\varphi) = m^2\,\varphi^2/2$  by implementing the dynamical renormalization group  \cite{drg,drg1,cao,greendrg}. We compare directly the phenomenological simplification with a local friction term, with the correct results from the (non-local) self energy. One of our main results is that the dynamical renormalization group improved solutions of the equations of motion with radiative corrections feature Sudakov-type (double) logarithmic enhancement. In terms of the number of e-folds during slow roll inflation $N_e(t)$ we find that the inflaton condensate evolves as
     \be \varphi_{Isr}(t) = \varphi^{(0)}_{Isr}(t)\times \Bigg\{ \begin{array}{l}
                                                       e^{\Upsilon\,N_e(t)} ~~;~~   \Upsilon = \frac{m^2\Gamma}{9H^3}~~;~~ (\Gamma=phenomenological~ friction~term) \\
                                                       e^{\Upsilon N^2_e(t)}~~;~~ \Upsilon =  -\frac{\lambda^2}{24\pi^2 H^2}~~(\mathrm{bosons})~;~ \frac{y^2_R}{12\pi^2}~~(\mathrm{fermion})
                                                     \end{array}
\nonumber \ee where $\varphi^{(0)}_{Isr}(t)$ is the slow roll solution in absence of coupling to spectator fields. Clearly, the simple local friction term \emph{does not} describe correctly inflaton dynamics. For comparison, we also analyze the solutions in Minkowski space-time implementing the dynamical renormalization group resummation thereby explicitly exhibiting the effects of cosmological expansion.     }

     \item{We extend the optical theorem to a cosmological setting in a finite time domain to study particle production and apply it to the case of a massless scalar spectator field conformally coupled to gravity, which does not feature gravitational particle production, thus clearly separating   particle production from coupling to the inflaton from its gravitational counterpart. To leading order in the spectator-inflaton coupling $\lambda$ we obtain the distribution function of produced particles which is strongly peaked on \emph{superhorizon wavelengths} and find that the total number of scalar spectator particles produced during slow roll inflation is
         \be  {N}(t)= \frac{\lambda^2 \,V_{ph}(t)}{12\pi\,H}\, \Big(\varphi^{(0)}_{Isr}(t)\Big)^2  \nonumber \ee where $V_{ph}(t)$ is the \emph{physical} volume, exhibiting a rapid growth  $\propto e^{3 N_e(t)}$ of the number of particles produced \emph{during} inflation. We compare this result in near de Sitter, to the case of Minkowski space time where the distribution function is peaked at $k=m/2$   as a consequence of energy conservation. The striking difference between the distribution functions in de Sitter and Minkowski space-times is a direct result of the lack of energy conservation and a global timelike Killing vector in the cosmological background. }

     \item{  We introduce a non-perturbative    mean field framework for the self-consistent dynamics of the inflaton condensate and a conformally coupled massless scalar spectator field.   An integral equation for the exact solutions of the spectator mode functions is derived, whose solution can be obtained as  a Born series. We recognize that the Born approximation yields the one-loop (non-local) self-energy obtained via the Schwinger-Keldysh and linear response methods, thereby establishing a direct link between the non-perturbative self-consistent formulation and the perturbative approach to the equations of motion in the loop expansion. Particle production is studied via a Bogoliubov tranformation between the exact spectator mode functions and  the ``free field-particle basis'' from which we obtain the distribution function and particle number in the Born approximation in complete agreement with the results from the optical theorem.   }

 \end{itemize}

 The article is organized as follows: section (\ref{sec:model}) introduces the model to study the consequences of spectator- inflaton coupling to scalars and fermions. For coherence and consistency of presentation section (\ref{sec:quanti}) summarizes the necessary ingredients of field quantization of bosonic and fermionic fields in a fixed de Sitter cosmological background. In section (\ref{sec:eominf}) we derive the equations of motion implementing the Schwinger-Keldysh formulation of non-equilibrium quantum field theory, we also provide an  alternative derivation based on linear response  theory. In this section we discuss the phenomenological approach that simplifies the radiative corrections from spectator fields by a local friction term, obtain explicitly the self-energy kernels for the cases of massless scalars conformally coupled to gravity and massless fermions Yukawa coupled to the inflaton, renormalize the equations of motion, including field renormalization in the case of fermionic spectators and solve them perturbatively for a simple inflaton potential $V(\varphi) = m^2\,\varphi^2/2$ in all cases. The perturbative solutions reveal secular terms, there is a striking difference between the secular terms from the phenomenological friction term    and those of the radiative corrections from spectators which feature (double) Sudakov logarithms of   secular  growth for both scalar and fermion spectators. We also compare to the solution in Minkowski space time to exhibit the differences associated with cosmological expansion. In section (\ref{sec:drg}) we implement the dynamical renormalization group (DRG) program to resum the secular terms in the perturbative solutions in all cases, in this section we obtain some of our main results on the impact of radiative corrections from spectator fields on the evolution of the inflaton, obtaining the (DRG) improved asymptotic solution for the evolution of the inflaton condensate during slow roll, comparing explicitly to the phenomenological friction term and to Minkowski space-time. In this section we show unambiguosly that the simple friction term does \emph{not} yield the correct dynamics. In section (\ref{sec:decay}) we study the production of spectator particles via their coupling to the inflaton condensate focusing on massless scalar spectators conformally coupled to gravity to highlight production via coupling to the inflaton rather than gravitational. This case illuminates the main aspects without the technical complications of spinors and field renormalization. We introduce a generalization of the optical theorem for couplings to condensates in a finite time domain and extend it to cosmology. We obtain the distribution function of the produced particles and the total number of particles produced comparing to Minkowski space time to clearly exhibit the effects of expansion and lack of a global timelike Killing vector. In section (\ref{subsec:bogos}) we introduce a non-perturbative mean field framework to study the self-consistent evolution of the inflaton \emph{and} a massless scalar spectator (conformally coupled to gravity), and obtain an integral equation (Lippman-Schwinger) for the \emph{exact} mode functions of the spectator fields which can be solved in a Born series. The exact mode functions are related to the free field ``particle'' states by a Bogoliubov transformation from which the distribution function of produced particles is obtained. The Born approximation reproduces the results from the self-energy radiative corrections and is in complete agreement with the results of the optical theorem for the distribution function and total number of produced particles, thereby establishing a direct link between the perturbative and non-perturbative formulations and a confirmation of equations of motion and the consequences of spectator fields. Section (\ref{sec:discussion}) presents a discussion of various subtle aspects and  summarizes the general lessons learned. Our conclusions are summarized in section (\ref{sec:conclusions}). Several appendices analyze correlation functions of spectator fields related to the self-energy kernels and provide a derivation of the optical theorem with condensates in a finite time domain.

\section{Inflaton and spectators: a model}\label{sec:model}

We consider a spatially flat de Sitter cosmology with Hubble expansion rate $H$ and defined by the  Friedmann-Robertson-Walker (FRW)   metric in comoving coordinates given by
\be  g_{\mu \nu} = \textrm{diag}(1, -a^2, -a^2, -a^2)~~;~~ a(t) = e^{Ht} \,. \label{frwmetric}\ee
It is convenient to pass from comoving time $t$,   to conformal time $\eta$ with $d\eta = dt/a(t)$, in terms of which the metric becomes
\be  g_{\mu \nu} = C^2(\eta)\, \textrm{diag}(1,-1,-1,-1) \, , \label{conformalmetric} \ee with
\be C(\eta) = -\frac{1}{H\eta} ~~;~~ -\infty \leq \eta \leq 0 \,.\label{Cds}\ee

The inflaton field is described by a real scalar field  $\phi_I(\vx,t)$, the spectator fields are taken to be a real scalar  $\phi_s(\vx,t)$ and a fermion $\Psi(\vx,t)$    with action given by\cite{BD,parker,fulling}

\bea  S   & = &    \int d^4 x \sqrt{|g|} \Bigg\{\frac{1}{2} g^{\mu\nu}\,\partial_\mu \phi_I \,\partial_\nu \phi_I-V(\phi_I) + \frac{1}{2} g^{\mu\nu}\,\partial_\mu \phi_s \partial_\nu \phi_s-\frac{1}{2} \big[M^2_s -\xi_s\,R\big]\phi^2_s   -   \lambda \, \phi_I \, \phi^2_s \nonumber \\ & + & \overline{\Psi}  \Big[i\,\gamma^\mu \;  \mathcal{D}_\mu -m_f-y\, \phi_I \Big]\Psi     \Bigg\}\,,
\label{action}\eea  where
\be R= 12 H^2 \,,\label{ricci}\ee is the Ricci scalar in de Sitter space-time,  and $\xi_s$ is a coupling to gravity, with $\xi_s=0, 1/6$ corresponding to minimal or conformal coupling of the spectator bosonic field, respectively.  We   consider   the inflaton to be  minimally coupled to gravity. Our aim is to  obtain the equation of motion  including radiative corrections from spectator fields for the homogeneous inflaton condensate, identified as the expectation value of the inflaton field in a translational invariant coherent state
\be \langle \phi_I (\vx, t)  \rangle \equiv \varphi_I(t) \,. \label{cond}\ee

The Dirac $\gamma^\mu$ are the curved space-time $\gamma$ matrices
and the fermionic covariant derivative is given
by\cite{weinbergbook,BD,duncan,casta,parker,baacke}
\bea
\mathcal{D}_\mu & = &  \partial_\mu + \frac{1}{8} \;
[\gamma^c,\gamma^d] \;  e^\nu_c  \; \left(D_\mu e_{d \nu} \right)
\cr \cr %\label{fercova}\\
D_\mu e_{d \nu} & = & \partial_\mu e_{d \nu} -\Gamma^\lambda_{\mu
\nu} \;  e_{d \lambda} \nonumber
\eea
\noindent where the vierbein field $e^\mu_a$ is defined as
$$
g^{\mu\,\nu} =e^\mu_a \;  e^\nu_b \;  \eta^{a b} \; ,
$$
\noindent $\eta_{a b}$ is the Minkowski space-time metric, greek indices refer to
curved space time coordinates and latin indices to the local
 Minkowski space time coordinates. The curved space-time  matrices $\gamma^\mu$ are given in terms of
the Minkowski space-time ones $\gamma^a$  by
\be
\gamma^\mu = \gamma^a e^\mu_a \quad , \quad
\{\gamma^\mu,\gamma^\nu\}=2 \; g^{\mu \nu}  \; .
\ee

 In conformal time the vierbeins $e^\mu_a$ are particularly simple
\be
 e^\mu_a = C^{-1}(\eta)\; \delta^\mu_a ~~;~~ e^a_\mu = C(\eta) \; \delta^a_\mu
\ee
\noindent and the Dirac Lagrangian density simplifies to the
following expression
\be \label{ecdi}
\sqrt{-g} \; \overline{\Psi}\Bigg(i \; \gamma^\mu \;  \mathcal{D}_\mu
\Psi -m_f-y\,\phi \Bigg)\Psi  =
(C^{\frac{3}{2}}\overline{\Psi}) \;  \Bigg[i \;
{\not\!{\partial}}-(m_f+y\,\phi) \; C(\eta) \Bigg]
\left(C^{\frac{3}{2}}{\Psi}\right)
\ee
\noindent where $i {\not\!{\partial}}=\gamma^a \partial_a$ is the usual Dirac
differential operator in Minkowski space-time in terms of flat
space time $\gamma^a$ matrices.

Introducing the conformally rescaled fields
\be C(\eta) \phi_I(\vx,t) = \chi_I(\vx,\eta) ~~;~~C(\eta) \phi_s(\vx,t) = \chi_s(\vx,\eta)  ~~;~~ C^{\frac{3}{2}}(\eta){\Psi(\vx,t)}= \psi(\vx,\eta) \label{rescaledfields}\ee
 and neglecting surface terms which do not contribute to the equations of motion, the action becomes
   \be  S    =
  \int d^3x \; d\eta \,\mathcal{L}[\chi_I;\chi_s;\psi] \;, \label{rescalagds}\ee
  where
  \be \mathcal{L}[\chi_I;\chi_s;\psi]=   \mathcal{L}_I[\chi_I]+ \mathcal{L}_0[\chi_s]+\mathcal{L}_0[\psi]+\mathcal{L}_{int}[\chi_I,\chi_s,\psi]  \,,\label{Ltot} \ee with
  \bea \mathcal{L}_I[\chi_I] & = & \frac{1}{2}\,
{\chi'_I}^2-\frac{1}{2}\,(\nabla \chi_I)^2 - \widetilde{V}(\chi_I;\eta)  \,, \label{l0chi}\\
\mathcal{L}_0[\chi_s] & = & \frac{1}{2} \big[
{\chi'_s}^2- (\nabla \chi_s)^2- \mathcal{M}^2_s(\eta)\,\chi^2_s\big] \,, \label{lchis}\\
\mathcal{L}_0[\psi] & = & \overline{\psi} \;  \Big[i \;
{\not\!{\partial}}+ \frac{m_f}{H\eta}    \Big]
 {\psi}  \,,\label{l0psi}\\ \mathcal{L}_{int}[\chi_I,\chi_s,\psi] & = & -\chi_I(\vx,\eta) \, \mathcal{O}(\vx,\eta)  \; , \label{lI}\eea where to consolidate notation  we  introduced the composite operator
 \be \mathcal{O}(\vx,\eta) = \lambda\, C(\eta)\,:\chi^2_s(\vx,\eta): + y :\overline{\psi}(\vx,\eta)\,\psi(\vx,\eta): \,,\label{opeO}\ee  and normal ordered the $\chi_I;\chi_s;\psi$ interactions in the interaction picture of the free spectator fields $\chi_s;\psi$.

  The    potential $\widetilde{V}$ is given by
 \be \widetilde{V}(\chi_I;\eta) = (C(\eta))^4 \,V\Big(\frac{\chi_I}{C(\eta)}\Big)  - \frac{\chi^2_I}{\eta^2}\,    \,,\label{tilpot}\ee
  and
 \be
\mathcal{M}^2_s (\eta)  = \Big[\frac{M^2_s}{H^2}+12\Big(\xi_s -
\frac{1}{6}\Big)\Big]\frac{1}{\eta^2} \,. \label{masschi2}\ee

\textbf{Slow roll:} Within the slow roll paradigm of inflation, the slow roll conditions for the inflaton condensate $\varphi_I$
\be \frac{1}{2} \Big(\frac{ \dot{\varphi}_I}{2 M_{pl}H}\Big)^2 \ll 1 ~~;~~ \Big| \frac{\ddot{\varphi}_I}{H \dot{\varphi}_I}\Big| \ll 1 \,,\label{slowroll}\ee imply that the homogeneous inflaton condensate  during this
inflationary stage $\varphi_I \simeq \mathrm{constant}$, therefore the rescaled field $\chi_I$  features a homogeneous condensate $\X(\eta)= \langle \chi_I(\vx,\eta)\rangle$  that behaves as
\be \X(\eta) =  -\frac{\varphi_I(\eta)}{H\,\eta} \,,\label{conchisr}\ee namely the homogeneous condensate contribution to the interaction term (\ref{lI}) acts as a ``pump'' field with $C(\eta) \X(\eta) \propto 1/\eta^2~;~ \X(\eta) \propto 1/\eta$ so that the interaction of spectator fields with the inflaton condensate   become stronger at later time. This feature will become important when we analyze the contributions from spectator fields to the equation of motion of the inflaton, in particular, in  comparing to a local ``friction'' term, and studying the production of spectator fields during inflation as a consequence of this ``pump''.

\section{Quantization of free spectator fields}\label{sec:quanti}

In anticipation of obtaining the equations of motion implementing the Schwinger-Keldysh\cite{schwinger,keldysh,maha,jordan,beilok} or ``in-in'' formulation, we summarize the main aspects of the quantization of spectator fields that will be necessary to obtain the correlation functions that enter in the self-energy radiative corrections.

\subsection{Bosonic spectator}\label{subsec:bosespec}
The free field Heisenberg field equation of motion for the bosonic spectator field $\chi_s$ is
\be \Big[ \frac{\partial^2}{\partial \eta^2} - \nabla^2 +\mathcal{M}^2_s(\eta)\Big]\chi_s(\vx,\eta) = 0 \,,\label{heiseombos}\ee where $\mathcal{M}^2_s(\eta)$ is given by equation (\ref{masschi2}). The Heisenberg free field $\chi_s(\vx,\eta)$  and its canonical momentum obtained from the Lagrangian density (\ref{lchis}), $ \pi_s(\vx,\eta)= d\chi_s(\vx,\eta)/d\eta$ are expanded  in mode functions in a comoving volume $V$ as
\be
\chi_s(\vx,\eta)  =  \frac{1}{\sqrt{V}}\,\sum_{\vq} \Big[a_{\vq}\,g_q(\eta)+ a^\dagger_{-\vq}\,g^*_q(\eta) \Big]\,e^{i\vq\cdot\vx}\,, \label{chiex} \ee
\be \pi_s(\vx,\eta) =  \frac{1}{\sqrt{V}}\,\sum_{\vq} \Big[a_{\vq}\,\frac{d}{d\eta}g_q(\eta)+ a^\dagger_{-\vq}\,\frac{d}{d\eta}g^*_q(\eta) \Big]\,e^{i\vq\cdot\vx}\,, \label{piex} \ee   where  $g_q(\eta)$ are solutions of the mode equations
\be \Big[ \frac{d^2}{d \eta^2} +q^2 -\frac{1}{\eta^2}\Big(\nu^2_s -\frac{1}{4} \Big)  \Big]g_q(\eta) = 0 \,,  \label{gmodes}\ee where
\be \nu^2_{s} = \frac{9}{4}- \Big(\frac{M^2_{s}}{H^2}+12\, \xi_s \Big)  \,.
\label{nusa} \ee

We choose Bunch-Davies boundary conditions
\be g_q(\eta)_{~\overrightarrow{-q\eta \rightarrow \infty}}~\frac{e^{-iq\eta}}{\sqrt{2q}}\,,\label{bdbc}\ee yielding
\be g_q(\eta)= \frac{1}{2}\,e^{i\frac{\pi}{2}(\nu_s+\frac{1}{2})}\,\sqrt{-\pi\,\eta}\,H^{(1)}_{\nu_s}(-q\eta)\,,\label{gqeta}\ee  which imply canonical commutation relations for $a_{\vq}, a^\dagger_{\vq}$ and the Bunch-Davis vacuum state $|0_s\rangle$ is such that
\be a_{\vq} |0_s\rangle  =0 \,. \label{bdvac}\ee

The normal ordered composite operator $:\chi^2_s(\vx,\eta):$ is \emph{defined} in the free field interaction picture by ordering the annihilation and creation operators in the product of the quantized fields (\ref{chiex}) so that the annihilation operators $a_{\vq}$ always stand to the right of creation operators $a^\dagger_{\vq}$, yielding
\be \bra{0_s}:\chi^2_s(\vx,\eta):\ket{0_s} =0\,. \label{bosono}\ee

  Quantization with non-Bunch Davies boundary conditions can be studied similarly with straightforward generalizations in terms of  linear combinations of the mode functions (\ref{gqeta}) and its complex conjugate,   here we consider this simpler case to highlight the main physical consequences.

\subsection{Fermionic spectator}\label{subsec:ferspec}

 After the conformal rescaling (\ref{rescaledfields}), the Dirac equation for the free Fermi field  $\psi$ is
\be  \Big[i \;
{\not\!{\partial}}- M_\psi(\eta)    \Big]
 {\psi}  = 0 ~~;~~M_\psi(\eta) = -   \frac{m_f}{H\eta} \label{diraceqn}\ee
For Dirac fermions the solution $ \psi({\vec x},\eta) $ is expanded  as
\be
\psi_D(\vec{x},\eta) =    \frac{1}{\sqrt{V}}
\sum_{\vec{k},\lambda}\,   \left[b_{\vec{k},\lambda}\, U_{\lambda}(\vec{k},\eta)\,e^{i \vec{k}\cdot
\vec{x}}+
d^{\dagger}_{\vec{k},\lambda}\, V_{\lambda}(\vec{k},\eta)\,e^{-i \vec{k}\cdot
\vec{x}}\right] \; ,
\label{psiex}
\ee
where the spinor mode functions $U_\lambda,V_\lambda$ obey the  Dirac equations
\bea
\Bigg[i \; \gamma^0 \;  \partial_\eta - \vec{\gamma}\cdot \vec{k}
-M_\psi(\eta) \Bigg]U_\lambda(\vec{k},\eta) & = & 0 \label{Uspinor} \\
\Bigg[i \; \gamma^0 \;  \partial_\eta + \vec{\gamma}\cdot \vec{k} -M_\psi(\eta)
\Bigg]V_\lambda(\vec{k},\eta) & = & 0 \label{Vspinor}
\eea

We choose to work with the standard Dirac representation of the Minkowski space time $\gamma^a$ matrices. It proves
convenient to write
\bea
U_\lambda(\vec{k},\eta) & = & \Bigg[i \; \gamma^0 \;  \frac{d}{d\eta} -
\vec{\gamma}\cdot \vec{k} +M_\psi(\eta)
\Bigg]f_k(\eta)\, \mathcal{U}_\lambda \label{Us}\\
V_\lambda(\vec{k},\eta) & = & \Bigg[i \; \gamma^0 \;  \frac{d}{d\eta} +
\vec{\gamma}\cdot \vec{k} +M_\psi( \eta)
\Bigg]h_k(\eta)\,\mathcal{V}_\lambda \label{Vs}
\eea
\noindent with $\mathcal{U}_\lambda;\mathcal{V}_\lambda$ being
constant spinors  obeying
\be
\gamma^0 \; \mathcal{U}_\lambda  =  \mathcal{U}_\lambda
\label{Up} \qquad , \qquad
\gamma^0 \;  \mathcal{V}_\lambda  =  -\mathcal{V}_\lambda
\ee
The mode functions $f_k(\eta);h_k(\eta)$ obey the following
equations of motion
\bea \left[\frac{d^2}{d\eta^2} +
k^2+M^2_\psi(\eta)-i \; M'_\psi(\eta)\right]f_k(\eta) & = & 0 \,, \label{feq}\\
\left[\frac{d^2}{d\eta^2} + k^2+M^2_\psi(\eta)+i \; M'_\psi(\eta)\right]h_k(\eta)
& = & 0 \,.\label{geq}
\eea where $' \equiv \frac{d}{d\eta}$.
We choose Bunch-Davies boundary conditions for the solutions, namely
\be f_k(\eta) ~~{}_{\overrightarrow{-k\eta \rightarrow \infty}}~~ e^{-ik\eta}~~;~~ h_k(\eta) ~~{}_{\overrightarrow{-k\eta \rightarrow \infty}}~~ e^{ik\eta} \,, \label{bdfketa}\ee which leads to the choice
\be h_k(\eta)= f^*_k(\eta)\,, \label{choice} \ee  and $f_k(\eta)$ is a solution of
\be \left[\frac{d^2}{d\eta^2} +
k^2+
\frac{1}{\eta^2}\Big[\frac{m^2_f}{H^2}-i\frac{m_f}{H}\Big]\right]f_k(\eta)   =   0 \,. \label{eqnfketa}\ee   we find
\be f_k(\eta) = e^{i\frac{\pi}{2}(\nu_\psi+1/2)}\,\, \sqrt{\frac{-\pi k \eta}{2}}\,\,H^{(1)}_{\nu_\psi}(-k\eta)~~;~~ \nu_\psi = \frac{1}{2}+i\frac{m_f}{H}\,. \label{fketasolu}\ee
The fermionic Bunch-Davies vacuum $\ket{0_F}$ is defined such that
\be b_{\vk,\lambda}\ket{0_F}=0~~;~~ d_{\vk,\lambda}\ket{0_F}=0\,. \label{bdfervac}\ee

Two limits are noteworthy: the sub-Hubble limit $(-k\eta) \rightarrow \infty$ of these modes is given by (\ref{bdfketa}) and also of interest is their super-Hubble behavior $(-k\eta) \rightarrow 0$, given by
\be f_k(\eta) \propto (-H\eta)^{-im_f/H} \propto e^{i m_f t}\,, \label{superhubf}\ee remarkably, up to a constant the super-Hubble fermionic modes behave just as the long-wavelength limit of \emph{negative energy states} in Minkowski space-time. In contrast to the bosonic case,  the amplitude of the mode functions remains bound and of order unity for super-Hubble wavelengths.

Introducing
\be \Omega(k,\eta) =  i \frac{f'_k(\eta)}{f_k(\eta)}+M_\psi(\eta) \label{wofketa}\ee where $' = d/d \eta$, the Dirac spinors are found to be
\be U_\lambda(\vk,\eta) = N_k \,f_k(\eta)\, \left(
                            \begin{array}{c}
                              \Omega(k,\eta)\, \, \chi_\lambda \\
                              \vec{\sigma}\cdot \vec{k}\, \,\chi_\lambda \\
                            \end{array}
                          \right) ~~;~~ \chi_1 = \left(
                                                   \begin{array}{c}
                                                     1 \\
                                                     0 \\
                                                   \end{array}
                                                 \right) \;; \; \chi_2 = \left(
                                                                       \begin{array}{c}
                                                                         0 \\
                                                                         1 \\
                                                                       \end{array}
                                                                     \right) \,,
 \label{Uspinorsol} \ee  and

 \be V_\lambda(\vk,\eta) = N_k \,f^*_k(\eta)\, \left(
                            \begin{array}{c}
                               \vec{\sigma}\cdot \vec{k}\, \, \varphi_\lambda \\
                               \Omega^*(k,\eta)\, \, \varphi_\lambda  \\
                            \end{array}
                          \right) ~~;~~ \varphi_1 = \left(
                                                   \begin{array}{c}
                                                    0 \\
                                                     1 \\
                                                   \end{array}
                                                 \right) \;; \; \varphi_2 = -\left(
                                                                       \begin{array}{c}
                                                                         1 \\
                                                                         0 \\
                                                                       \end{array}
                                                                     \right) \,.
 \label{Vspinorsol} \ee  These spinors are ortho-normalized
 \be U^\dagger_\lambda(\vk,\eta)\, U_\lambda(\vk,\eta) = V^\dagger_\lambda(\vk,\eta) \,V_\lambda(\vk,\eta) = 1 ~~;~~ U^\dagger_\lambda(\vk,\eta) V_{\lambda'}(-\vk,\eta) = 0 ~~\forall \lambda, \lambda' \,,\label{orthospinors}\ee from which it follows that
 \be |N_k|^2 \Bigg[\Big(if'_k(\eta)+M_\psi(\eta) f_k(\eta)\Big)\,\Big(-i{f'_k(\eta)}^*+M_\psi(\eta) f^*_k(\eta)\Big)+k^2 f^*_k(\eta)f_k(\eta) \Bigg] =1\,. \label{norma} \ee Using equation (\ref{feq}) it is straightforward to find that the bracket is indeed $\eta$ independent, and evaluating as $\eta \rightarrow -\infty$ we find (up to an irrelevant phase)
 \be N_k = \frac{1}{k\sqrt{2}}\,. \label{Nofk}\ee Furthermore it is straightforward to confirm that  the $U$ and $V$ spinors obey the charge conjugation relation
 \be i\gamma^2 U^*_\lambda(\vk,\eta) = V_\lambda(\vk,\eta) ~~:~~ i\gamma^2 V^*_\lambda(\vk,\eta) = U_\lambda(\vk,\eta) ~~;~~ \lambda = 1,2 \,. \label{chargeconj}\ee

 The case of Majorana fermions can be obtained straightforwardly from the construction above.   Majorana (charge self-conjugate) fields obey\footnote{We set the Majorana phase to zero as it is not relevant for the discussion.}
 \be \psi^c(\vx,\eta) = C (\overline{\psi}(\vx,\eta))^T  = \psi(\vx,\eta) ~~;~~ C = i\gamma^2\gamma^0 \,,\label{majorana} \ee the charge conjugation properties (\ref{chargeconj}) immediately lead to the quantized   Majorana fields
\be
\psi_M(\vec{x},\eta) =    \frac{1}{\sqrt{V}}
\sum_{\vec{k},\lambda}\,   \left[b_{\vec{k},\lambda}\, U_{\lambda}(\vec{k},\eta)\,e^{i \vec{k}\cdot
\vec{x}}+
b^{\dagger}_{\vec{k},\lambda}\, V_{\lambda}(\vec{k},\eta)\,e^{-i \vec{k}\cdot
\vec{x}}\right] \; ,
\label{psiexmajo}
\ee
In the case of Majorana fields the free-field fermionic part of the Lagrangian must be multiplied by a factor $1/2$ since a Majorana field has half the number of degrees of freedom of the Dirac field. In the study that follows we will focus solely on Dirac fields, with a straightforward extrapolation to the Majorana case, henceforth the label ``D'' for the quantum fields is dropped.

With the ortho-normalization conditions (\ref{orthospinors}) the annihilation ($b,d$) and creation ($b^\dagger, d^\dagger$) operators obey canonical anticommutation relations, yielding the equal time anticommutation relations
\be \big\{\psi^\dagger(\vx,\eta),\psi(\vx',\eta) \big\} = \delta^{(3)}(\vx-\vx') \,. \label{anticomm}\ee
Just as in the bosonic case the composite operator $:\overline{\psi}(\vx,\eta)\psi(\vx,\eta):$ is \emph{defined} in the interaction picture of free fields such that in the product of the quantized operators (\ref{psiex}) the annihilation operators $b,d$ always appear to the right of creation operators ($b^\dagger,d^\dagger$), so that
\be \bra{0_F}:\overline{\psi}(\vx,\eta)\psi(\vx,\eta):\ket{0_F} =0\,. \label{ferno}\ee

\section{Inflaton equation of motion}\label{sec:eominf}

The action (\ref{rescalagds}-\ref{lI}) for the conformally rescaled fields in conformal time, yields the canonical momenta conjugate to the fields $\chi_{I,s}$
\be \pi_I = \chi'_I ~~;~~ \pi_s = \chi'_s \label{canonical} \ee
  with the equal time commutation relations for bosonic fields given by
 \be  [\pi_I(\vx,\eta),\chi_I(\vx',\eta)]   =  -i\,\delta^{(3)}(\vx-\vx')~~;~~  [\pi_s(\vx,\eta),\chi_s(\vx',\eta)] = -i\,\delta^{(3)}(\vx-\vx')\,,\label{comrel}\ee
 and the (conformal) Hamiltonian
\bea H(\eta) & = &  \int \Bigg\{ \frac{1}{2}\Big[ {\pi^2_I(\vx,\eta)} +  (\nabla \chi_I)^2 \Big] +\widetilde{V}(\chi_I;\eta) + \frac{1}{2} \Big[  {\pi^2_s(\vx,\eta)} +   (\nabla \chi_s)^2 +\mathcal{M}^2_s(\eta)\chi^2_2(\vx,\eta)\Big] \nonumber \\ & + & \psi^\dagger(\vx,\eta) \Big( -i \vec{\alpha}\cdot \vec{\nabla} + \gamma^0 M_\psi(\eta) \Big)\,\psi(\vx,\eta)+ \chi_I(\vx,\eta)\,\mathcal{O}(\vx,\eta)   \Bigg\}\,d^3x\,.  \label{conham}\eea

 It is straightforward to show that the canonical (anti) commutation relations (\ref{anticomm},\ref{comrel})  yield the Heisenberg field equations in conformal time as Hamiltonian equations of motion with the \emph{time dependent Hamiltonian} (\ref{conham}), and that these equations of motion are the same as those obtained from the variation of the action (\ref{action}) or (\ref{rescalagds})  after conformal rescaling. Therefore, the conformal Hamiltonian (\ref{conham}) is the generator of time evolution. Hence for any operator $ {\mathcal{A}}$   in the Heisenberg picture, Heisenberg's Hamilton's equation of motion become
\be \frac{d}{d\eta} {\mathcal{A}}(\vx,\eta) = i \big[ H(\eta),  {\mathcal{A}}(\vx,\eta)\big] \,,\label{Heiseom}\ee whose solution is
\be  {\mathcal{A}}(\vx,\eta_0) = U^{-1}(\eta,\eta_0)\, {\mathcal{A}}(\vx,\eta_0)\,U(\eta,\eta_0) \,,\label{solhem}\ee where the unitary time evolution operator is given by
\be U(\eta,\eta_0)= T\Big( e^{-i\int^\eta_{\eta_0}\,H(\eta^{'})\,d\eta^{'}} \Big)~~;~~ U^{-1}(\eta,\eta_0) = \widetilde{T}\Big( e^{i\int^\eta_{\eta_0}\,H(\eta^{'})\,d\eta^{'}} \Big) \,,\label{Uop}\ee  and $T,\widetilde{T}$ are the time ordering and anti-ordering symbols respectively. With an initial state described by a density matrix $\rho(\eta_0)$, normalized such that $\mathrm{Tr} \rho(\eta_0) =1$, expectation values of a Heisenberg field operator $ {\mathcal{A}}$ are given by
\be \langle \mathcal{A}(\vx,\eta) \rangle = \mathrm{Tr}\mathcal{A}(\vx,\eta) \,\rho(\eta_0) \,.\label{exval}\ee  Expectation values and correlation functions   are obtained via functional derivatives of the generating functional\cite{beilok,boytad}
\be \mathcal{Z}[J^+,J^-] \equiv \mathrm{Tr}\Big[ U(\eta,\eta_0;J^+)\,\rho(\eta_0)\,U^{-1}(\eta,\eta_0;J^-)\Big] \,,\label{ZJ}\ee with respect to the external sources $J^\pm$, where
\be   U(\eta,\eta_0;J^+) = \mathrm{T} \Big( e^{-i \int^\eta_{\eta_0}H(\eta';J^+)} \Big)~~;~~  U^{-1}(\eta,\eta_0;J^-)= \widetilde{T}\Big( e^{i\int^\eta_{\eta_0}\,H(\eta^{'};J^-)\,d\eta^{'}} \Big)\ee  with the Hamiltonian including the coupling  of operators $\mathcal{A}$ to the external sources
\be H(\eta,J^{\pm}) \equiv H(\eta) + \int d^3 x J^{\pm}(\vx,\eta)\,\mathcal{A}(\vx,\eta) \,.\label{Hjs}\ee

For example\cite{beilok}
\bea \langle   \mathcal{A}^+(\vx_1,\eta_1)\mathcal{A}^+(\vx_2,\eta_2) \rangle & \equiv & \langle T \Big(\mathcal{A}(\vx_1,\eta_1)\mathcal{A}(\vx_2,\eta_2)\Big)\rangle  =   \mathrm{Tr}  \Big(T \mathcal{A}(\vx_1,\eta_1)\mathcal{A}(\vx_2,\eta_2)\Big) \rho(\eta_0) \nonumber \\ & = &  -\frac{\delta^2\,\mathcal{Z}[J^+,J^-]}{ \delta J^+(\vx_1,\eta_1)\delta J^+(\vx_2,\eta_2) } \Big|_{J^+=J^-=0}\,,  \label{timor} \eea
\bea \langle   \mathcal{A}^-(\vx_1,\eta_1)\mathcal{A}^-(\vx_2,\eta_2) \rangle & \equiv & \langle \widetilde{T} \Big(\mathcal{A}(\vx_1,\eta_1)\mathcal{A}(\vx_2,\eta_2)\Big)\rangle  =   \mathrm{Tr}   \rho(\eta_0) \Big( \widetilde{T} \mathcal{A}(\vx_1,\eta_1)\mathcal{A}(\vx_2,\eta_2)\Big)\nonumber \\ & = &  -\frac{\delta^2\,\mathcal{Z}[J^+,J^-]}{ \delta J^-(\vx_1,\eta_1)\delta J^-(\vx_2,\eta_2) } \Big|_{J^+=J^-=0}\,,  \label{antimor} \eea

\bea
\langle \mathcal{A}^-(\vx_2,\eta_2)\mathcal{A}^+(\vx_1,\eta_1)\rangle & \equiv & \langle \mathcal{A}(\vx_2,\eta_2)\mathcal{A}(\vx_1,\eta_1)\rangle =  \mathrm{Tr}      \mathcal{A}(\vx_1,\eta_1)\, \rho(\eta_0)\,\mathcal{A}(\vx_2,\eta_2) \nonumber \\ & = &  \frac{\delta^2\,\mathcal{Z}[J^+,J^-]}{ \delta J^+(\vx_1,\eta_1)\delta J^-(\vx_2,\eta_2) }\Big|_{J^+=J^-=0}\,, \label{timo}\eea

\bea
\langle \mathcal{A}^+(\vx_2,\eta_2)\mathcal{A}^-(\vx_1,\eta_1)\rangle & \equiv & \langle \mathcal{A}(\vx_1,\eta_1)\mathcal{A}(\vx_2,\eta_2)\rangle =  \mathrm{Tr}      \mathcal{A}(\vx_2,\eta_2)\, \rho(\eta_0)\,\mathcal{A}(\vx_1,\eta_1) \nonumber \\ & = &  \frac{\delta^2\,\mathcal{Z}[J^+,J^-]}{ \delta J^+(\vx_2,\eta_2)\delta J^-(\vx_1,\eta_1) }\Big|_{J^+=J^-=0}\,,\label{antimo}\eea  where the \emph{anti time ordered} correlation
\be \langle \widetilde{T} \Big(\mathcal{A}(\vx_1,\eta_1)\mathcal{A}(\vx_2,\eta_2)\Big)\rangle  = \langle   \mathcal{A}(\vx_1,\eta_1)\mathcal{A}(\vx_2,\eta_2)\rangle \,\Theta(\eta_2-\eta_1) + \langle   \mathcal{A}(\vx_2,\eta_2)\mathcal{A}(\vx_1,\eta_1)\rangle \,\Theta(\eta_1-\eta_2)\,.\label{defanti}\ee

  An important result is that
\bea  && \langle  \mathcal{A}^+(\vx,\eta)\rangle   \equiv     \mathrm{Tr}      \mathcal{A}(\vx,\eta)\, \rho(\eta_0)  = -i\frac{\delta\,\mathcal{Z}[J^+,J^-]}{ \delta J^+(\vx,\eta)} \Big|_{J^+=J^-=0} \nonumber \\ & = &  \langle  \mathcal{A}^-(\vx,\eta)\rangle   \equiv    \mathrm{Tr}       \rho(\eta_0) \,  \mathcal{A}(\vx,\eta)= i\frac{\delta\,\mathcal{Z}[J^+,J^-]}{ \delta J^-(\vx,\eta)} \Big|_{J^+=J^-=0} \,. \label{exvalj}\eea

 Referring  to the fields $\chi_{I,s},\psi^\dagger,\psi$ collectively as $\{\chi\}$, and the sources coupled to them as $\{J\}$, with the understanding that the sources coupled to $\overline{\psi},\psi$ are Grassman valued, the generating functional (\ref{ZJ}) in the field representation can be written in a functional integral representation
\bea \mathcal{Z}[\{J^+\},\{J^-\}] & = & \int D\{\chi_f\} D\{\chi_i\} D\{\chi'_i\} \,\bra{\{\chi_f\}} U(\eta,\eta_0;\{J^+\}) \ket{\{\chi_i\}}\,\bra{\{\chi_i\}}\rho(\eta_0)\times\nonumber \\ & & \ket{\{\chi'_i\}}\bra{\{\chi'_i\}}U^{-1}(\eta,\eta_0;\{J^-\})\ket{\{\chi_f\}}\,,\label{parti} \eea where the functional integrals for the fermionic degrees of freedom are in terms of Grassman valued fields. In turn the field matrix elements of the evolution operators can be written as path integrals, namely
\bea  \bra{\{\chi_f\}} U(\eta,\eta_0;\{J^+\}) \ket{\{\chi_i\}} & \equiv &  \int \mathcal{D}\{\chi^+\} \,e^{i\int  \mathcal{L}[\{\chi^+\};\{J^+\}]\,  d^4x} \,,\label{lplus} \\ \bra{\{\chi'_i\}} U^{-1}(\eta,\eta_0;\{J^-\}) \ket{\{\chi_f\}} & \equiv &  \int \mathcal{D} \{\chi^-\} \,e^{-i\int \mathcal{L}[\{\chi^-\};\{J^-\}]\,  d^4x}    \,,\label{lmin}\eea with boundary conditions
\be   \{\chi^+\}(\eta_0) = \{\chi_i\}~;~\{\chi^+\}(\eta)=\{\chi_f\}~~;~~ \{\chi^-\}(\eta_0) = \{\chi'_i\}~;~\{\chi^-\}(\eta)=\{\chi_f\}  \,,\label{bcs}\ee the functional integral over $\{\chi_f\}$ represents the trace.
The effective Lagrangian densities on either branch ($\pm$) are given by
\be \mathcal{L}[\{\chi^{\pm}\};J^{\pm}] = \mathcal{L}[\{\chi^{\pm}\}]-\{J^{\pm}\}\,\{\chi^\pm\}  \,,\label{lagspm}  \ee
  where $\mathcal{L}[\{\chi\}]$ is given by (\ref{Ltot} -\ref{opeO}) and used the shorthand notation  $\{J^{\pm}\}\,\{\chi^\pm\}\equiv J^\pm_I\,\chi^\pm_I + J^\pm\,\chi^\pm_s + \cdots$.

   Finally, the functional and path integral representation of the generating functional becomes
\be \mathcal{Z}[\{J^+\},\{J^-\}] =   \int D\{\chi_f\} D\{\chi_i\} D\{\chi'_i\} \int \mathcal{D}\{\chi^+\} \mathcal{D}\{\chi^-\}\, e^{i \int \Big[\mathcal{L}[\{\chi^+\};\{J^+\}]-\mathcal{L}[\{\chi^-\};\{J^-\}]\Big] d^4x} \,\rho(\eta_0) \,,\label{pathint} \ee with the boundary conditions on the fields $\chi^\pm$ given by eqns. (\ref{lplus},\ref{lmin}) and the notation $\int d^4x \equiv \int^\eta_{\eta_0} d\eta' \int d^3 x$. This is the in-in or Schwinger-Keldysh formulation of non-equilibrium quantum field theory\cite{schwinger,keldysh,maha,jordan}.

Our objective is to obtain the   equation of motion for the homogeneous  expectation value of the inflaton $\chi_I(\vx,\eta)$ (or $\phi_I(\vx,\eta)=\chi_I(\vx,\eta)/C(\eta)$), namely
\be \langle \chi_I(\vx,\eta) \rangle \equiv \mathrm{Tr} \, \chi_I(\vx,\eta) \rho(\eta_0) \equiv \X(\eta) \,,\label{Xv}\ee  including radiative corrections from spectator fields up to one loop order.

We consider $\X$ to be spatially homogeneous, consistently with homogeneity and isotropy of the cosmology, hence only the zero momentum component of $\chi_I$ acquires an expectation value. The equation of motion for $\X$ is obtained from the identity (\ref{exvalj}) which implies that $\langle \chi^+_I \rangle = \langle \chi^-_I \rangle = \X$,  separating the condensate from the fluctuations by  writing
\be \chi^\pm_I(\vx,\eta) = \X(\eta) + \delta^\pm(\vx,\eta)\,, \label{shift}\ee    in the Lagrangians $\mathcal{L}[\{\chi^\pm\},J^\pm]$ in equation (\ref{lagspm}) and requesting that the fluctuation field $\delta(\vx,\eta)$ features a vanishing expectation value
\be \langle \delta^\pm(\vx,\eta)\rangle =0\,,\label{zerodelta}\ee  to \emph{all orders} in perturbation theory\cite{boytad}. The initial density matrix $\rho(\eta_0)$ is taken to be
\be \rho(\eta_0) = \ket{0_\delta;0_s;0_F}\bra{0_\delta;0_s;0_F}\,,\label{inirhovac}\ee where all the vacua are the Bunch Davies vacua for each field $\delta,\chi_s,\psi$ respectively.

\begin{figure}[ht!]
\begin{center}
\includegraphics[height=3in,width=4in,keepaspectratio=true]{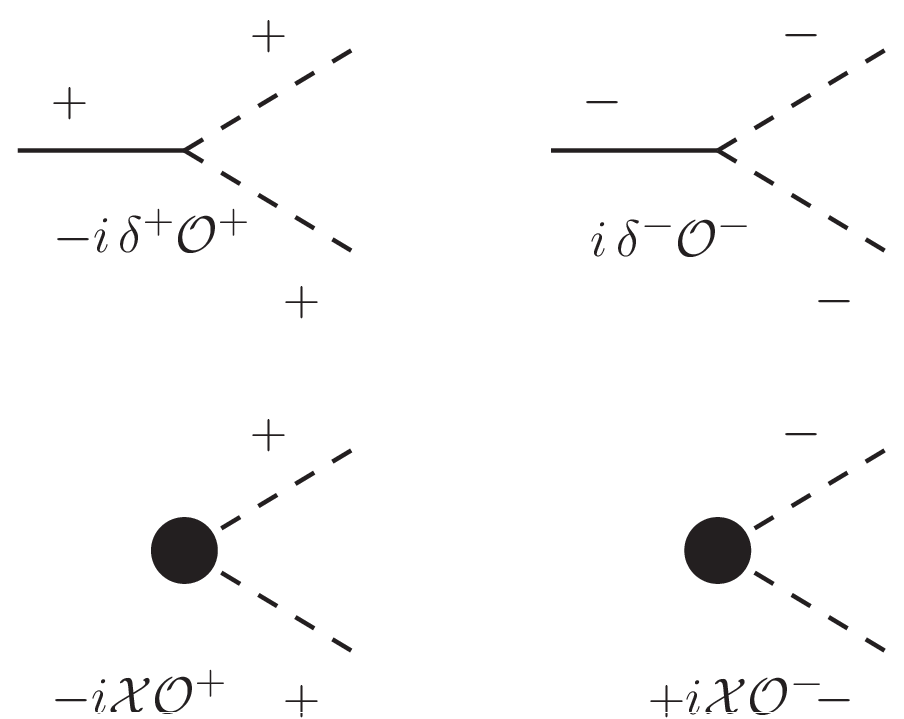}
\caption{Interaction vertices with the composite operator $\mathcal{O}^\pm$ from the Lagrangian (\ref{shiftedlag}). The solid lines correspond to the fluctuations of the inflaton field $\delta^\pm$ the dashed lines represent \emph{both} $:\chi^2_s$ and $:\overline{\psi} \psi:$ for the respective branches ($\pm$).  The dark dot stands for the inflaton condensate $\X$ . }
\label{fig:vertices}
\end{center}
\end{figure}

Upon integration by parts and neglecting surface terms, and coupling sources only to the fluctuating fields $\chi^\pm,\delta^\pm,\psi^\pm$,  we obtain (primes denote $\partial/\partial \eta$)
\bea && i\int  \Big[\mathcal{L}[\{\chi^+\},J^+] - \mathcal{L}[\{\chi^-\},J^-]\Big]  d^4x = \nonumber \\
&& + i \int \Bigg\{\mathcal{L}_0[\delta^+;J^+_\delta]  -\mathcal{L}_0[\delta^-;J^-_\delta] \Bigg\}\,d^4 x \nonumber \\
&& i \int \Bigg\{\mathcal{L}_0[\chi^+_s,J^+_s]+\mathcal{L}_0[\psi^+;J^+_{\psi}] - \mathcal{L}_0[\chi^-_s;J^-_s]-\mathcal{L}_0[\psi^-;J^-_\psi] \Bigg\}\,d^4 x \nonumber \\
&& -i\int  \Big[\X^{''}(\eta)+\frac{d}{d\X}\widetilde{V}(\X;\eta)\Big]\delta^+(\vx,\eta)\,d^4 x -  \Big( \delta^+ \rightarrow \delta^- \Big)\nonumber \\
&&-i  \int  \,\Bigg\{\Big(\X+\delta^+\Big)\,  \mathcal{O}^+(\vx,\eta)- \Big(\X+\delta^-\Big)\,  \mathcal{O}^-(\vx,\eta)\Bigg\}\,d^4x \,,\label{shiftedlag}
 \eea
 where we have kept only quadratic terms in the inflaton fluctuations $\delta$, with
 \be \mathcal{L}_0[\delta] = \frac{1}{2}\Bigg[\Big(\frac{\partial \, \delta}{\partial \eta } \Big)^2- (\nabla \delta)^2 - \frac{d^2}{d\X^2}\widetilde{V}(\X;\eta)\,\delta^2 \Bigg]\,, \label{lzerodelta}\ee and included sources $J_\delta;J_s;J_\psi$   to obtain a perturbative expansion via functional derivatives with respect to these sources. The sources are linearly coupled to the various fields, for fermions $\overline{J}^\pm_\psi;J^\pm_\psi$  are Grassman valued.
The interaction vertices associated with the composite operator $\mathcal{O}^\pm$ on the respective forward and backward branches are depicted in figure (\ref{fig:vertices}).

 Following the method introduced in ref.\cite{boytad,cao} the strategy to obtain the equation of motion for the inflaton condensate is to treat \emph{both} last terms in (\ref{shiftedlag}), namely the linear terms in $\delta^\pm$ along with the interaction vertices with the composite operators $\mathcal{O}^\pm$  as perturbations, and request that the vanishing conditions (\ref{zerodelta}) are fulfilled order by order in perturbation theory.

 To zeroth order in the couplings $\lambda,y$, we obtain
 \be \langle \delta^+(\vec{y},\eta')\rangle  = -i \int \Bigg\{ \Big[\langle \delta^+(\vec{y},\eta')\delta^+(\vec{x},\eta)\rangle -\langle \delta^+(\vec{y},\eta')\delta^-(\vec{x},\eta)\rangle\Big] \Big[\X^{''}(\eta)+\frac{d}{d\X}\widetilde{V}(\X;\eta)\Big]\Bigg\}\,d^4x=0 \label{zerothord} \,,  \ee since the correlation functions
 $\langle \delta^+(\vec{y},\eta')\delta^\pm(\vec{x},\eta)\rangle$ are independent,
  it follows that
 \be  \X^{''}(\eta)+\frac{d}{d\X}\widetilde{V}(\X;\eta) =0 \,,\label{zerotheom}\ee it is straightforward to confirm that the same equation of motion is obtained from the condition $\langle \delta^-(\vec{y},\eta')\rangle =0$. Because perturbation theory is carried out in the free field theory of the $\delta^\pm,\chi^\pm$ fields, these must always appear in pairs in correlation functions, therefore the next contribution is of second order in the interaction vertices corresponding to a one-loop self-energy, which is given by
 \bea && \langle \delta^+(\vec{y},\eta_1)\rangle^{(1\,\mathrm{loop})}   =    - \,\int d^4x \langle \delta^+(\vec{y},\eta_1)\delta^+(\vec{x},\eta)\rangle \times \label{1lup} \\ & & \int d^4x'  \,\Big[\langle \mathcal{O}^+(\vx,\eta)\,\mathcal{O}^+(\vx',\eta')\rangle-  \langle \mathcal{O}^+(\vx,\eta)\,\mathcal{O}^-(\vx',\eta') \rangle\Big]\X(\eta') \,d^4x'\,,\nonumber\eea where we have only considered the contribution from the correlation function $\langle \delta^+(\vec{y},\eta_1)\delta^+(\vec{x},\eta)\rangle\ $ since the other $\langle \delta^+(\vec{y},\eta_1)\delta^-(\vec{x},\eta)\rangle\ $  yields the same equation, consistently with the zeroth order case as can be straightforwardly confirmed by using the identities (\ref{timor}-\ref{defanti}).

    \begin{figure}[ht!]
\begin{center}
\includegraphics[height=3in,width=4in,keepaspectratio=true]{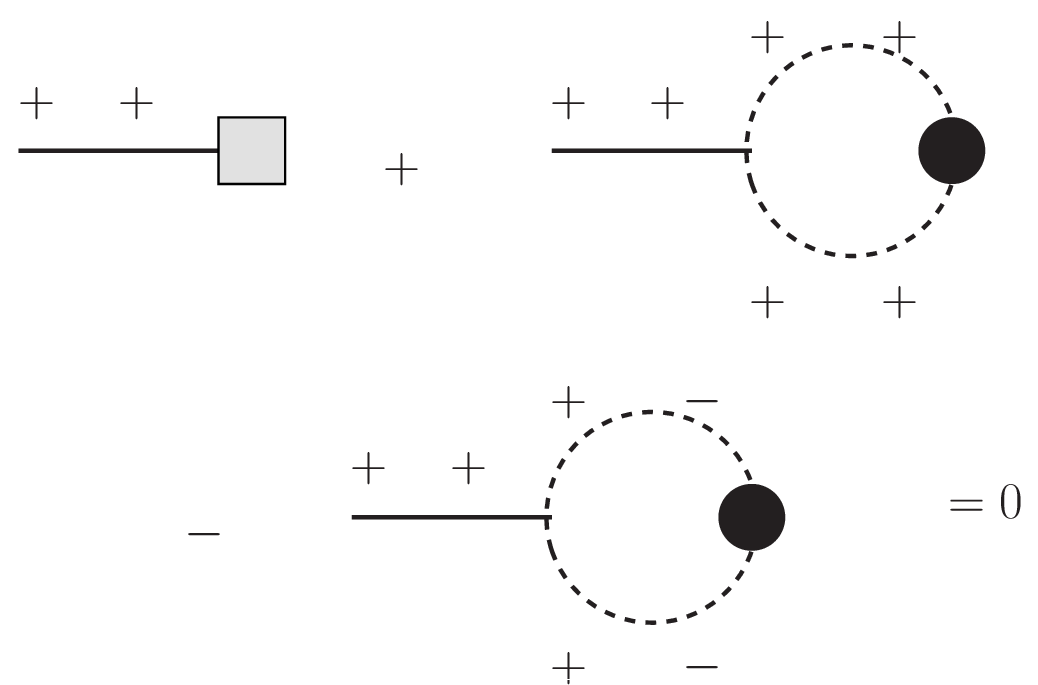}
\caption{Representation of the diagrams leading to the equation of motion (\ref{1lupeom}). The solid line is the propagator of the inflaton fluctuation $\langle \delta^+ \delta^+\rangle$, the gray square is the tree level term $\Big[\X^{''}(\eta)+\frac{d}{d\X}\widetilde{V}(\X;\eta)\Big]$, the dark dot stands for $\X$, the dashed lines represent both $:\chi^2_s$ and $:\overline{\psi} \psi:$ for the respective branches ($\pm$). The loops represent the self energy $\Sigma$.}
\label{fig:eom}
\end{center}
\end{figure}

 Combining the zeroth order term (\ref{zerothord}) and the one loop contribution yields the   equation of motion up to one loop order with the contribution from spectator fields linearized in $\X$
  \be \X^{''}(\eta)+\frac{d}{d\X}\widetilde{V}(\X;\eta)+\int^{0}_{\eta_0} \int  \Sigma(\vx,\vx';\eta,\eta')\,\X(\eta')\,d\eta'\,d^3x' =0 \,,\label{1lupeom}\ee where the self-energy, depicted by the loops in fig. (\ref{fig:eom}),  is given by
  \be \Sigma(\vx,\vx';\eta,\eta') = -i \,\Big[\langle \mathcal{O}^+(\vx,\eta)\,\mathcal{O}^+(\vx',\eta')\rangle-  \langle \mathcal{O}^+(\vx,\eta)\,\mathcal{O}^-(\vx',\eta') \rangle\Big]\,.\label{sigma}  \ee  The correlation functions defining the self-energy $\Sigma$ are given by
  \bea \langle \mathcal{O}^+(\vx,\eta)\,\mathcal{O}^+(\vx',\eta')\rangle & = & \langle \mathcal{O}(\vx,\eta)\,\mathcal{O}(\vx',\eta')\rangle\,\Theta(\eta-\eta')+ \langle \mathcal{O}(\vx',\eta') \mathcal{O}(\vx,\eta) \rangle\,\Theta(\eta'-\eta) \,\label{plusplus}\\
  \langle \mathcal{O}^+(\vx,\eta)\,\mathcal{O}^-(\vx',\eta')\rangle & = & \langle  \mathcal{O}(\vx',\eta') \mathcal{O}(\vx,\eta) \rangle \,,\label{plusminus}\eea from which it follows that
  \be \Sigma(\vx,\vx';\eta,\eta') = -i \,\Big[\langle \mathcal{O}(\vx,\eta)\,\mathcal{O}(\vx',\eta')\rangle -  \langle \mathcal{O}(\vx',\eta')\,\mathcal{O}(\vx,\eta) \rangle\Big]\,\Theta(\eta-\eta')\,,\label{sigmaret}  \ee is the retarded self energy. The correlation functions of the composite operator $\mathcal{O}$ given by (\ref{opeO}) are given by
  \bea  \langle \mathcal{O}(\vx,\eta)\,\mathcal{O}(\vx',\eta')\rangle  & = &  \lambda^2\,C(\eta)\,C(\eta') \bra{0_s}:\chi^2_s(\vx,\eta):\,:\chi^2_s(\vx',\eta'): \ket{0_s} \nonumber \\ & + &  y^2\,\bra{0_F}:\overline{\psi}(\vx,\eta)\psi(\vx,\eta)::\overline{\psi}(\vx',\eta')\psi(\vx',\eta'):\ket{0_F}\,,\label{corresO}   \eea where we have used (\ref{bosono},\ref{ferno}).

  The result (\ref{corresO}) implies that the self energy is a sum of the bosonic and fermionic spectators self-energies respectively,
\be  \Sigma(\vx,\vx';\eta,\eta') =  \Sigma_s(\vx,\vx';\eta,\eta')+ \Sigma_f(\vx,\vx';\eta,\eta')\,,\label{sumsig}\ee where using the results of appendix (\ref{app:corres}), we find
\be \Sigma_s(\vx,\vx';\eta,\eta') = -2i\lambda^2 C(\eta) C(\eta')\Big[ G^>(\vx,\vx';\eta,\eta') - G^<(\vx,\vx';\eta,\eta')\Big]\,\Theta(\eta-\eta') \,,\label{sigbos}\ee
\be \Sigma_f(\vx,\vx';\eta,\eta') = -iy^2 \Big[\mathcal{G}^>_f(\vx,\vx';\eta,\eta')-\mathcal{G}^<_f(\vx,\vx';\eta,\eta')\Big]\,\Theta(\eta-\eta') \,,\label{sigfer}\ee
The bosonic ($G^>,G^<$) and fermionic ($\mathcal{G}^>_f,\mathcal{G}^<_f$) correlation functions are obtained in appendix (\ref{app:corres}), these are shown to be functions of $\vx-\vx'$ by translational invariance   of the spatially flat FRW metric, therefore the spatial integral in equation (\ref{1lupeom}) can be done straightforwardly since the condensate $\X$ is homogeneous, yielding
\be \X^{''}(\eta)+\frac{d}{d\X}\widetilde{V}(\X;\eta)+\int^{\eta}_{\eta_0}   \widetilde{\Sigma}(\eta,\eta')\,\X(\eta')\,d\eta'  =0 \,,\label{infeom} \ee this is the final form of the equation of motion for the homogeneous inflaton condensate with radiative corrections from spectator fields up to one loop order and linearized in $\X(\eta)$. The one loop self energy is given by
\be \widetilde{\Sigma}(\eta,\eta') \equiv \widetilde{\Sigma}_s(\eta,\eta') + \widetilde{\Sigma}_f(\eta,\eta') \,,\label{totwitsig}\ee where (see appendix (\ref{app:corres}))
\bea  \widetilde{\Sigma}_s(\eta,\eta') & = & -2i\lambda^2 C(\eta) C(\eta') \int  \Big[g^2_k(\eta)\,g^{*2}_k(\eta')- g^2_k(\eta')\,g^{*2}_k(\eta)\Big]   \frac{d^3k}{(2\pi)^3}  \label{wtsigs} \\ \widetilde{\Sigma}_f(\eta,\eta') & = & -2i y^2 \int   \Big[  f^2_k(\eta)f^{*2}_k(\eta') \,\Omega_k(\eta)\Omega^*_k(\eta')- f^2_k(\eta')f^{*2}_k(\eta) \,\Omega_k(\eta')\Omega^*_k(\eta) \Big] \frac{d^3k}{k^2\,(2\pi)^3}.\label{wtsigf} \eea
These are the general forms of the scalar and fermionic self-energies, where  $g_k(\eta);f_k(\eta),\Omega_k(\eta)$ are given by equations (\ref{gqeta},\ref{fketasolu},\ref{wofketa}) respectively.
\subsection{Inflaton equation of motion from linear response}\label{subsec:eomLR}
The equation of motion for the inflaton condensate (\ref{infeom}) can be obtained in an alternative manner by implementing the theory of linear response, providing a complementary
derivation which   confirms the result from the in-in formulation and provides   further insight. We begin with the Heisenberg field equation for the inflaton field $\chi_I$ as obtained from the full Lagrangian density (\ref{Ltot}) after conformal rescaling, namely
\be \chi^{''}_I(\vx,\eta)-\nabla^2\chi_I(\vx,\eta)+\frac{d \widetilde{V}}{d\chi_I} = - \mathcal{O}(\vx,\eta) \,\label{heisinf}\ee  along with the Heisenberg equations of motion for the spectator fields that enter in the composite operator $\mathcal{O}$. Implementing the shift
\be \chi_I(\vx,\eta) = \X(\eta) + \delta(\vx,\eta) \,\label{shiftI}\ee as in the Schwinger-Keldysh formulation (\ref{shift}), yielding for the equation of motion (\ref{heisinf})
\be \X''(\eta) +  \frac{d \widetilde{V}}{d\chi_I}\Big|_{\chi_I = \X}+ \delta''(\vx,\eta) - \nabla^2\delta(\vx,\eta) +  \frac{d^2 \widetilde{V}}{d^2\chi_I}\Big|_{\chi_I = \X}\,\delta(\vx,\eta) + \cdots = - \mathcal{O}(\vx,\eta) \,\label{shiftedeom}\ee where the dots stand for terms higher order in $\delta(\vx,\eta)$. Taking the expectation value of this equation in the Bunch Davis vacuum for all fields $\ket{0_\delta;0_s;0_F}$ with
\be \bra{0_\delta;0_s;0_F}\delta(\vx,\eta)\ket{0_\delta;0_s;0_F} =0 \,,\label{exvalbd}\ee yields
\be \X''(\eta) +  \frac{d \widetilde{V}}{d\chi_I}\Big|_{\chi_I = \X} = - \langle \mathcal{O}(\vx,\eta)\rangle ~~;~~ \langle \mathcal{O}(\vx,\eta)\rangle \equiv
\bra{0_\delta;0_s;0_F}\mathcal{O}(\vx,\eta)\ket{0_\delta;0_s;0_F} \,. \label{exvaleomh}\ee

We   neglected the contributions $\propto \bra{0_\delta;0_s;0_F}\delta^2(\vx,\eta)\ket{0_\delta;0_s;0_F}  \cdots $ because we focus solely on the contributions from spectator fields. The contribution from the inflaton fluctuations will be discussed  in section (\ref{sec:discussion}).

 The composite operator in the Heisenberg picture is
\be   {\mathcal{O}}(\vx,\eta) = U^{-1}(\eta,\eta_0)\, {\mathcal{O}}(\vx,\eta_0)\,U(\eta,\eta_0) \,,\label{opeOhem}\ee where the unitary time evolution operator $U(\eta,\eta_0)$ given by eqn. (\ref{Uop})  obeys
\be i \frac{d}{d\eta} U(\eta,\eta_0) = H(\eta) U(\eta,\eta_0) ~~;~~ U(\eta_0,\eta_0) =1\,.\label{eomU}\ee It is convenient to pass to the interaction picture, by writing in obvious notation
\be H(\eta) \equiv H_0(\eta) + H_i(\eta) \,,\label{Hsplit}\ee where the total Hamiltonian $H(\eta)$ is given by equation (\ref{conham}), $H_0$ is the free field part and $H_i$ is the interaction term which after the shift (\ref{shiftI}) becomes to leading order in couplings and consistently neglecting the inflaton fluctuations
\be H_i(\eta) = \int d^3 x \X(\eta) \mathcal{O}(\vx,\eta) \,,\label{Hishift}\ee the expectation value of the inflaton $\X(\eta)$ acts as a c-number external source- a ``pump'' term. Writing
\be U(\eta,\eta_0) = U_0(\eta;\eta_0)\,U_I(\eta;\eta_0) ~~;~~  U_0(\eta_0;\eta_0)=1~;~ \,U_I(\eta_0;\eta_0)=1 \,,\label{Usplit}\ee with
\be i \frac{d}{d\eta} U_0(\eta,\eta_0) = H_0(\eta) U_0(\eta,\eta_0) \,, \label{eomUzero}\ee it follows that
\be {\mathcal{O}}(\vx,\eta) = U^{-1}_I(\eta,\eta_0)\, {\mathcal{O}^{(0)}}(\vx,\eta)\,U_I(\eta,\eta_0)\,,\label{Ointpic}\ee where
\be  {\mathcal{O}^{(0)}}(\vx,\eta) = U^{-1}_0(\eta,\eta_0)\, {\mathcal{O}}(\vx,\eta_0)\,U_0(\eta,\eta_0)\,,\label{Ofreefield}\ee  evolves in time with free field time evolution with the free field Hamiltonian $H_0$. The evolution equation for $U_0(\eta,\eta_0)$ (\ref{eomUzero})  combined with the evolution equation (\ref{eomU}) yields
\be   i \frac{d}{d\eta} U_I(\eta,\eta_0) = H_I(\eta) U_I(\eta,\eta_0)~~;~~ H_I(\eta) = U^{-1}_0(\eta,\eta_0) H_i(\eta) U_0(\eta,\eta_0) = \int d^3 x \X(\eta) \mathcal{O}^{(0)}(\vx,\eta) \,. \label{eomUI}\ee The solution of (\ref{eomUI}) in perturbation theory is
\be U_I(\eta,\eta_0) = 1 -i \int^\eta_{\eta_0}  \int d^3 x' \X(\eta')\, \mathcal{O}^{(0)}(\vx',\eta')\,d\eta' + \cdots \,.\label{UIsol} \ee where the dots stand for higher orders in the mean field $\X$ and in the couplings $\lambda,y$ which are included in the operator $\mathcal{O}$. Therefore the Heisenberg field operator (\ref{Ointpic}) to linear order in the mean field $\X(\eta)$ is given by
\be \mathcal{O}(\vx,\eta) = {\mathcal{O}^{(0)}}(\vx,\eta_0) + i \int^\eta_{\eta_0}  \int d^3 x' \X(\eta')\,\Big[\mathcal{O}^{(0)}(\vx',\eta'),\mathcal{O}^{(0)}(\vx,\eta)\Big] \,d\eta'\,,\label{OlinX}\ee from which, the expectation value right hand side of the equation of motion for the inflaton condensate (\ref{exvaleomh}) is given by
\be \langle \mathcal{O}(\vx,\eta) \rangle = \langle \mathcal{O}^{(0)}(\vx,\eta) \rangle + i \int^\eta_{\eta_0}  \int d^3 x' \X(\eta')\,\langle \Big[\mathcal{O}^{(0)}(\vx',\eta'),\mathcal{O}^{(0)}(\vx,\eta)\Big]\rangle   \,d\eta'\,,\label{exvalOlinX}\ee the first term on the right hand side $ \langle \mathcal{O}^{(0)}(\vx,\eta) \rangle=0 $ because the composite operator $\mathcal{O}^{(0)}(\vx,\eta)$ is normal ordered with respect to the Bunch-Davies vacuum state $ \ket{0_\delta;0_s;0_F}$, therefore the equation of motion for the condensate becomes
\be  \X^{''}(\eta)+\frac{d}{d\X}\widetilde{V}(\X;\eta)-i\,  \int^\eta_{\eta_0}  \int d^3 x'  \,\langle \big[\mathcal{O}^{(0)}(\vx,\eta),\mathcal{O}^{(0)}(\vx',\eta') \big]\rangle \,  \X(\eta')  \,d\eta' =0 \,,  \label{eomLR} \ee which is precisely the equation of motion (\ref{1lupeom}) with the self-energy (\ref{sigmaret}) obtained with the Schwinger-Keldysh (in-in) formulation up to one loop.

\subsection{Friction term:}
The effect of spectator fields upon the inflaton condensate is usually described \emph{phenomenologically} via a local friction term in the equation of motion for the conformally unscaled condensate
\be \ddot{\varphi_I}+ \frac{d V(\varphi_I)}{d\varphi_I}+ (3H+\Gamma) \, \dot{\varphi_I} =0 \,,\label{fric}\ee   where the dots stand for derivatives with respect to comoving time. The phenomenological ``friction'' term $\Gamma$ is assumed  to arise from the decay of the inflaton into the spectator fields, and  is usually identified with the decay rate obtained in Minkowski space time, which for a   particle of mass $m$ decaying into two massless bosonic or fermionic  ones are given respectively by
\be \Gamma = \frac{\lambda^2}{8\pi m} ~(\mathrm{bosonic~decay~products}) ~~;~~ \Gamma= \frac{y^2 m}{8\pi^2}~(\mathrm{fermionic~decay~ products}) \,. \label{Minkrate}\ee
One of our objectives is to compare the time evolution of the inflaton condensate with the local friction term (\ref{fric}) and with the non-local one loop self-energy (\ref{totwitsig}). In order to be able to obtain the solutions of the equations of motion in both cases and to compare them, we consider the simplest potential
\be V(\varphi_I) = \frac{1}{2} \,m^2 \,\varphi^2_I  \,, \label{simpot}\ee  yielding
\be  \widetilde{V}(\chi_I;\eta) = \frac{1}{2}\, \Big[\frac{m^2}{H^2}-2\Big]\frac{\chi^2_I}{\eta^2} \,,\label{masit} \ee which allows us to focus on the friction and self-energy corrections neglecting possible non-linearities from the potential $\widetilde{V}(\chi_I;\eta)$, thereby allowing us to obtain exact solutions comparing them explicitly.

For the case of the local friction term   the equation of motion (\ref{fric}) with the potential (\ref{simpot})  can be solved exactly, namely
\be \varphi_I(t) = A\,e^{-\Omega_+ t} + B\,e^{-\Omega_- t}~~;~~ \Omega_\pm = \frac{3\,\widetilde{H}}{2} \,\Bigg[1 \pm \sqrt{1- \frac{4m^2}{9\widetilde{H}^2}}\, \Bigg] \,,\label{solfric}\ee where
\be \widetilde{H} = H\,\Big(1 + \frac{\gamma}{3} \Big) ~~;~~ \gamma = \frac{\Gamma}{H} \,. \label{Htil}\ee Under the conditions $m/H \ll 1; \gamma  \ll 1$ the slow roll solution is
\be \varphi^{sr}_I(t) \simeq B e^{-\widetilde{\kappa} Ht} ~~;~~ \widetilde{\kappa} = \frac{\kappa}{\big(1+ \frac{\gamma}{3} \big)}\simeq \kappa\,\big(1-\frac{\gamma}{3}\big) ~~;~~ \kappa = \frac{m^2}{3H^2} \ll 1\,. \label{slosol}\ee

Passing to conformal time and conformally rescaled field, the equation of motion with the friction term (\ref{fric})  is given by
\be \frac{d^2}{d\eta^2} \,\X(\eta) + \Big( \frac{m^2}{H^2} - 2 \Big)\frac{ \X(\eta)}{\eta^2} -\frac{\gamma}{\eta^2}\,\frac{d}{d\eta} \Big(\eta \X(\eta)\Big) =0 \,,\label{fricon} \ee with general solution
\be \X(\eta) = \mathcal{A}\,\eta^{\alpha_+} + \mathcal{B} \,\eta^{\alpha_-} \,,\label{Xetasr}\ee with
\be \alpha_\pm = \frac{1}{2} \Bigg\{ (1+\gamma) \pm  \big(3+  {\gamma}\big) \sqrt{1- \frac{4m^2}{9\widetilde{H}^2} }\,      \Bigg\} \,.\label{alfas}\ee For $m^2/H^2 \ll 1$ the slow roll solution is obtained setting $\mathcal{A}=0$ and given by
\be \X^{sr}(\eta) = \mathcal{B} \,\eta^{\alpha_-}\simeq \mathcal{B} \,\eta^{-(1-\kappa)}\,\eta^{-\kappa \gamma/3} \,,\label{Xsr}\ee for which
\be \varphi_{Isr}(\eta) \simeq -\mathcal{B} H \eta^{\kappa(1-\gamma/3)} \simeq  \varphi^{(0)}_{Isr}(\eta)\,\eta^{-\frac{\kappa \gamma}{3}}\,\,,\label{slorovarfi}\ee remains nearly constant for $\kappa,\gamma \ll 1$ and $\varphi^{(0)}_{Isr}(\eta)$ is the solution in absence of friction.

The main point of obtaining these solutions is to compare with the result from the correct equation of motion (\ref{infeom}) for bosonic and fermionic spectators.
 With the objective of comparing the dynamics obtained via the phenomenological local friction term and that from the non-local self energy kernel, it is instructive to solve
 the equation of motion in perturbation theory in terms of the small dimensionless ratio $\gamma = \Gamma/H$, writing
 \be \X(\eta) = \X^{(0)}(\eta) + \gamma \X^{(1)}(\eta) + \cdots \,,\label{Xfricex}\ee and solving (\ref{fric}) order by order, yields the set of equations
 \bea  \Big[\frac{d^2}{d\eta^2}  + \frac{ 1}{\eta^2}\Big( \frac{m^2}{H^2} - 2 \Big) \Big]\, \X^{(0)}(\eta) & = & 0 \label{zerofric}\\
  \Big[\frac{d^2}{d\eta^2}  + \frac{ 1}{\eta^2}\Big( \frac{m^2}{H^2} - 2 \Big) \Big]\, \X^{(1)}(\eta)  & = & \frac{1}{\eta^2} \frac{d}{d\eta} \Big(\eta \X^{(0)}(\eta)\Big) \,\label{onefric}\\ \vdots & = & \vdots \eea The general solution of the zeroth order equation (\ref{zerofric}) is
 \be \X^{(0)}(\eta) = \mathcal{A}\,\eta^{\beta_+} + \mathcal{B} \,\eta^{\beta_-} \,,\label{zerofrisol}   \ee with
 \be \beta_\pm = \frac{1}{2} \Bigg\{ 1 \pm  3 \sqrt{1- \frac{4m^2}{9{H}^2} }\,      \Bigg\} \,,\label{betas}\ \ee for $m^2/3H^2 = \kappa \ll1$ the zeroth order slow roll solution corresponds to $\mathcal{A}=0$ and is given by
 \be \X^{(0)}_{sr}(\eta) =  \X^{(0)}_{sr}(\eta_0) \,\Big( \frac{\eta}{\eta_0}\Big)^{\beta_-}  \,,\label{zerothsr}\ee  for $\kappa \ll 1$ it follows that
 \be \beta_- \simeq -(1-\kappa)\,, \label{vetamin}\ee so that during slow roll
 \be \varphi^{(0)}_{Isr}(\eta) = \X^{(0)}_{sr}(\eta_0)\,(-H\eta) \propto \eta^{\kappa} \simeq \mathrm{constant} \,.\label{confit}\ee

 The first order equation (\ref{onefric}) becomes
 \be    \Big[\frac{d^2}{d\eta^2}  + \frac{ 1}{\eta^2}\Big( \frac{m^2}{H^2} - 2 \Big) \Big]\, \X^{(1)}(\eta)  =  \X^{(0)}_{sr}(\eta_0) \,\Big( \frac{\eta}{\eta_0}\Big)^{\beta_-}\,\frac{1}{\eta^2}\, (1+\beta_-)   \,.\label{fistsr} \ee This equation is solved by using the Green's function of the differential operator on its left hand side
 \be \mathbb{G}(\eta,\eta') = \frac{1}{\Delta} \Big[ \eta^{\beta_+}\,\eta^{'\,\beta_-} - \eta^{\beta_-}\,\eta^{'\,\beta_+}\Big]\Theta(\eta-\eta')~~;~~ \Delta = \beta_+ - \beta_- \,,\label{grin}\ee
 \be \X^{(1)}(\eta) = \X^{(0)}_{sr}(\eta_0) \,\Big( \frac{1}{\eta_0}\Big)^{\beta_-} \, (1+\beta_-)\, \int^0_{\eta_0} \mathbb{G}(\eta,\eta')\, \eta^{'\,(\beta_--2)}\,d\eta' \simeq  -\X^{(0)}_{sr}(\eta_0)\,\frac{\kappa}{3}\, \,\Big( \frac{\eta}{\eta_0}\Big)^{\beta_-}\, \ln\Big( \frac{\eta}{\eta_0}\Big)\,, \label{X1sol}\ee where we have kept the leading order in $\kappa \ll 1$ and neglected subdominant (non-logarithmic) terms in the limit $\eta \rightarrow 0$. Combining with the zeroth order slow roll solution (\ref{zerothsr})
  we find
 \be \X^{sr}(\eta) =  \X^{(0)}_{sr}(\eta_0)\,\Big(\frac{\eta}{\eta_0}\Big)^{\beta_-}\, \Big[ 1- \frac{\kappa \gamma}{3} \, \ln\big( \frac{\eta}{\eta_0}\big) +\cdots \Big] \,.\label{soluln}\ee

This is precisely what one obtains by expanding   the slow roll solution (\ref{Xsr}) in the small ratios $\kappa;\gamma \ll1 $. This form of the solution will allow us to compare directly to the perturbative solution of (\ref{infeom}).

We now consider the radiative corrections from spectator fields determined by the last term in the equation of motion (\ref{infeom}) by analyzing separately the contributions from bosonic and fermionic fields to the self energy in various relevant cases. In order  to compare to the exact result with the phenomenological friction term, we consider the simplest potential (\ref{simpot}), for which the equation of motion (\ref{infeom}) simplifies to
\be \X^{''}(\eta) + \Big( \frac{m^2}{H^2} - 2 \Big)\frac{ \X(\eta)}{\eta^2}+ \int^{\eta}_{\eta_0}   \widetilde{\Sigma}(\eta,\eta')\,\X(\eta')\,d\eta'  = 0 \,. \label{lineom}\ee

\subsection{Bosonic fields:}
The quantization of bosonic and fermionic fields in sections (\ref{subsec:bosespec},\ref{subsec:ferspec}) and the one-loop self-energies (\ref{wtsigs},\ref{wtsigf}) are general, however in this study we focus on the case when the scalar spectator fields are massless and conformally coupled to gravity, namely with  $M_s=0 ~;~ \xi_s =\frac{1}{6}~;~ \nu_s = \frac{1}{2}$ in equations (\ref{gmodes},\ref{nusa}). This case  corresponds to the mode functions
\be g^{(0)}_q(\eta) = \frac{e^{-iq\eta}}{\sqrt{2q}} \,,\label{gqzero}\ee these are the ``closest'' to the Minkowski mode functions. With these mode functions the quantized scalar spectator free field and its canonical conjugate are given by
\bea \chi_s(\vx,\eta) & = & \frac{1}{V}\sum_{\vk} \frac{1}{\sqrt{2k}}\Big[a_{\vk} \,  e^{-ik\eta}  + a^\dagger_{-\vk} \, e^{ik\eta} \Big]\,e^{i\vk\cdot\vx}\,,\label{chizero}\\
\pi_s(\vx,\eta) & = & \frac{-i}{V}\sum_{\vk} \frac{k}{\sqrt{2k}}\Big[a_{\vk} \,  e^{-ik\eta}  - a^\dagger_{-\vk} \, e^{ik\eta} \Big]\,e^{i\vk\cdot\vx}\,.\label{pizero}
\eea

This case is of particular relevance for comparison with   Minkowski space time,  and effectively describes the limit of deeply sub-horizon modes even for  the most general case with $M_s \neq 0; \xi_s \neq 1/6$. Furthermore, as discussed in references\cite{parker1,kolb,parker,ford,BD,fulling,long}, there is no gravitational particle production for conformally coupled massless bosonic fields,   allowing us to understand particle production of these fields solely from their interaction with the inflaton.

The bosonic spectator contribution to the self-energy is given by eqn. (\ref{wtsigs}) with the above form of the mode functions, namely
 \be \widetilde{\Sigma}_s(\eta,\eta') = -\frac{2i\lambda^2}{H^2\,\eta\,\eta'} \,\int \Big[ \frac{e^{-2i(\eta-\eta')}-e^{2i(\eta-\eta')}}{4k^2}\Big]\,\frac{d^3k}{(2\pi)^3}\,.\label{sigbosi} \ee

 The momentum integral in (\ref{sigbosi}) is carried out by introducing convergence factors $\varepsilon \rightarrow 0^+$
\be \frac{1}{8\pi^2} \int^{\infty}_0 \Big[e^{-2ik(\eta-\eta' -i\varepsilon)}- e^{2ik(\eta-\eta' +i\varepsilon)}  \Big] dk = \frac{-i}{8\pi^2} \Bigg[ \frac{\eta-\eta'}{(\eta-\eta')^2+\varepsilon^2}\Bigg]\,,\label{integbos}\ee yielding
\be \widetilde{\Sigma}_s(\eta,\eta')  =  -\frac{\lambda^2}{4\pi^2 H^2\,\eta\,\eta'}\,\Bigg[ \frac{\eta-\eta'}{(\eta-\eta')^2+\varepsilon^2}\Bigg] \equiv \frac{\alpha^2}{\eta\,\eta'} \,\frac{d}{d\eta'} \, \ln\Big[\frac{(\eta-\eta')^2+\varepsilon^2}{(-\eta^*)^2}  \Big] ~~;~~ \alpha^2 = \frac{\lambda^2}{8\pi^2 H^2}\,,\label{tilsigbos}     \ee and we introduced the renormalization scale $(-\eta^*)^2$ to render dimensionless the logarithm, of course nothing depends on this scale. The integral of the self energy in the equation of motion (\ref{lineom}) becomes
\bea \int^{\eta}_{\eta_0}   \widetilde{\Sigma}_s (\eta,\eta')\,\X(\eta')\,d\eta'  & = & {\alpha^2} \,\ln\Big( \frac{\varepsilon^2}{(-\eta^*)^2}\Big)\,\frac{\X(\eta)}{\eta^2} -  \frac{\alpha^2}{\eta\,\eta_0} \,\ln\Big[\frac{(\eta-\eta_0)^2+\varepsilon^2}{(-\eta^*)^2}  \Big]\,\X(\eta_0) \nonumber \\
& -& \frac{\alpha^2}{\eta}\,\int^{\eta}_{\eta_0}  \ln\Big[\frac{(\eta-\eta')^2}{(-\eta^*)^2}  \Big]\,\frac{d}{d\eta'} \Big( \frac{\X(\eta')}{\eta'}\Big)\, d\eta' \,,\label{radbos}\eea in the integral term in the second line we have taken $\varepsilon \rightarrow 0$ since the short distance singularity as $\eta'\rightarrow \eta$ is integrable. The first term on the right hand side is a renormalization of the inflaton mass as can be immediately seen in the equation of motion (\ref{lineom})
\be \frac{m^2}{H^2}+  \frac{\lambda^2}{8\pi^2 H^2} \,\ln\Big( \frac{\varepsilon^2}{(-\eta^*)^2}\Big) \equiv \frac{m^2_R}{H^2}\,.\label{renmass1}\ee Although nothing depends on the renormalization scale $\eta^*$, it is convenient to choose it as $\eta^* = \eta_0$, since with this choice the second boundary term on the right hand side of (\ref{radbos}) is $\propto \alpha^2/\eta^2_0$  in the limit $\eta \rightarrow 0$ it  remains constant and   is negligible for $ -H\eta_0 \gg 1$ therfore it can be safely neglected. After mass renormalization, and neglecting the boundary term, the equation of motion (\ref{lineom}) becomes
\be   \Big[\frac{d^2}{d\eta^2}  + \frac{ 1}{\eta^2}\Big( \frac{m^2_R}{H^2} - 2 \Big) \Big]\, \X(\eta)=  \frac{\alpha^2}{\eta}\,\int^{\eta}_{\eta_0}  \ln\Big[\frac{(\eta-\eta')^2}{(-\eta_0)^2}  \Big]\,\frac{d}{d\eta'} \Big( \frac{\X(\eta')}{\eta'}\Big)\, d\eta'\, ,\label{eominte}\ee for which we seek a perturbative solution of the form
\be \X(\eta) = \X^{(0)}(\eta) + \alpha^2 \X^{(1)}(\eta) + \cdots \label{perbos}\ee  leading to a hierarchy of equations
\bea  \Big[\frac{d^2}{d\eta^2}  + \frac{ 1}{\eta^2}\Big( \frac{m^2_R}{H^2} - 2 \Big) \Big]\, \X^{(0)}(\eta) & = & 0 \,, \label{zerothX}\\
 \Big[\frac{d^2}{d\eta^2}  + \frac{ 1}{\eta^2}\Big( \frac{m^2_R}{H^2} - 2 \Big) \Big]\, \X^{(1)}(\eta) & = &     \frac{1}{\eta}\,\int^{\eta}_{\eta_0}  \ln\Big[\frac{(\eta-\eta')^2}{(-\eta_0)^2}  \Big]\,\frac{d}{d\eta'} \Big( \frac{\X^{(0)} (\eta')}{\eta'}\Big)\, d\eta'\,,   \label{firsX}\\
 \vdots & = & \vdots \label{highord} \eea The zeroth order equation is the same as (\ref{zerofric}) with slow roll solution given by (\ref{zerothsr}) but with $\beta_-$ (\ref{betas}) in terms of the renormalized mass $m_R$. With the zeroth order slow roll solution (\ref{zerothsr}), the source on the right hand side of (\ref{firsX}) is simple to evaluate, in the limit $\eta/\eta_0 \rightarrow 0$  equation (\ref{firsX}) becomes
 \be   \Big[\frac{d^2}{d\eta^2}  + \frac{ 1}{\eta^2}\Big( \frac{m^2_R}{H^2} - 2 \Big) \Big]\, \X^{(1)}(\eta)  =  \frac{2}{\eta^2} \,\X^{(0)}_{sr}(\eta_0)\,\Big(\frac{\eta}{\eta_0}\Big)^{\beta_-} \,\Big[ \ln\Big(\frac{\eta}{\eta_0}\Big) -1\Big] \,, \label{sourcefisX} \ee where to leading order in $\alpha^2;\kappa$ we have set $\kappa \simeq 0$ in terms inside the bracket. Comparing the first order equations (\ref{fistsr},\ref{sourcefisX}) is already evident that the radiative contribution from the self-energy ie \emph{not} equivalent to a local friction term, radiative corrections from spectators feature a stronger divergence as $\eta/\eta_0 \rightarrow 0$.

 With the Green's function (\ref{grin}) we find to leading  order in $\kappa$ and leading logarithms as $\eta/\eta_0 \rightarrow 0$
 \be \X^{(1)}(\eta) = - \frac{1}{3}\,\X^{(0)}_{sr}(\eta_0)\,\Big(\frac{\eta}{\eta_0}\Big)^{\beta_-} \Big[  \ln^2\Big( \frac{\eta}{\eta_0} \Big)-\frac{4}{3}\,\ln\Big( \frac{\eta}{\eta_0} \Big)+\cdots\Big] \,. \label{X1bos}\ee Therefore up to order $\alpha^2$ the solution is
 \be \X_{sr}(\eta) = \X^{(0)}_{sr}(\eta_0)\,\Big(\frac{\eta}{\eta_0}\Big)^{\beta_-}\Bigg\{1 - \frac{\alpha^2}{3}\,\Big[  \ln^2\Big( \frac{\eta}{\eta_0} \Big)-\frac{4}{3}\,\ln\Big( \frac{\eta}{\eta_0} \Big)+\cdots\Big] + \cdots  \Bigg\}\,,\label{solubosfin}\ee which is obviously very different from the solution (\ref{soluln}) with a local friction term.

\subsection{Fermion fields:}\label{subsec:fermions}

We now consider the fermion self energy with  massless Fermi fields, setting $m_f=0$ in equations (\ref{diraceqn},\ref{eqnfketa}), yielding
\be f_k(\eta) = e^{-ik\eta}\,\label{fofkzero}\ee this is the ``closest'' to Minkowski space time. Again in this case, there is no gravitational production of particles, just as in the massless conformally coupled bosonic case, and allows us to compare directly to Minkowski space time, exhibiting in the simplest and most clear manner the effects of cosmological expansion in the radiative corrections to the equations of motion of the inflaton.   Replacing this solution along with $\Omega_k(\eta)= k $ into the fermion self-energy (\ref{wtsigf}) yields

\bea  \widetilde{\Sigma}_f(\eta,\eta')   & = &   -2i y^2 \int   \Big[e^{-2ik(\eta-\eta'-i\varepsilon)}- e^{2ik(\eta-\eta'+i\varepsilon)}   \Big] \frac{d^3k}{(2\pi)^3}   =  \frac{y^2}{4\pi^2}\,\frac{d^2}{d\eta^{'\,2}}\,\Big[ \frac{\eta-\eta'}{(\eta-\eta')^2+\varepsilon^2} \Big] \nonumber \\ & = & -\frac{y^2}{8 \pi^2}\,\frac{d^3}{d\eta^{'\,3}}\,\ln\Big[ \frac{(\eta-\eta')^2+\varepsilon^2}{(-\eta_0)^2} \Big]\,,  \label{sigferzero}\eea where we introduced a convergence factor $\varepsilon \rightarrow 0^+$    as in the bosonic case (\ref{integbos}) and set the renormalization scale  to $\eta_0$.

Integrating by parts and neglecting boundary terms at the initial time $\eta_0$ that vanish   $\propto 1/\eta^2_0,1/\eta_0$ as $\eta \rightarrow 0, \eta_0 \rightarrow -\infty$, we find
\be \int^{\eta}_{\eta_0}  \widetilde{\Sigma}_f(\eta,\eta') \,\X(\eta')\,d\eta' = -\frac{y^2}{4\pi^2}\,\frac{\X(\eta)}{\varepsilon^2}+ \frac{y^2}{8\pi^2}\,\ln\Big[ \frac{\eta^2_0}{\varepsilon^2}\Big]\,\frac{d^2}{d\eta^2}\X(\eta)+\frac{y^2}{8\pi^2}\,\int^{\eta}_{\eta_0} \ln\Big[ \frac{(\eta-\eta')^2}{(-\eta_0)^2} \Big]\,\frac{d^3}{d\eta^{\,' 3}}\X(\eta')\,d\eta' \,,\label{intsigfer}\ee where in the last term we have set $\varepsilon \rightarrow 0^+$ since the logarithmic singularity as $\eta' \rightarrow \eta$ is integrable. The first two terms in the right hand side of equation (\ref{intsigfer}) require renormalization, interpreting $\epsilon \equiv 1/\Lambda$ and $-\eta_0 = 1/\mu$ with $\Lambda$ an ultraviolet (short distance) cutoff and $\mu$ a renormalization scale, the first term suggests a mass renormalization $\simeq \Lambda^2 \X(\eta)$ as is the case in Minkowski space time if the momentum integral in the self-energy contribution from the fermion loop is cut off with $\Lambda$ (see the next subsection below). However in de Sitter space-time in conformal time and upon conformal rescaling, a bosonic mass term must be $\propto 1/\eta^2$, whereas the first term does not feature such dependence on $\eta$. This is an artifact of the regularization procedure, for example, in dimensional regularization there are no powers of an ultraviolet cutoff, all ultraviolet divergences appear as poles in $D-4$ with  $D$ the space-time dimension, a single pole in $D-4$ is associated with a logarithmic divergence. We will neglect this divergence, by assuming that a proper counterterm has been introduced in the Lagrangian to cancel the first term in (\ref{intsigfer}). We now write the equation of motion (\ref{lineom}) as
\be \X^{''}(\eta) Z^{-1}_{\chi} + \Big( \frac{m^2}{H^2} - 2 \Big)\frac{ \X(\eta)}{\eta^2}  =  -\frac{y^2}{8\pi^2}\,\int^{\eta}_{\eta_0} \ln\Big[ \frac{(\eta-\eta')^2}{(-\eta_0)^2} \Big]\,\frac{d^3}{d\eta^{\,' 3}}\X(\eta')\,d\eta' \,,\label{eomfirm}   \ee where we have introduced the wavefunction (field) renormalization
\be Z^{-1}_{\chi} = 1+\frac{y^2}{8\pi^2}\,\ln\Big[ \frac{\eta^2_0}{\varepsilon^2}\Big]\,.  \label{zeta}\ee Renormalization proceeds by defining the renormalized fields, mass and coupling, to do this consistently, we restore the inflaton coupling to gravity $m^2 \rightarrow m^2+ (\xi_I-1/6)R$ and define the renormalized variables

\be \X_R(\eta) = \frac{\X(\eta)}{\sqrt{Z_\chi}}~~;~~ m^2_R + (\xi_R-1/6)R  = Z_{\chi}\,\big(m^2 +(\xi_I-1/6)R\big) ~~;~~   y_R = \sqrt{Z_{\chi}}\,y \Rightarrow y\,\X= y_R\,\X_R\,,\label{rentot}\ee and choose the renormalized $\xi_R=0$ as minimal coupling. The renormalized equation of motion now becomes
\be \X^{''}_R(\eta)   + \Big( \frac{m^2_R}{H^2} - 2 \Big)\frac{ \X_R(\eta)}{\eta^2}  =  -\frac{y^2_R}{8\pi^2}\,\int^{\eta}_{\eta_0} \ln\Big[ \frac{(\eta-\eta')^2}{(-\eta_0)^2} \Big]\,\frac{d^3}{d\eta^{\,' 3}}\X_R(\eta')\,d\eta' \,.\label{reneomfer} \ee We follow the procedure outlined above for bosonic fields,    and write
\be \X_R(\eta) = \X^{(0)}_R(\eta) +  \frac{y^2_R}{8\pi^2}\,\X^{(1)}(\eta) + \cdots \,,\label{chiren}\ee yielding a hierarchy of equations similar to (\ref{zerothX}-\ref{highord}), the zeroth order slow roll solution is given by (\ref{zerothsr}) but with $\beta_-$ in terms of the renormalized mass $m_R$. The first order correction obeys
\be  \Big[\frac{d^2}{d\eta^2}  + \frac{ 1}{\eta^2}\Big( \frac{m^2_R}{H^2} - 2 \Big) \Big]\, \X^{(1)}_R(\eta) =  -  \X^{(0)}_{sr}(\eta_0) \,\Big( \frac{1}{\eta_0}\Big)^{\beta_-}\, \beta_- \,(\beta_- -1)\,(\beta_- - 2)\, \int^{\eta}_{\eta_0} \ln\Big[ \frac{(\eta-\eta')^2}{(-\eta_0)^2}\Big] \,\eta^{'\,\beta_- -3} d\eta' \,,\label{chi1fer}\ee

In the limit $\eta/\eta_0 \rightarrow 0$ the integral on the right hand side can be done straightforwardly, and to leading order, taking $\kappa \simeq 0$ yields the first order equation
\be  \Big[\frac{d^2}{d\eta^2}  + \frac{ 1}{\eta^2}\Big( \frac{m^2_R}{H^2} - 2 \Big) \Big]\, \X^{(1)}_R(\eta) =  -  \frac{4}{\eta^2} \,\X^{(0)}_{sr}(\eta_0)\,\Big(\frac{\eta}{\eta_0}\Big)^{\beta_-} \,\Big[ \ln\Big(\frac{\eta}{\eta_0}\Big) - \frac{3}{2}\Big] \,.\label{chi1ferfin}\ee Again, using the Green's function (\ref{grin}), we find to leading order and logarithms in the limit $\eta_0/\eta \rightarrow \infty$
\be \X^{(1)}_R(\eta) =  \frac{2}{3}\,\X^{(0)}_{sr}(\eta_0)\,\Big(\frac{\eta}{\eta_0}\Big)^{\beta_-} \Big[  \ln^2\Big( \frac{\eta}{\eta_0} \Big)-\frac{7}{3}\,\ln\Big( \frac{\eta}{\eta_0} \Big)+\cdots\Big] \,. \label{X1fer}\ee Therefore up to $\mathcal{O}(y^2_R)$ the solution for the fermionic case is
\be    \X_{sr}(\eta) = \X^{(0)}_{sr}(\eta_0)\,\Big(\frac{\eta}{\eta_0}\Big)^{\beta_-}\Bigg\{1 + \frac{y^2_R}{12\pi^2}\,\Big[  \ln^2\Big( \frac{\eta}{\eta_0} \Big)-\frac{7}{3}\,\ln\Big( \frac{\eta}{\eta_0} \Big)+\cdots\Big] + \cdots  \Bigg\}\,,\label{soluferfin}\ee

\subsection{Minkowski space-time:}\label{subsec:mink}

It is illuminating to compare to the case of Minkowski space time, where a local friction term is a reasonable approximation to the dynamics in weak coupling and at long time to manifestly exhibit how cosmological expansion modifies the dynamics.

The Minkowski space-time case is obtained by   replacing
\be \eta \rightarrow t~;~ \eta_0 \rightarrow t_0 ~ ; ~ C(\eta) \rightarrow 1\, , \label{minkrepla}\ee the latter now implies time translational invariance. The equation of motion for the condensate now reads
\be \frac{d^2}{dt^2} \X(t) + m^2 \,  \X(t)+ \int^t_{t_0} \widetilde{\Sigma}(t-t') \,\X(t')\,dt' =0 \,, \label{minkeom}\ee we treat   the bosonic and fermionic cases separately to highlight the subtle aspects of renormalization for the fermionic spectator field.

\subsubsection{Bosonic spectator:}\label{subsecboso}

The bosonic self-energy is simply given by eqn. (\ref{tilsigbos}) with the replacements (\ref{minkrepla}),   namely
\be \widetilde{\Sigma}(t-t') = - \frac{\lambda^2}{4\pi^2}\, \frac{t-t'}{(t-t')^2+\epsilon^2}= \frac{\lambda^2}{8\pi^2}\,\frac{d}{dt'}\,\ln\Big[\frac{(t-t')^2+\epsilon^2 }{t^2_*}\Big]~~;~~ \epsilon \rightarrow 0\,,\label{SEMink}\ee  where $t_*$ is a   scale introduced to render dimensionless the argument of the logarithm. Integrating by parts
\be \int^t_{t_0} \widetilde{\Sigma}(t-t') \,\X(t')\,dt' = \frac{\lambda^2}{8\pi^2}\,\Bigg\{\ln\Big[ \frac{\epsilon^2}{t^2_*}\Big]\,\X(t)- \ln\Big[\frac{(t-t_0)^2+\epsilon^2 }{t^2_*}\Big]\,\X(t_0)-\int^t_{t_0} \ln\Big[\frac{(t-t')^2+\epsilon^2 }{t^2_*}\Big]\,\frac{d}{dt'}\X(t')\,dt'\Bigg\}\,,\label{intsigbos}\ee the first (local) term is a renormalization of the mass, defining
\be \Lambda = 1/\epsilon~;~ \mu = 1/t_* \,, \label{parasbos}\ee and the renormalized mass by
\be m^2_R = m^2-\frac{\lambda^2}{4\pi^2}\,\ln\Big[\frac{\Lambda}{\mu} \Big] \,,\label{mRbosi}\ee
  in terms of which the equation of motion (\ref{minkeom}) becomes
\be \frac{d^2}{dt^2} \X(t) + m^2_R \,  \X(t)=  \frac{\lambda^2}{8\pi^2}\Bigg\{ \ln\Big[\frac{(t-t_0)^2+\epsilon^2 }{t^2_*}\Big]\,\X(t_0)+\int^t_{t_0} \ln\Big[\frac{(t-t')^2+\epsilon^2 }{t^2_*}\Big]\,\frac{d}{dt'}\X(t')\,dt'\Bigg\}\,. \label{nueombos}\ee

 Just as in the cosmological case, we seek a perturbative solution of (\ref{minkeom})
  \be \X(t) = \X^{(0)}(t)+\frac{\lambda^2}{8\pi^2} \,\X^{(1)}(t) + \cdots \label{minkpersol}\ee yielding the hierarchy
  \bea \Big[ \frac{d^2}{d t^2} + m^2_R\Big]\,\X^{(0)}(t) & = & 0 \,,\nonumber\\
\Big[ \frac{d^2}{d t^2} + m^2_R\Big]\,\X^{(1)}(t)  & = &   \ln\Big[\frac{(t-t_0)^2+\epsilon^2 }{t^2_*}\Big]\,\X^{(0)}(t_0)+\int^t_{t_0} \ln\Big[\frac{(t-t')^2+\epsilon^2 }{t^2_*}\Big]\,\frac{d}{dt'}\X^{(0)}(t')\,dt'  \nonumber\\
 \vdots ~~~~~~ & = & ~~~~~~\vdots \eea

The solution to the zeroth order equation   is
\be \X^{(0)}(t) = X\,e^{-im_Rt}+X^*\,e^{im_Rt} \,,\label{X0sol} \ee where $X$ is a complex amplitude. In this case it is more convenient to integrate by parts the right hand side of the first order equation, yielding
\be \Big[ \frac{d^2}{d t^2} + m^2_R\Big]\,\X^{(1)}(t)   =    \ln\Big[\frac{\epsilon^2}{t^2_*}\Big]\,\X^{(0)}(t)+ 2\int^t_{t_0} \Bigg[\frac{(t-t')}{(t-t')^2+\epsilon^2}\Big]\, \X^{(0)}(t')\,dt'\,.\label{firstordbos}\ee  The integral on the right hand side is straightforward yielding
\be \Big[ \frac{d^2}{d t^2} + m^2_R\Big]\,\X^{(1)}(t)   =  2\,\Bigg\{X\,e^{-im_R t} \mathcal{F}(t) + X^* \,e^{im_R t}\,\mathcal{F}^*(t)   \Bigg\} \,,\label{fireq}\ee where

\be \mathcal{F}(t) =  \ln\Big( \frac{\mu}{m_R}\Big)-\gamma + Ci[m_RT] + i \, Si[m_RT] ~~;~~ T= t-t_0 \,,\label{finft} \ee where $\mu = 1/t_*$,  $\gamma=0.577\cdots$ is   Euler's constant, and $Ci,Si$ are the cosine and sine integral functions respectively, which feature the behavior
\be Ci[m_RT]~~{}_{\overrightarrow{m_RT \gg 1 }} ~~ 0 ~~;~~ Si[m_RT]~~{}_{\overrightarrow{m_RT \gg 1 }} ~~ \frac{\pi}{2} \,.\label{asyfot}\ee Therefore, in the long time limit
\be \mathcal{F}(t) ~~{}_{\overrightarrow{m_RT \gg 1 }} ~~  \ln\Big( \frac{\mu}{m_R}\Big)-\gamma + i\,\frac{\pi}{2} \,.\label{flontime}\ee We note that after mass renormalization, the right hand side of the first order equation is ultraviolet finite.

The solution to the first order equation (\ref{fireq}) is
\be \X^{(1)}(t) = 2\,\int^\infty_{t_0}  G_R(t-t')\,\Big[X\,e^{-im_Rt'} \, \mathcal{F}(t')+X^*\, e^{im_Rt'}\,\mathcal{F}^*(t') \Big] dt' \,,\label{X1solmin}\ee where the retarded Green's function of the differential operator on the left hand side of (\ref{fireq}) is
\be G_R(t-t') = \frac{1}{m_R}\,\sin[m_R(t-t')]\,\Theta(t-t') \,.\label{Gretmink}\ee Therefore up to order $ {\lambda}^2$ we find
\bea && \X(t)  =  X\,e^{-im_Rt}\,\Big[1+ i \frac{ \lambda^2}{8\pi^2\,m_R}\,\int^t_{t_0}\mathcal{F}(t')\,dt' \Big] + X^*\,e^{im_Rt}\,\Big[1- i \frac{ \lambda^2}{8\pi^2\,m_R}\,\int^t_{t_0} \mathcal{F}^*(t')\,dt' \Big] \nonumber \\
& - & i\,\frac{ \lambda^2}{8\pi^2\,m_R}\, X \,e^{im_Rt}\,  \int^t_{t_0}  e^{-2im_Rt'} \mathcal{F}(t')\,dt'   + i\,\frac{ \lambda^2}{8\pi^2\,m_R} X^*\,e^{-im_Rt}\, \int^t_{t_0} e^{2im_Rt'} \mathcal{F}^*(t')\,dt' \,.\label{X1fini}\eea     In the long time limit the first line features   secular terms that grow linearly in time from the integrals as a consequence of the long time limit (\ref{flontime}), whereas the contributions from the second line are not secular, rapidly oscillating  and bounded in time in this limit.

\subsubsection{Fermionic spectator:}\label{subsecfer}

The fermionic self energy is obtained from (\ref{sigferzero}) by the replacement (\ref{minkrepla}), yielding
  \bea  \widetilde{\Sigma}_f(t-t')     =     \frac{y^2}{4\pi^2}\,\frac{d^2}{dt^{'\,2}}\,\Big[ \frac{t-t'}{(t-t')^2+\epsilon^2} \Big]   =   -\frac{y^2}{8 \pi^2}\,\frac{d^3}{dt^{'\,3}}\,\ln\Big[ \frac{(t-t')^2+\epsilon^2}{t^2_*} \Big]\,.  \label{sigfermink}\eea Integrating by parts, and neglecting boundary terms $\propto 1/(t-t_0),1/(t-t_0)^2 \rightarrow 0$ as $t-t_0 \rightarrow \infty$,  we find
  \bea  && \int^t_{t_0} \widetilde{\Sigma}_f(t-t') \,X(t')\,dt'  =  \frac{y^2}{4\pi^2} \Bigg\{ -\frac{\X(t)}{\epsilon^2} + \ln \Big[ \frac{\Lambda}{\mu}\Big]\,\frac{d^2\X(t)}{dt^2}  \nonumber \\ & + & \frac{1}{2}\,\ln\Big[  \frac{(t-t_0)^2+\epsilon^2}{t^2_*} \Big]\,\frac{d^2\X(t)}{dt^2}\Big|_{t_0} + \frac{1}{2} \int^t_{t_0} \ln\Big[  \frac{(t-t')^2+\epsilon^2}{t^2_*} \Big]\,\frac{d^3 \X(t')}{dt^{\,' 3}} \,dt' \Bigg\} \,,\label{intsigmin} \eea where we used the definitions  (\ref{parasbos}).

  Therefore, the equation of motion (\ref{minkeom}) becomes
  \be  Z^{-1}_{\X}\frac{d^2}{dt^2} \X(t) + m^2_1 \,  \X(t)  = -\frac{y^2}{8\pi^2} \Bigg\{  \ln\Big[  \frac{(t-t_0)^2+\epsilon^2}{t^2_*} \Big]\,\frac{d^2\X(t)}{dt^2}\Big|_{t_0}+  \int^t_{t_0} \ln\Big[  \frac{(t-t')^2+\epsilon^2}{t^2_*} \Big]\,\frac{d^3 \X(t')}{dt^{\,' 3}} \,dt' \Bigg\}\,, \label{ferminkeom}   \ee where
  \be m^2_1 = m^2- \frac{y^2}{4\pi^2\,\epsilon^2} ~~; ~~ Z^{-1}_{\X} = 1+ \frac{y^2}{4\pi^2}\, \ln \Big[ \frac{\Lambda}{\mu}\Big]\,.\label{ferparas}\ee Introducing the renormalized field,  mass and Yukawa coupling
  \be \X_R(t) = \frac{\X(t)}{\sqrt{Z_{\X}}}~~;~~ m^2_R = m^2_1\,Z_{\X}~~;~~ y_R = \sqrt{Z_{\X}}\,y \Rightarrow y\X(t) = y_R \X_R(t) \,, \label{minkrens}\ee the equation of motion (\ref{ferminkeom}) is fully renormalized and becomes
  \be   \Big[\frac{d^2}{dt^2}+ m^2_R \Big] \X_R(t)    = -\frac{y^2_R}{8\pi^2} \Bigg\{  \ln\Big[  \frac{(t-t_0)^2+\epsilon^2}{t^2_*} \Big]\,\frac{d^2\X_R(t)}{dt^2}\Big|_{t_0}+  \int^t_{t_0} \ln\Big[  \frac{(t-t')^2+\epsilon^2}{t^2_*} \Big]\,\frac{d^3 \X_R(t')}{dt^{\,' 3}} \,dt' \Bigg\}\,. \label{ferminkeomren}   \ee Proposing a perturbative expansion
  \be \X_R(t) = \X_R^{(0)}(t) + \frac{y^2_R}{8\pi^2} \X^{(1)}_R(t) +\cdots \,,\label{expafer}\ee yields the hierarchy
  \bea \Big[ \frac{d^2}{d t^2} + m^2_R\Big]\,\X^{(0)}_R(t) & = & 0 \,,\nonumber\\
\Big[ \frac{d^2}{d t^2} + m^2_R\Big]\,\X^{(1)}_R(t)  & = &    -\ln\Big[  \frac{(t-t_0)^2+\epsilon^2}{t^2_*} \Big]\,\frac{d^2\X^{(0)}_R(t)}{dt^2}\Big|_{t_0}-  \int^t_{t_0} \ln\Big[  \frac{(t-t')^2+\epsilon^2}{t^2_*} \Big]\,\frac{d^3 \X^{(0)}_R(t')}{dt^{\,' 3}} \,dt'  \nonumber\\
 \vdots ~~~~~~ & = & ~~~~~~\vdots \eea
 The zeroth order solution is
 \be  \X^{(0)}_R(t) = X_R \,e^{-im_Rt}+X^*_R\,e^{im_Rt} \,,\label{X0solfer} \ee and using the zeroth order equation to write $d^2\X^{(0)}_R(t)/dt^2 = -m^2_R \X^{(0)}_R(t)$, integrating by parts and using the results leading up to equation (\ref{fireq}), we find, remarkably

\be \Big[ \frac{d^2}{d t^2} + m^2_R\Big]\,\X^{(1)}_R(t)   =  2\,m^2_R \,\Bigg\{X_R\,e^{-im_R t} \mathcal{F}(t) + X^*_R \,e^{im_R t}\,\mathcal{F}^*(t)   \Bigg\} \,,\label{fireqfer}\ee where $\mathcal{F}(t)$ is given by (\ref{finft}), yielding, similarly to the result (\ref{X1fini})
\bea && \X_R(t)  =  X_R\,e^{-im_Rt}\,\Big[1+ i \,\frac{ y^2_R\,m_R}{8\pi^2}\,\int^t_{t_0}\mathcal{F}(t')\,dt' \Big] + X^*_R\,e^{im_Rt}\,\Big[1- i \,\frac{ y^2_R\,m_R}{8\pi^2}\,\int^t_{t_0} \mathcal{F}^*(t')\,dt' \Big] \nonumber \\
& - & i\,\frac{ y^2_R\,m_R}{8\pi^2}\, X_R \,e^{im_Rt}\,  \int^t_{t_0}  e^{-2im_Rt'} \mathcal{F}(t')\,dt'   + i\,\frac{ y^2_R\,m_R}{8\pi^2}\,X^*_R\,e^{-im_Rt}\, \int^t_{t_0} e^{2im_Rt'} \mathcal{F}^*(t')\,dt' \,.\label{X1finifer}\eea

    As in   the solution (\ref{X1fini}), in the long time limit the first line features   secular terms that grow linearly in time   the second line contributions are not secular, rapidly oscillating  and bounded in time in this limit. We note that the final expressions are the same for bosons and fermions with the only replacement
     \be \frac{ \lambda^2}{8\pi^2\,m_R} ~(\mathrm{bosons}) \Rightarrow \frac{ y^2_R\,m_R}{8\pi^2}~(\mathrm{fermions}) \,.\label{coupes}\ee

     The similarities and differences between the Minkowski and de Sitter space results are truly noteworthy: the similarity of the self energy kernels by the replacement $t \leftrightarrow \eta$ in both cases, bosons and fermions is a consequence of the mode functions (\ref{gqzero}) for the   massless bosons   conformally coupled to gravity and (\ref{fofkzero}) for massless fermions, \emph{this is one of the main reasons for choosing these cases}. The differences are manifestly in the vertices, including $C(\eta)$ in the bosonic case and the time evolution of the unperturbed condensate which is given by (\ref{zerothsr}) in de Sitter and (\ref{X0sol}) in Minkowski space-times.   These differences bear a striking impact in the evolution of the condensate: whereas in Minkowski space time the secular terms arising from radiative corrections either for bosons or fermions grow \emph{linearly} in time and feature the \emph{same sign}, leading to the expectation that the De Sitter counterpart would grow as $\simeq |\ln(\eta)|$ because $\eta \propto e^{-Ht}$,  instead the secular terms in de Sitter grow as $\ln^2(\eta)$ in the long time ($\eta \rightarrow 0$) limit. We refer to this behavior as \emph{Sudakov} (double) logarithms in analogy with similar logarithms arising from infrared physics in gauge theories\cite{peskin}. In de Sitter space time the origin of these Sudakov logarithms is also strong ``infrared'' physics and a consequence of the growing behavior of the slow roll solution (\ref{zerothsr}) at long time, $\eta\rightarrow 0$.  {  This aspect merits emphasizing: whereas the self-energy kernel features the same type of logarithms in de Sitter and Minkowski space time,  the origin of the Sudakov enhancement is the behavior of the inflaton condensate $\propto 1/\eta$ as $\eta \rightarrow 0$, as compared to the oscillatory behavior of the condensate in Minkowski space time. }

     These Sudakov-type logarithms have also been found in
     studies of correlation functions and loop contributions in de Sitter space time\cite{woodard,tsamis,glavan}. Furthermore, unlike in Minkowski space time where bosonic and fermionic secular contributions feature the same sign, in de Sitter space time they feature opposite signs.

     This explicit comparison between Minkowski and de Sitter space times should unambiguously dispel the implicit assumption in the literature of the validity of the extrapolation of S-matrix results,   for example using a decay rate obtained in Minkowski space time, to cosmology. Using results from S-matrix theory in Minkowski space time within the cosmological realm  must be thoroughly and skeptically scrutinized. The results of this section with a direct comparison between Minkowski and de Sitter space-times  indicate that in general such assumptions are unwarranted.

\section{  Dynamical Renormalization Group   resummation:}\label{sec:drg}

The final results, either during the inflationary stage (\ref{solubosfin},\ref{soluferfin}) or Minkowski space-time (\ref{X1fini},\ref{X1finifer}) all suggest that radiative corrections to the evolution of the condensate are manifest as \emph{renormalization of the amplitudes that feature secular terms that grow in time}. In the Minkowski space-time results this is evident in the first line of the perturbative solutions (\ref{X1fini},\ref{X1finifer}), since the function $\mathcal{F}(t)$ becomes constant for $m_R t \gg 1$ (see equation (\ref{finft})), therefore the integrals $\int^t_{t_0} \mathcal{F}(t') dt'$ in the first lines of (\ref{X1fini},\ref{X1finifer}) grow linearly in time at long time. In the inflationary stage, the logarithms in the first order corrections in (\ref{solubosfin},\ref{soluferfin}) grow as $\eta \rightarrow 0$, with $\ln^2(\eta/\eta_0) \simeq 3600$  for $\simeq 60$ e-folds of inflation.

These secular terms result in a breakdown of the perturbative expansion at long time, and in all cases, these terms multiply the initial amplitude, suggesting to absorb these secular terms in a time dependent renormalization of the amplitude. The dynamical renormalization group, introduced as a dynamical resummation of secular divergences in the theory of amplitude equations for pattern formation\cite{drggold}, and adapted to study dynamical processes in non-equilibrium quantum field theory\cite{drg1,drg} and in cosmology\cite{cao,greendrg} provides a systematic, non-perturbative framework that leads to a resummation of the secular terms in the perturbative solution yielding improved and controlled asymptotics.

\subsection{Minkowski space-time}\label{subsec:drgmink}

We begin the discussion by focusing on the Minkowski space-time case, as it will yield familiar results that provide a benchmark for the method. Let us focus on the first lines of equations (\ref{X1fini},\ref{X1finifer})  wherein the brackets feature secular terms that grow linearly in time, neglecting the second lines in these equation  which yield finite contributions at long time, for bosons and fermions
\be   \X(t)  =  X\,e^{-im_Rt}\,\Big[1+ i g^2\,\int^t_{t_0}\mathcal{F}(t')\,dt' \,\Big] + X^*\,e^{im_Rt}\,\Big[1- i \,g^2\,\int^t_{t_0} \mathcal{F}^*(t')\,dt' \,\Big]\,,\label{secular} \ee with
 \be  g^2 = \frac{ \lambda^2}{8\pi^2\,m_R} ~\mathrm{bosons}~~~;~~ g^2 = \frac{ y^2_R\,m_R}{8\pi^2} ~\mathrm{fermions} \,.\label{gsqu}\ee   The second term in (\ref{secular}) is simply the complex conjugate of the first, hence consider the first term for the analysis. The bracket suggests to absorb the secular term in a time dependent renormalization of the amplitude $X$, this is implemented via the dynamical renormalization group as follows\cite{drggold,drg1,drg}. Let us introduce an arbitrary renormalization time scale $\overline{t}$ and a ``wave function renormalization'' $\mathcal{Z}[\overline{t}]$
and write the amplitude $X$ as
\be X = X[\overline{t}] \,\mathcal{Z}[\overline{t}]\,,\label{renamp}\ee expanding $\mathcal{Z}$ as
 \be \mathcal{Z}= 1+ g^2\, z_1[\overline{t}] + \cdots \label{zexpa} \ee Inserting into (\ref{secular}) yields
 \be   \X(t)  =  X[\overline{t}]\,e^{-im_Rt}\,\Big[1+  g^2 \,\Big( i\, \int^t_{t_0}\mathcal{F}(t')\,dt' +  z_1[\overline{t}]\Big)  \, +\cdots \Big] + c.c. \,,\label{secular2} \ee
choosing
\be z_1[\overline{t}] = -i  \int^{\overline{t}}_{t_0}\mathcal{F}(t')\,dt'\,,\label{zeta1}\ee it follows that
\be  \X(t)  =  X[\overline{t}]\,e^{-im_Rt}\,\Big[1+  g^2\,i\, \int^t_{\overline{t}}\mathcal{F}(t')\,dt'   +\cdots \Big] + c.c. \,,\label{secular3} \ee
and the perturbative expansion has been improved by choosing $\overline{t}$ close to $t$. Since the scale $\overline{t}$ is arbitrary and the solution $\X(t)$ does not depend on this scale, it obeys the \emph{dynamical renormalization group equation}\cite{drggold,drg,drg1}
\be \frac{d}{d\overline{t}}\X(t) =0 \,,\label{drgeqn}\ee which up to $\mathcal{O}(\lambda^2)$ becomes
 \be \frac{d}{d\overline{t}}X[\overline{t}] = i g^2\,\mathcal{F}[\overline{t}]X[\overline{t}] \,.\label{drgeq2}\ee with solution
 \be X[\overline{t}] = X[\overline{t}_i]\,e^{i g^2\,\int^{\overline{t}}_{\overline{t}_i}\mathcal{F}(t')\,dt'} \,.\label{drgsol}\ee Now, choosing $\overline{t}_i =0;\overline{t}=t$, taking the long time limit $m_R \, t \gg 1$  where $\mathcal{F}[t]$ is given by (\ref{flontime}) and inserting the (DRG) improved amplitude (\ref{drgsol}) into (\ref{secular2}) we obtain the dynamical renormalization group improved solution
 \be \X(t) = X(0) e^{-i\widetilde{m}_R t}\,e^{-\frac{\Gamma}{2} t} + c.c. \,,\label{finsolut}\ee
 where

 \be \widetilde{m}_R = m_R-g^2 \Big(\ln\Big( \frac{\mu}{m_R}\Big)-\gamma\Big)~~;~~ \Gamma = \pi \, g^2 = \Bigg\{ \begin{array}{c}
                                                                                                    \frac{ \lambda^2}{8\pi\,m_R} ~~\mathrm{bosons} \\
                                                                                                    \frac{ y^2_R\,m_R}{8\pi}~~\mathrm{fermions}
                                                                                                  \end{array}
  \,.\label{decay}\ee This is the correct solution in Minkowski space time, the decay width $\Gamma$ is precisely the transition probability per unit time for a particle of mass $m_R$ to decay into a pair of massless bosons or fermions. The oscillation frequency $\widetilde{m}_R$ is simply related to $m_R$ by a finite renormalization. It is a straightforward exercise to confirm that for $\Gamma \ll m_R$ the final solution (\ref{finsolut}) is a solution of the equation of motion with a local friction term, namely
  \be \ddot{\X}(t) + \widetilde{m}^2_R \,\X(t) + \Gamma \dot{\X}(t) =0 \,.\label{frikeq}\ee

\subsection{During slow roll inflation}\label{subsec:infla}

During the slow roll inflationary stage, the solution for the inflaton condensate are given by equations (\ref{solubosfin},\ref{soluferfin}) respectively for (massless conformally coupled) bosons and fermions. Keeping the leading Sudakov logarithmic secular term in the limit $\eta \rightarrow 0$, both cases yield
\be  \X_{sr}(\eta) = \X^{(0)}_{sr}(\eta_0)\,\Big(\frac{\eta}{\eta_0}\Big)^{\beta_-}\,\Big\{1 + \Upsilon \,   \ln^2\Big( \frac{\eta}{\eta_0} \Big)    + \cdots  \Big\} ~~;~~ \Upsilon = \Bigg\{\begin{array}{c}
           -\frac{\lambda^2}{24\pi^2 H^2}~~\mathrm{bosons} \\
           \frac{y^2_R}{12\pi^2}~~\mathrm{fermions}
         \end{array}
\,, \label{inflagensol}\ee by introducing
\be \mathcal{H}(\eta) = \frac{2}{\eta} \, \ln\Big( \frac{\eta}{\eta_0} \Big) \,,\label{hofeta}\ee we can rewrite the general solution (\ref{inflagensol}) as
\be  \X_{sr}(\eta) = \X^{(0)}_{sr}(\eta_0)\,\Big(\frac{\eta}{\eta_0}\Big)^{\beta_-}\,\Big\{1 + \Upsilon \,\int^{\eta}_{\eta_0} \mathcal{H}(\eta')\,d\eta' + \cdots \Big\}\,,
\label{newinflasol}\ee written in this form, we can simply adapt the steps described above. We write the initial amplitude
\be  \X^{(0)}_{sr}(\eta_0) \equiv \mathcal{Q}[\overline{\eta}]\,\mathcal{Z}[\overline{\eta}] ~~;~~ \mathcal{Z}[\overline{\eta}] = 1 + \Upsilon\, z_1[\overline{\eta}]+ \cdots \,\label{reninfla}\ee and choose
\be z_1[\overline{\eta}]= - \int^{\overline{\eta}}_{\eta_0} \mathcal{H}(\eta')\,d\eta'\,,\label{zuno}\ee therefore
\be \X_{sr}(\eta) = \mathcal{Q}[\overline{\eta}]\,\Big(\frac{\eta}{\eta_0}\Big)^{\beta_-}\,\Big\{1 + \Upsilon \,\int^{\eta}_{\overline{\eta}} \mathcal{H}(\eta')\,d\eta' + \cdots \Big\}\,.
\label{newinflasol2}\ee  since $\X_{sr}(\eta)$ does not depend on the arbitrary scale $\overline{\eta}$ the (DRG) equation   now yields
\be \frac{d}{d\overline{\eta}}\X_{sr}(\eta)=0 ~~\Rightarrow \frac{d}{d\overline{\eta}}\mathcal{Q}(\overline{\eta}) = \Upsilon \,\mathcal{H}(\overline{\eta})\,\mathcal{Q}(\overline{\eta})\,,\label{drginfla} \ee with solution
\be \mathcal{Q}(\overline{\eta})= \mathcal{Q}(\overline{\eta}_i)\,e^{\Upsilon \, \int^{\overline{\eta}}_{\overline{\eta}_i}\,\mathcal{H}(\eta')\,d\eta'}\,.\label{drginflasol}\ee Now taking $\overline{\eta } = \eta~~;~~\overline{\eta }_i = \eta_0 $ and identifying  $\mathcal{Q}(\eta_0) \equiv \X_{sr}(\eta_0) $ the (DRG) improved solution for the inflaton condensate is given by
\be \X_{sr}(\eta) = \X_{sr}(\eta_0)\,\Big(\frac{\eta}{\eta_0}\Big)^{\beta_-}  \, e^{ \Upsilon\, \ln^2 \big(\frac{\eta}{\eta_0} \big) }\,,\label{inflasolfinal}\ee

Following the same steps for the solution with the phenomenological friction term, equation (\ref{soluln}) we obtain

\be  \X_{sr}(\eta) = \X_{sr}(\eta_0)\,\Big(\frac{\eta}{\eta_0}\Big)^{\beta_-}\,\Big(\frac{\eta}{\eta_0}\Big)^{-\frac{\kappa \gamma}{3}}\,.\label{fricdrg}  \ee

We can now rescale back to the inflaton condensate $\varphi_I(\eta)$ during slow roll related to $\X_{sr}(\eta)$ by equation (\ref{conchisr}) writing the Sudakov logarithms in terms of the
number of efolds $N_e(t) = H(t-t_0)$ since the beginning of slow roll inflation at $\eta_0$ (comoving time $t_0$), as
\be N_e \equiv \ln\Big( \frac{\eta_0}{\eta}\Big) \,,\label{Nefold}\ee and summarize our results for the dynamics of the inflaton condensate during the slow roll stage in comoving time $t$ as
\be \varphi_{Isr}(t) = \varphi^{(0)}_{Isr}(t)\times \Bigg\{ \begin{array}{l}
                                                       e^{\frac{\kappa\gamma}{3}\,N_e(t)} ~~;~~   \frac{\kappa\gamma}{3} = \frac{m^2\Gamma}{9H^3}~~;~~ (\Gamma ~ = ~phenomenological~ friction~term) \\
                                                       e^{\Upsilon N^2_e(t)}~~;~~ \Upsilon =  -\frac{\lambda^2}{24\pi^2 H^2}~~(\mathrm{bosons})~;~ \frac{y^2_R}{12\pi^2}~~(\mathrm{fermion})
                                                     \end{array}
\label{finresu}\ee where $\varphi^{(0)}_{Isr}(t) $ is the slow roll solution in absence of interactions with spectator fields.

Clearly, the phenomenological friction term  \emph{does not} describe reliably the radiative corrections to the dynamics of the inflaton condensate in the expanding cosmology, a result that extends  those obtained in a radiation dominated cosmology in ref.\cite{cao}   to slow roll inflation.

By implementing the same method (DRG) both in Minkowski and de Sitter space times, we have unambiguously highlighted the main differences associated with cosmological expansion, and thereby established directly that assuming that the results from Minkowski space-time, such as a friction term in the equation of motion based on the S-matrix decay rate are valid in cosmology, is  in general unwarranted.

\section{Particle production from inflaton condensate}\label{sec:decay}
The results    in Minkowski space-time clearly indicate that the decay of the inflaton amplitude (\ref{finsolut}), and the friction term in the equation of motion for the amplitude (\ref{frikeq}) are determined by the decay rate $\Gamma$, which is the total probability \emph{per unit time} for the decay process of a single particle of mass $m_R$ at rest into two massless particles. This indicates that the dissipative radiative contributions from spectator fields is associated with the production of spectator particles. Particle production as a consequence of dynamical evolution of condensates has recently been studied in Minkowski space time\cite{nathan}. In this section we address the important aspect of production of spectator particles as a consequence  of their coupling to the inflaton condensate \emph{during the inflationary, nearly de Sitter era}, in contrast to previous studies of particle production during post-inflationary reheating\cite{rehe}-\cite{rehe7}.

{  In a cosmological space time the kinematics of particle production is not restricted by energy conservation because of a lack of a time-like Killing vector, however in a spatially flat FRW cosmology, which is the scenario considered here, spatial translational invariance implies total momentum conservation. In the case of particles produced by the homogeneous inflaton condensate, the total momentum is zero, therefore for the cases under consideration particles are produced as back-to-back pairs of total zero momentum.}

In this and next sections we focus solely on the case of massless scalar spectators conformally coupled to gravity which do  not feature gravitational production and  illustrate the main physical aspects in a simple and clear manner, unobscured by the technical complications of spinors and field renormalization associated with fermion spectator fields which will be studied elsewhere.

  Production of spectator particles via their coupling to the inflaton condensate is analyzed from two complementary perspectives: \textbf{i:)} a generalization of the optical theorem from S-matrix theory to a finite time domain in a cosmological background, \textbf{ii:)} as a weak coupling limit of a non-perturbative mean field approach\cite{nathan} wherein particle production is associated with a  Bogoliubov transformation between the free field and interacting basis.
\subsection{Optical theorem in a finite time domain:}\label{subsec:othm}
In appendix (\ref{subsec:OT}) we provide a generalization of the optical theorem for a quantum field coupled to a  condensate in a finite time domain with an explicit analysis up to second order in Minkowski space-time, and its extrapolation to the cosmological context in conformal time. The main result is given by equation (\ref{desitprod}), which relates the total pair production probability to the correlation function of the composite operator $\mathcal{O}(\vx,\eta)$ (\ref{opeO}), in the free field interaction picture.

 These results are general depending only on the separation of the interaction Hamiltonian and the definition of interaction picture, which only requires a mode expansion in a ``free field basis'', and do not depend on whether fields are minimally or conformally coupled to gravity.

There is an important \emph{difference} between the probability of pair production determined by the optical theorem (\ref{desitprod}) and the retarded self energy that determines the radiative correction to the equation of motion of the inflaton from spectator fields (\ref{sigmaret}):
\bea \mathcal{P}_{0 \rightarrow 2}[\eta,\eta_0] & \propto &   \bra{0}\Big[ {\mathcal{O}^{(0)}}(\vx_1,\eta_1){\mathcal{O}^{(0)}}(\vx_2,\eta_2) +{\mathcal{O}^{(0)}}(\vx_2,\eta_2){\mathcal{O}^{(0)}}(\vx_1,\eta_1)\Big]\ket{0} \nonumber \\ \Sigma(\vx_1,\vx_2;\eta,\eta') & \propto &  -i\, \bra{0}\Big[ {\mathcal{O}^{(0)}}(\vx_1,\eta_1){\mathcal{O}^{(0)}}(\vx_2,\eta_2) -{\mathcal{O}^{(0)}}(\vx_2,\eta_2){\mathcal{O}^{(0)}}(\vx_1,\eta_1)\Big]\ket{0} \,, \label{diffe}\eea therefore, while the main ingredients, namely the correlation functions of the composite operators, are similar, in general there is no \emph{direct} quantitative relationship between the radiative corrections to the equations of motion and particle production. For the case under consideration, of a massless scalar field conformally coupled to gravity, we can simply use the     mode functions (\ref{gqzero}) and the result (\ref{desitprodfin}) from appendix (\ref{subsec:OT}) to obtain
\be  \mathcal{P}_{0 \rightarrow 2}[\eta,\eta_0] = \frac{2\lambda^2 }{H^2} \, V\, \int^\eta_{\eta_0} d\eta_1 \int^\eta_{\eta_0} d\eta_2\,\Big(\frac{\X(\eta_1)}{\eta_1}\Big)\,
\Big(\frac{\X(\eta_2)}{\eta_2}\Big) \int \frac{e^{-2ik(\eta_1-\eta_2-i\epsilon)}}{4k^2}\,\frac{d^3k}{(2\pi)^3} \,,\label{prodp}\ee where we introduced a convergence factor $\epsilon \rightarrow 0^+$. The momentum integral is straightforward, only the term that is symmetric under $\eta_1 \leftrightarrow \eta_2$ survives, yielding   $\delta(\eta_1-\eta_2)/16\pi$ with the final result,
\be  \mathcal{P}_{0 \rightarrow 2}[\eta,\eta_0] = \frac{\lambda^2 }{8\pi\,H^2} \, V\, \int^\eta_{\eta_0} \Big( \frac{\X(\eta')}{\eta'}\Big)^2 d\eta'\,.\label{2parpro}\ee Replacing $\X(\eta)$ by the slow roll solution $\X^{(0)}_{sr}(\eta)$ given by equation (\ref{zerothsr}) to leading order in the coupling, using the rescaling relation (\ref{rescaledfields}) for the scalar spectator, and  taking  the long time limit $\eta_0/\eta \rightarrow \infty$ we find
\be  \mathcal{P}_{0 \rightarrow 2}[\eta,\eta_0] = \frac{\lambda^2 \,V}{8\pi\,H}\,\frac{\Big(\varphi^{(0)}_{Isr}(\eta)\Big)^2 \,C^3(\eta)}{1-2\beta_-} \approx \frac{\lambda^2 \,V_{ph}(\eta)}{24\pi\,H}\, \Big(\varphi^{(0)}_{Isr}(\eta)\Big)^2  \,,\label{parprods} \ee where $\varphi^{(0)}_{Isr}(t)$ is the (nearly constant) slow roll solution of the (unscaled)  inflaton condensate, we used the slow roll condition $m/H \ll 1$,  and $V_{ph}(\eta) = V\,C^3(\eta)$ is the physical spatial volume. There is a simple relation between $\mathcal{P}_{0\rightarrow 2}$ and the total number of   spectator particles  produced via the interaction with the condensate, namely
\be \mathcal{N}(\eta,\eta_0) = \bra{0}\widehat{N}(\eta,\eta_0)\ket{0}     \,,\label{parnum}\ee where $\widehat{N}[t,t_0]$ is the Heisenberg operator
\be \widehat{N}[\eta,\eta_0] = U^{-1}_I(\eta,\eta_0) \,\widehat{N} \,  U_I(\eta,\eta_0)~~;~~  \widehat{N} = \sum_{\vk} a^\dagger_{\vk}\,a_{\vk} \,. \label{heisN}\ee and $U_I(\eta,\eta_0)$ is the time evolution operator in interaction picture (\ref{UIsol},\ref{timeord}). Writing $U_I(\eta,\eta_0) \equiv 1 + i\mathcal{T}[\eta,\eta_0]$ as in (\ref{defmat}) with $t,t_0 \rightarrow \eta,\eta_0$,   expanding up to second order as in appendix (\ref{subsec:OT}), and using that
\be \bra{0}a^\dagger_{\vk} =0 ~~;~~ a_{\vk}\ket{0} =0 \,,\label{vacN}\ee we find to leading order in the coupling
\be \mathcal{N}[\eta,\eta_0] = \bra{0}\mathcal{T}^{\dagger(1)}[\eta,\eta_0]\,\widehat{N}\,\mathcal{T}^{(1)}[\eta,\eta_0]\ket{0} \,,\label{secordN}\ee introducing the identity
in the two particle space as in equation (\ref{prodprob})  in appendix (\ref{subsec:OT}) and recognizing that $\widehat{N}\ket{\vk_1,\vk_2} = 2 \ket{\vk_1,\vk_2} $ we finally find the relation
\be \mathcal{N}(\eta,\eta_0) = 2 \,\mathcal{P}_{0\rightarrow 2}[\eta,\eta_0] =  \frac{\lambda^2 \,V_{ph}(\eta)}{12\pi\,H}\, \Big(\varphi^{(0)}_{Isr}(\eta)\Big)^2  \,.\label{totNoft}\ee

An important corollary of this discussion is that the number of particles produced does not reflect directly the Sudakov-type logarithmic growth of the inflaton condensate, but it \emph{grows} $\propto {C}^3(\eta) = e^{3N_e(t)}$, this is a consequence of the quantitative difference between the self-energy and production probability kernels.

\subsection{Distribution functions: de Sitter vs. Minkowski}\label{subsec:distfun}
With the relation (\ref{totNoft}), a consequence of the optical theorem, and the result (\ref{prodp})  it follows that
\be \mathcal{N}(\eta,\eta_0)=   V\, \int \frac{d^3k}{(2\pi)^3} \frac{ \lambda^2 }{H^2\,k^2} \Bigg| \int^\eta_{\eta_0} d\eta_1 \Big(\frac{\X(\eta_1)}{\eta_1}\Big)\,e^{-2ik\eta_1} \Bigg|^2 \ee

therefore we \emph{define} the distribution function as
\be \mathds{F}[k;\eta] \equiv \frac{(2\pi)^3}{V}\,\frac{d\mathcal{N}(\eta,\eta_0) }{d^3k} =  \frac{ \lambda^2 }{H^2\,k^2} \Bigg| \int^\eta_{\eta_0} d\eta_1 \Big(\frac{\X(\eta_1)}{\eta_1}\Big)\,e^{-2ik\eta_1} \Bigg|^2 \,.\label{fdist}\ee

\begin{figure}[!ht]
\begin{center}
\includegraphics[height=3in,width=4in,keepaspectratio=true]{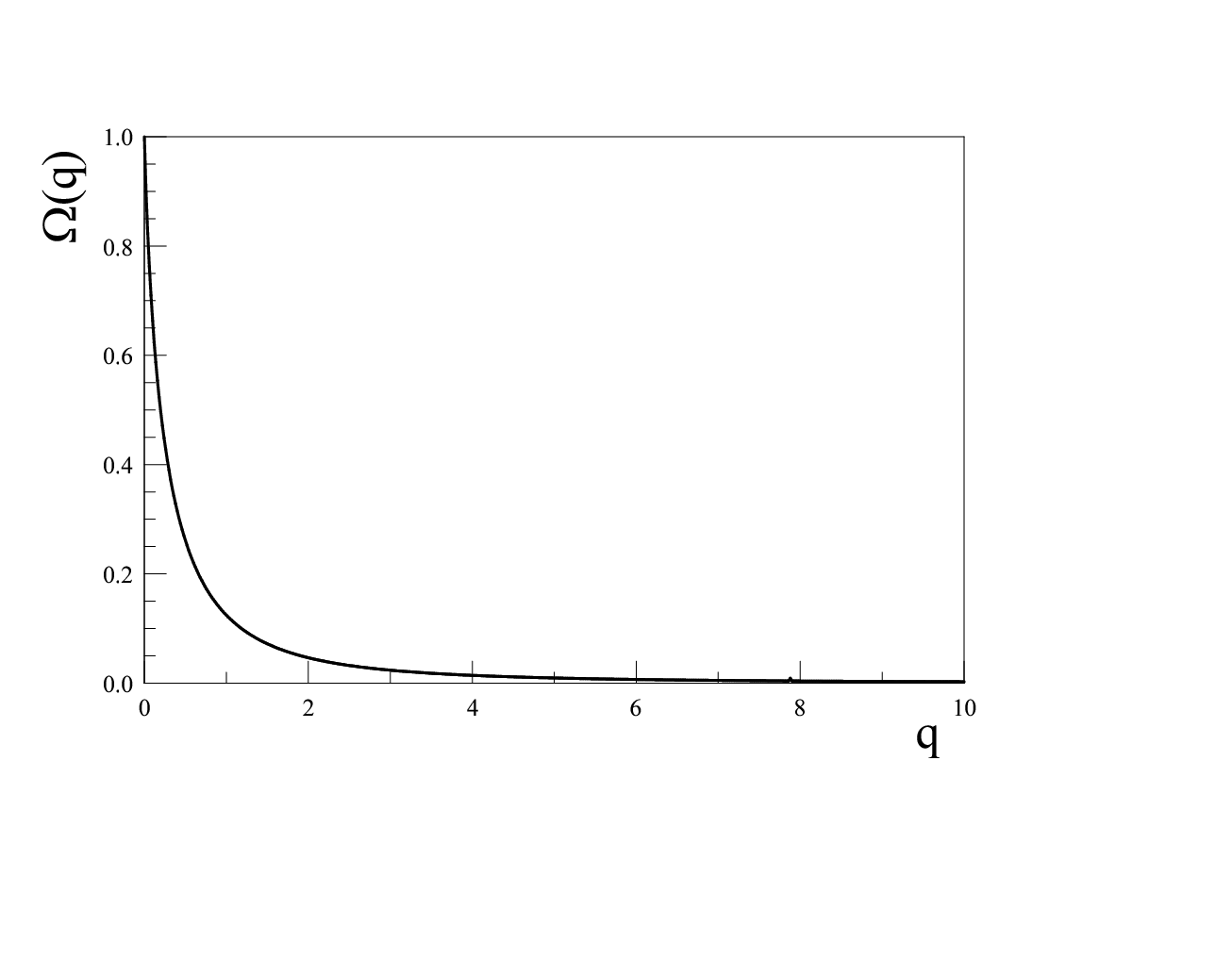}
\caption{The function $\Omega(q)$ (\ref{omedis}) for $\beta_-=-1$ vs. $q$. }
\label{fig:omeg}
\end{center}
\end{figure}

With $\X(\eta)$ given to leading order by the slow roll solution (\ref{zerothsr}), rescaling $\eta_1 = \eta\,z$, taking $\eta_0/\eta \rightarrow \infty$ and using the scaling relation (\ref{rescaledfields}) for the scalar spectator field, the distribution function becomes
\be \mathds{F}(k;\eta)= \frac{\lambda^2\,\Big(\varphi^{(0)}_{Isr}(\eta)\Big)^2 }{H^2\,k^2_{ph}(\eta)}\,\,\Omega\Big(\frac{k_{ph}(\eta)}{H}\Big) \,,\label{fdis}\ee where $k_{ph}(\eta)$ is the physical wavevector $k/C(\eta)$, and the function $\Omega(q)$ is given by
\be \Omega(q) = \Big| \int^\infty_1 z^{\beta_--1}\,e^{2iqz}\,dz  \Big|^2 \equiv \frac{\Big|\Gamma[\beta_-,-2iq]\Big|^2}{\big (2q\big)^{2\beta_-}}~~;~~ \Omega(0) = 1/\beta^2_-\,,\label{omedis}\ee and $\Gamma[a,x]$ is the incomplete Gamma function.  During slow roll with $\varphi^{(0)}_{Isr}$ approximately constant, the distribution function is solely a function of $k_{ph}(\eta)$. For slow roll with $m/H \ll1$ we can replace $\beta_- \rightarrow -1$, and while the analytic solution (\ref{omedis}) does not seem very illuminating,  figure (\ref{fig:omeg}) displays $\Omega(q)$ for this case.

Remarkably, the distribution function is strongly peaked at \emph{superhorizon wavelengths} $k_{ph}(\eta)/H \rightarrow 0$.

In Minkowski space time the relation $\mathcal{N}(t) = 2 \mathcal{P}_{0 \rightarrow 2}(t)$ of course still holds, therefore equations (\ref{prodprob},\ref{rhs}) yield the distribution function
\be \mathds{F}_{Mink}[k;t] \equiv \frac{(2\pi)^3}{V}\,\frac{d\mathcal{N}(t) }{d^3k} =  \frac{ \lambda^2 }{ k^2}\, \Bigg| \int^t_{t_0} d\eta_1 \X(t_1) \,e^{-2ik t_1} \Bigg|^2 \,.\label{fdistmink}\ee With the zeroth order solution (\ref{X0sol}) the time integral yields
\be \Bigg| \int^t_{t_0} d\eta_1 \X(t_1) \,e^{-2ik t_1} \Bigg|^2  =  |X|^2\, \Bigg[ \frac{\sin\big[(t-t_0)\,(k-m_R/2) \big]}{(k-m_R/2)}\Bigg]^2 + \cdots  \label{minkdis}\ee where the dots
stand for terms that are subleading in the limit $m_R(t-t_0)\rightarrow \infty$. The term within brackets is the familiar resonant term from Fermi's Golden rule: strongly peaked at $k = m_R/2$, and  in the limit $t-t_0 \rightarrow \infty$
\be  {\Bigg[ \frac{\sin\big[(t-t_0)\,(k-m_R/2) \big]}{(k-m_R/2)}\Bigg]^2} {}_{~~ \overrightarrow{(t-t_0)\rightarrow \infty }}  ~~   \pi\,(t-t_0)\,\delta(k-m_R/2)\,,\label{FGR} \ee which is obviously the statement of energy conservation: the total momentum of the pair of particles is zero but the momentum of each particle is constrained by energy conservation to $k=m_R/2$. This is in striking contrast with the distribution function (\ref{fdis}) during slow roll inflation displayed in fig.(\ref{fig:omeg}). The   difference is a consequence of the fact that whereas the momentum of each particle in the pair (with total zero momentum) is restricted to $k= m_R/2$ by energy conservation in Minkowski space-time,   there is no such constraint in the expanding cosmology because energy is \emph{not} conserved, a consequence of the lack of a global time-like Killing vector.

\section{Equivalence with Mean field dynamics}\label{subsec:bogos}

The results obtained above are confirmed by a complementary approach: the linearized (weak coupling) limit of the mean field equation of motion\cite{nathan}. Considering solely a scalar spectator field (massless conformally coupled to gravity), its interaction with the condensate is given by
\be \mathcal{L}_{int}[\chi_s] = - \lambda \,C(\eta)\,\X(\eta)\,\chi^2_s(\vx,\eta) \label{efmas}\ee
and can be interpreted as a time dependent mass term with  an effective Lagrangian density for the spectator field
\be \mathcal{L}_{eff}[\chi_s] =   \frac{1}{2} \big[
{\chi'_s}^2- (\nabla \chi_s)^2-\lambda\,\sigma(\eta)\,\chi^2_s\big]~~;~~\sigma(\eta) = 2\,C(\eta)\,\X(\eta)  \,,\label{leffs}\ee  yielding a Heisenberg field equation similar to (\ref{heiseombos}),
\be \Big[ \frac{\partial^2}{\partial \eta^2} - \nabla^2 +\lambda \, \sigma(\eta)\Big]\chi_s(\vx,\eta) = 0     \,.\label{meffbos}\ee Expanding $\chi_s(\vx,t)$ and its canonical momentum $\pi_s(\vx,\eta)$ as
\bea
\chi_s(\vx,\eta)  & = &   \frac{1}{\sqrt{V}}\,\sum_{\vk} \Big[c_{\vk}\,g_k(\eta)+ c^\dagger_{-\vk}\,g^*_k(\eta) \Big]\,e^{i\vk\cdot\vx}\,, \label{nuchiex} \\
\pi_s(\vx,\eta)  & = &   \frac{1}{\sqrt{V}}\,\sum_{\vk} \Big[c_{\vk}\,g^{'}_k(\eta)+ c^\dagger_{-\vk}\,g^{*\,'}_k(\eta) \Big]\,e^{i\vk\cdot\vx}\,, \label{nupiex} \eea where $' \equiv d/d\eta$ and
  the mode functions $g_k(\eta)$ now obey the equations of motion
  \be \Big[ \frac{d^2}{d \eta^2} +k^2 + \lambda\,\sigma(\eta)  \Big]g_k(\eta) = 0 \,,  \label{nugmodes}\ee with Wronskian condition
\be W[g,g^*] = g^*_k(\eta) \frac{d}{d\eta}\,g_k(\eta) - g_k(\eta) \frac{d}{d\eta}\,g^*_k(\eta)  = -i \,,\label{Wrons}\ee which guarantees that the annihilation and creation
operators $c_{\vk},c^\dagger_{\vk}$ are time independent and obey canonical commutation relations. The \emph{interacting} Bunch-Davies vacuum is defined such that
\be c_{\vk}\ket{\widetilde{0}} = 0 \,.\label{intBD} \ee

Taking now the expectation value in the interacting vacuum  $\ket{\widetilde{0}}$ the equation of motion for the inflaton condensate (\ref{exvaleomh}) with $\widetilde{V}(\chi_I;\eta)$ given by (\ref{masit}) now becomes
\be  \X^{''}(\eta) + \Big( \frac{m^2}{H^2} - 2 \Big)\frac{ \X(\eta)}{\eta^2}= -\lambda \, C(\eta) \, \bra{\widetilde{0}}:\chi^2_s(\vx,\eta):\ket{\widetilde{0}}\,,\label{nueq} \ee the normal ordering prescription  stands for subtracting from the expectation value   the  free field result, namely with the field expansion (\ref{nuchiex}) and the definition (\ref{intBD})
\be \bra{\widetilde{0}}:\chi^2_s(\vx,\eta):\ket{\widetilde{0}} \equiv \frac{1}{V}\,\sum_{\vk} \Big[|g_k(\eta)|^2 - |g^{(0)}_k(\eta)|^2\Big] \,,\label{nunord}\ee  where the mode functions $g^{(0)}_k(\eta)$ are those of free fields (\ref{gqzero}),  therefore the expectation value of the normal ordered operator vanishes in the free field case. The equations of motion in this mean field framework become the self-consistent set of equations
\be  \X^{''}(\eta) + \Big( \frac{m^2}{H^2} - 2 \Big)\frac{ \X(\eta)}{\eta^2}= -\lambda \, C(\eta) \,\frac{1}{V}\,\sum_{\vk} \Big[|g_k(\eta)|^2 - |g^{(0)}_k(\eta)|^2\Big]\,,\label{minfi} \ee along with the mode equations (\ref{nugmodes}) and the Wronskian condition (\ref{Wrons}). This self-consistent mean field formulation is the cosmological extension of the framework proposed in ref.\cite{nathan} to study non-perturbatively   the dynamical evolution of homogeneous condensates in Minkowski space-time.

We now establish contact between this non-perturbative mean field theory  with the equation of motion up to one loop discussed in the previous section by \emph{linearizing} the solution of the mode functions, so that the right hand side of (\ref{minfi}) is linear in $\X(\eta)$.

 This is achieved by re-writing equation (\ref{nugmodes}) as a Lippman-Schwinger integral equation\cite{BDie,zel,ford2} by using the
Green's function of the differential operator for $\lambda =0$, and the boundary condition that $g_k(\eta) \rightarrow g^{(0)}_k(\eta)$ as $\lambda \rightarrow 0$,
\be g_{k}(\eta) = g^{(0)}_k(\eta) - \lambda \,\int^\eta_{\eta_0} \frac{\sin[k(\eta-\eta_1)]}{k}\,\sigma(\eta_1)\,g_{k}(\eta_1)\,d\eta_1 \equiv g^{(0)}_k(\eta)+\widetilde{g}_k(\eta) \,,\label{LSeq}\ee where
\be g^{(0)}_k = \frac{e^{-ik\eta}}{\sqrt{2k}}\,,\label{gzero}\ee is the solution for $\lambda =0$.

We note that (\ref{LSeq}) yields the unique solution of the differential equation (\ref{nugmodes}) with initial conditions $g_k(\eta_0) = e^{-ik\eta_0}/\sqrt{2k}; g'_k(\eta_0) = -i\,k\,g_k(\eta_0)$ \emph{to all orders in the coupling} thereby automatically fulfilling the Wronskian condition (\ref{Wrons}). These are the same as the initial conditions  of  the \emph{free field }mode functions at the initial time $\eta_0$.

Incidentally,  we note that if we consider that $\varphi_{Isr}$ is a \emph{constant} during slow roll, then using the rescaling (\ref{rescaledfields}) it follows that $\lambda \,\sigma(\eta) = 2 \lambda \, \varphi_{Isr}/H^2\eta^2$, therefore, comparing with the results (\ref{gmodes}-\ref{bdvac}), we find that the exact solution of the mode equations (\ref{nugmodes}) with Bunch-Davies boundary conditions is given by eqn. (\ref{bdvac}) with $\nu^2_s = \frac{1}{4}- \frac{2 \lambda\,\varphi_{Isr}}{H^2}$. Bunch-Davies boundary conditions are ensured by the integral equation (\ref{LSeq}) in the limit $\eta_0 \rightarrow -\infty$ since in this limit the solution $g_k(\eta) \rightarrow  g^{(0)}_k $ as $\eta \rightarrow \eta_0 \rightarrow -\infty$, which is precisely the Bunch-Davies boundary condition.

The integral equation  (\ref{LSeq}) yields the solution in general without assuming the constancy of $\varphi_{Isr}$ and it is more convenient than the exact solution because it lends itself to a Born series,
\be \widetilde{g}_k(\eta)= \lambda g^{(1)}_k(\eta) + \lambda^2 g^{(2)}_k(\eta) + \cdots \label{bornseries}\ee which allows us to compare with the results for particle production obtained via the optical theorem up to second order in the coupling. The first order (Born) approximation is given by
\be g^{(1)}_k(\eta) =  \frac{1}{iHk\sqrt{2k}}\Bigg\{ e^{ik\eta}\,\int^\eta_{\eta_0} e^{-2ik\eta_1}\,\Big(\frac{\X(\eta_1)}{\eta_1} \Big)\,d\eta_1 -e^{-ik\eta}\, \int^\eta_{\eta_0} \Big(\frac{\X(\eta_1)}{\eta_1} \Big)\,d\eta_1\Bigg\}\,,\label{born}\ee where to leading order in the coupling $\X(\eta) = \X^{(0)}_{sr}(\eta)$, given by equation (\ref{zerothsr}).

The linear response analysis in section (\ref{subsec:eomLR}) suggests that the full Heisenberg field (\ref{nuchiex}) is related to the free field $\chi^{(0)}_s(\vx,\eta)$ and its canonical momentum $\pi^{(0)}_s(\vx,\eta)$, given by (\ref{chizero},\ref{pizero}) respectively as\footnote{The evolution with $U_0(\eta,\eta_0)$ is already included in the time dependence of $\chi^{(0)}_s(\vx,\eta)$.}
\be \chi_s(\vx,\eta) = U^{-1}_I(\eta,\eta_0) \,  \chi^{(0)}_s(\vx,\eta) \,U_I(\eta,\eta_0) \,.  \label{unichis}\ee and similarly for its canonical momentum  $\pi_s(\vx,\eta)$ (\ref{nupiex}).  Since $U_I$ acts on the operators, it follows from (\ref{chizero},\ref{pizero}) that
\bea \chi_s(\vx,\eta) & \equiv &   \frac{1}{V}\sum_{\vk}\frac{1}{\sqrt{2k}} \Big[a_{\vk}(\eta) \,  {e^{-ik\eta}}  + a^\dagger_{-\vk}(\eta) \, {e^{ik\eta}}\Big]\,e^{i\vk\cdot\vx} \,,\label{chiexpu} \\
\pi_s(\vx,\eta) &\equiv &  \frac{-i}{V}\sum_{\vk}\frac{k}{\sqrt{2k}} \Big[a_{\vk}(\eta) \,  {e^{-ik\eta}}  - a^\dagger_{-\vk}(\eta) \, {e^{ik\eta}}\Big]\,e^{i\vk\cdot\vx} \,,\label{piexpu}\eea
with
\be  a_{\vk}(\eta) = U^{-1}_I(\eta,\eta_0) \,   a_{\vk}  \,U_I(\eta,\eta_0) \,.\label{trans}\ee The left hand side of equations (\ref{chiexpu},\ref{piexpu}) is given by the expansions (\ref{nuchiex},\ref{nupiex}), from which we obtain the relations
\bea a_{\vk}(\eta)  & = & c_{\vk}\,\alpha_{k}(\eta)+ c^\dagger_{-\vk} \,\beta_k(\eta) \,\label{bogo}\\
a^\dagger_{\vk}(\eta)  & = & c^\dagger_{\vk}\,\alpha^*_{k}(\eta)+ c_{-\vk} \,\beta^*_k(\eta) \,\label{bogocc}\eea with
\bea \alpha_k(\eta)  & = &  \sqrt{\frac{k}{2}}\,e^{ik\eta}\,\Big( g_k(\eta) + \frac{i}{k}\,g'_k(\eta)  \Big)\,,\label{alfabogo}\\
\beta_k(\eta)  & = &  \sqrt{\frac{k}{2}}\,e^{ik\eta}\,\Big( g^*_k(\eta) + \frac{i}{k}\,g^{*\,'}_k(\eta)  \Big)\,.\label{betabogo} \eea This is a   Bogoliubov transformation, the Wronskian condition (\ref{Wrons}) yields
\be |\alpha_k(\eta)|^2 - |\beta_k(\eta)|^2 =1\,,\label{bogouni}\ee implying that the transformation is unitary, namely canonical commutation relations for $c_{\vk},c^\dagger_{\vk}$ yield canonical commutation relations for $a_{\vk},a^\dagger_{\vk}$ and viceversa. The ``particle'' basis is that of free particles described by the mode functions $g^{(0)}_k(\eta)$.

 With the solution of the integral equation (\ref{LSeq}) it follows that
\bea \alpha_k(\eta) & = &  1+  \sqrt{\frac{k}{2}}\,e^{ik\eta}\,\Big( \widetilde{g}_k(\eta) + \frac{i}{k}\,\widetilde{g}^{\,'}_k(\eta)  \Big) \,,\label{alfaex}\\ \beta_k(\eta) & = & \sqrt{\frac{k}{2}}\,e^{ik\eta}\,\Big( \widetilde{g}^{\,*}_k(\eta) + \frac{i}{k}\,\widetilde{g}^{\,*\,'}_k(\eta)  \Big) \,,\label{betaex}\eea and the Born series (\ref{bornseries}) shows that $\alpha_k(\eta) \simeq 1 + \mathcal{O}(\lambda)~;~\beta_k(\eta) \simeq \mathcal{O}(\lambda)$.

We are now in position to make contact with the results of linear response in section (\ref{subsec:eomLR}) and the equation of motion  (\ref{lineom}) with self-energy radiative correction given by (\ref{sigbosi}).

To leading order in the coupling $\lambda$ we find
\be \bra{\widetilde{0}}:\chi^2_s(\vx,\eta):\ket{\widetilde{0}}   =   \frac{\lambda}{V}\,\sum_{\vk} 2\,\mathrm{Re}\Big[g^{(0)\,*}(\eta)\,g^{(1)}(\eta)\Big]\,, \ee with the Born approximation (\ref{born}) we find to leading order in $\lambda$
\bea -\lambda\,C(\eta) \, \bra{\widetilde{0}}:\chi^2_s(\vx,\eta):\ket{\widetilde{0}} & = &  \frac{2\,i\,\lambda^2 }{H^2\,\eta}\,\int^\eta_{\eta_0}\,d\eta_1 \int \Big[ \frac{e^{-2i(\eta-\eta_1)}-e^{2i(\eta-\eta_1)}}{4k^2}\Big]\,\frac{d^3k}{(2\pi)^3}\,\Bigg( \frac{\X(\eta_1)}{\eta_1}\Bigg)\nonumber \\& = &
-\int^\eta_{\eta_0} \widetilde{\Sigma}_s(\eta,\eta_1)\,\X(\eta_1)\,d\eta_1\,,\label{siginu}\eea where we recognize the bosonic self energy (\ref{sigbosi}). Therefore, equation (\ref{nueq}) to leading order in $\lambda$ yields the equation of motion (\ref{lineom}) including the one loop bosonic self-energy, thereby establishing a direct connection between the non-perturbative mean field formulation and the perturbative, linearized corrections to the equations of motion with radiative corrections from spectator loops. Yet another independent confirmation of the equations of motion with self-energy radiative corrections from spectators.

Furthermore, this equivalence highlights the \emph{backreaction} of spectator fields on the inflaton: the coupling to the condensate acts as a time dependent mass term therefore explicitly breaking the conformal coupling to gravity of the bosonic spectators leading to the production of spectator particles, which in turn   react back in the dynamics of the inflaton as a radiation reaction self-energy.

\vspace{1mm}

\textbf{Particle production:}

We can now calculate the number of particles in the interacting vacuum state $\ket{\widetilde{0}}$,
\be \bra{\widetilde{0}} a^\dagger_{\vk}(\eta)\, a_{\vk}(\eta) \ket{\widetilde{0}} = |\beta_k(\eta)|^2 = \frac{k}{2}\,\Big| \widetilde{g}^*_k(\eta) + \frac{i}{k}\,\widetilde{g}^{*\,'}_k(\eta) \Big|^2\,,\label{partnu} \ee in the Born approximation with $g^{(1)}_k(\eta)$ given by (\ref{born}), we find the \emph{distribution function}
\bea f(k;\eta) & \equiv & \bra{\widetilde{0}} a^\dagger_{\vk}(\eta)\, a_{\vk}(\eta) \ket{\widetilde{0}} = \frac{\lambda^2}{H^2k^2}\,\int^\eta_{\eta_0} d\eta_1 \,\int^\eta_{\eta_0} d\eta_2 \Big( \frac{\X(\eta_1)}{\eta_1}\Big)\,\Big( \frac{\X(\eta_2)}{\eta_2}\Big)\,e^{-2ik(\eta_1-\eta_2)}\nonumber \\ & = & \frac{\lambda^2}{H^2k^2}\,\Big| \int^\eta_{\eta_0} \,\Big( \frac{\X(\eta_1)}{\eta_1}\Big)\,e^{-2ik\,\eta_1} d\eta_1\Big|^2 = \mathds{F}[k;\eta]\,,\label{bbex} \eea where $ \mathds{F}[k;\eta]$  as the distribution function given by (\ref{fdist}) obtained via the optical theorem. Therefore, the distribution function, and consequently the total number of particles  obtained from the linearized leading order limit of the non-perturbative mean field framework is in complete agreement with the results from the one loop radiative corrections via the self-energy and the optical theorem and is given by (\ref{totNoft}).

These results establish a direct connection between the optical theorem and the Bogoliubov approach to particle production from the interaction with the inflaton condensate. Furthermore, the equivalence with the mean field formulation, at least in the weak field, linearized regime suggests this non-perturbative framework as a future  avenue to study the effects of spectator fields, namely their backreaction on the inflaton beyond weak coupling.

\section{Discussion:}\label{sec:discussion}

\textbf{General lessons:} Since we have studied specific models of spectators coupled to the inflaton, in particular massless scalars conformally coupled to gravity and Yukawa coupled massless fermions, we cannot claim ``universality'' of the results, however, there are several clear and general lessons that stand out and transcend the particular models.

\begin{itemize}
\item{We introduced three complementary methods to obtain consistently the equations of motion with   radiative  corrections from spectator fields: Schwinger-Keldysh (``in-in''),   linear response, and in the case of bosonic spectators linearization of a non-perturbative self-consistent dynamical mean field theory. Particle production as a consequence of the spectator-inflaton interactions are studied implementing a generalization of the optical theorem to a finite time domain as befits the cosmological setting, including the coupling to a homogeneous condensate. This method agrees with the result from Bogoliubov transformations in the dynamical mean field theory. These are general methods that are implementable in any model of spectator-inflaton coupling. The equivalence of the resulting equations of motion with these different approaches validates the general result (\ref{infeom}) for  the equations of motion of an homogeneous  inflaton condensate in the linearized regime in conformal time and after conformal rescaling.  }

\item{A perturbative solution of the equation of motion features secular terms that grow in time. A resummation of these by the dynamical renormalization group yields the long time behavior of the condensate which features strong dependence on the number of efolds during slow roll inflation. }

\item{ A phenomenological friction term $\Gamma \dot{\varphi}$ in the equation of motion for the inflaton condensate \emph{does not reliably describe the dissipative effects of spectator fields}. At least, the models studied provide an explicit example of this statement in the simplest manifestation of quantized fields which closely resemble Minkowski space time. This result extends to the inflationary stage those of reference\cite{cao} previously obtained in a radiation dominated cosmology. Of course this does not rule out the possibility that for \emph{some} models in some circumstances such a phenomenological description may be approximately valid. However, the cases studied in detail here, along with those obtained in ref.\cite{cao} provide definite examples where the simple friction term fails to describe the dynamics even qualitatively. Therefore in general, the validity of a local friction term cannot be taken for granted and must be scrutinized case by case.
    To be sure the safest avenue is to obtain the self-energy corrections from spectator fields to the equations of motion of the inflaton and analyze thoroughly its contribution, in particular the secular terms in a perturbative solution and an implementation of the (DRG) to provide an resumation of the secular terms. A benchmark test for the validity of a friction term would be a secular term that grows linearly with the number of e-folds $N_e$. }

\item{Particles, defined with respect to a free field basis, are produced \emph{during} inflation with a distribution function that is very different from the Minkowski case, which is constrained by energy conservation. For massless scalars conformally coupled to gravity the distribution function is strongly peaked on superhorizon wavelengths, the number of spectator particles grow in time  $N(t)\propto e^{3N_e(t)}$ with $N_e(t)$ the number of e-folds. The distribution function and total number of particles will depend on the particular model, however, the simple example studied here suggests that substantial spectator particle production  during inflation  as a consequence of coupling to the inflaton is in general a feature of the dynamical evolution at least for light scalar fields.  }

\end{itemize}

\vspace{1mm}

{  \textbf{On ``non-Universality'':} We have chosen a quadratic inflaton potential because it allows us to obtain an explicit solution to the equations of motion including the one-loop radiative corrections, therefore allowing a direct comparison with the phenomenological local friction term. This of course, is a particular simple model, chosen  with the sole purpose of solvability,  as such the results obtained,    for example the Sudakov logarithmic secular terms apply to this model but  may not describe the effect of radiative corrections with other choices. However, this choice, with a simple potential already shows unambiguously that the local friction term is not   adequate to describe the influence of  spectator fields. Of course with more complicated potentials and or non-minimal coupling to gravity, the first challenge is to solve exactly the equations of motion even without the corrections from spectator fields, a second challenge is to solve them exactly again with the local friction term, and finally to solve the equation of motion including the non-local one-loop correction and compare the results. Most certainly this is a daunting project, and, again while we cannot extrapolate the results of the simpler case to more complicated scenarios, one two of  the main lessons learned, are that the phenomenological local friction term does not capture the influence of spectators, and that there is profuse particle production during inflation remain, since the simpler, exactly solvable case is a clear   example of both features. The enhancements described by the Sudakov (double logarithms) have been obtained in exact de Sitter space time and it is not clear whether departures from exact de Sitter would result in a different behavior, a question that merits further study.  However, the main lesson is not the particularities of the enhancement but the more overarching result that a correct treatment of spectator fields must account for the correct self-energy contributions and that the phenomenological local friction term, in general, is not the correct description, furthermore, inflaton dynamics leads to substantial particle production during inflation. }

\vspace{1mm}

{  \textbf{Beyond linearized equations of motion:} One of the main objectives in this study is to obtain the \emph{linearized} equation of motion for the inflaton condensate including radiative corrections, because linearization allows us to directly compare to the phenomenological friction term $\Gamma \dot{\varphi}_I$ and to obtain explicitly the solution of the equation of motion. However, even up to one loop order, treating the inflaton condensate as a background for the spectator fields, would yield non-linear terms in the inflaton condensate, each power of the condensate multiplied by a corresponding coupling $\lambda$ for scalars or $y$ for Fermions. These non-linear terms may be obtained, for example in the case of scalars, by solving exactly for the mode functions $g_k(\eta)$, equations (\ref{meffbos}) and using this exact result in the mean field equations (\ref{nunord},\ref{minfi}), or alternatively by adapting the results of ref.\cite{candela}.  We solved the mode equations up to the linear term in the inflaton condensate which is  the Born approximation  (\ref{born}).  Linearization is reliable for weak couplings and small amplitudes, hence our results are valid only in these regimes, since the linear term in the inflaton condensate corresponds to the leading order in the couplings.  }

\vspace{1mm}

\textbf{Energy momentum tensor of spectators:} Although we studied spectator particle production  focusing on the distribution function and total number of produced particles, an expected consequence is that particle production will also impact the energy momentum tensor of spectator fields. An important aspect that remains to be studied is the covariant conservation of the \emph{full} energy momentum tensor, including the inflaton and spectators. Understanding this important aspect also entails a deeper study of renormalization aspects consistent with covariant conservation. These important questions will be addressed in future work.

\vspace{1mm}

\textbf{Isocurvature (entropy) perturbations:} The production of spectator particles growing in time suggests that their contribution to the total energy momentum tensor
will grow accordingly. In turn this implies that these degrees of freedom \emph{may} yield isocurvature perturbations, along with contributing to the entropy budget. Understanding this aspect clearly merits further  study, first by confirming covariant conservation as discussed above, followed by a thorough assessment of the energy density correlation functions \emph{including} the contributions from particle production. An important aspect to focus on is on the renormalization of these correlators since the energy momentum tensor is a dimension four operator its correlation functions feature short distance (ultraviolet) divergences that must be carefully renormalized.  These aspects remain to be studied and along with the spectator contributions to the  energy momentum tensor will be the subject of future work.

\vspace{1mm}

\textbf{Inflaton fluctuations:} We have neglected the contribution of inflaton fluctuations $\delta(\vx,\eta)$ since we focused on the contributions from spectator fields. Depending on the type of inflaton potential, its fluctuations \emph{may} contribute to the self-energy and the linearized equation of motion at one loop. If this is the case this contribution must be included along with those from spectators and treated on the same footing. However,   the inflaton is typically considered to be  minimally coupled to gravity, therefore the mode functions for the quantum fluctuations are given by equation (\ref{gqeta}), which   entails a very challenging technical endeavor to obtain the self-energy in an analytic form that would allow the implementation of the methods described above. Clearly the radiative corrections to the inflaton equation of motion  from its quantum fluctuations  merits a thorough study, which, however is outside the scope of this study and will remain the focus of future work.

\section{conclusions}\label{sec:conclusions}

Inflationary cosmology provides a very successful paradigm that solves many problems of the standard big bang cosmology, provides a mechanism for generating the primordial anisotropies that seed the (CMB) and is largely supported by the observational evidence from temperature fluctuations as measured by various ground and satellite observations. Yet there still are important and pressing questions on the dynamics during  the inflationary stage that merit a deeper understanding in the era of precision cosmology.  A theoretically appealing framework posits that a scalar field, the inflaton,  ``slow rolls'' along a potential landscape and its energy momentum tensor dominates the cosmological dynamics, while all other degrees of freedom, including those of the standard model of particle physics, and beyond,  are simple ``spectators'' whose contributions to the energy momentum tensor is subleading,  becoming dominant during the (pre) reheating stage post inflation. If these degrees of freedom become excited and dominant as a consequence of their coupling to the inflaton leading to profuse particle production \emph{after} inflation, it is a logical possibility that their coupling \emph{during} inflation leads to their back reaction on the dynamics of the inflaton.

In this article we set out to study precisely this back reaction along with some of its consequences. It is often assumed, without critical scrutiny, that the influence of spectator fields upon the equation of motion of the inflaton can be described in terms of a friction term $\simeq \Gamma \dot{\varphi}$ with $\Gamma$ associated with a decay rate obtained from S-matrix theory in Minkowski space time. A previous study\cite{cao} showed unambiguously that such a phenomenological description is \emph{not} warranted in a radiation dominated cosmology. Our main objectives in this study are: \textbf{i:)} to obtain the   equations of motion for the inflaton condensate including one-loop radiative corrections (self-energy) from spectator fields in the linearized approximation, obtain their solutions, and explicitly compare the dynamics to that obtained with the phenomenological friction term, \textbf{ii:)} the dissipative aspects of these radiative corrections are related to particle production, hence we seek to understand spectator particle production as a consequence of their coupling to the inflaton, focusing on obtaining the distribution function and total number of particles produced. We consider a massless scalar field conformally coupled to gravity and a msssless fermion field Yukawa coupled to the inflaton as a model for spectators which   do \emph{not} feature gravitational particle production, thereby allowing a clear understanding of their production from their coupling to the inflaton.

We obtained the equations of motion for the inflaton including radiative corrections from spectator fields (self-energy) up to one loop implementing two complementary methods: the Schwinger-Keldysh (``in-in'') and linear response formulations.  Within the models studied we obtained the fully renormalized equations of motion, including field renormalization in the case of fermions. Their perturbative solution features Sudakov-type logarithmic secular growth and implemented a dynamical renormalization group method to resum these secular divergences yielding an improved evolution of the inflaton condensate. By implementing the same methods in Minkowski space time we learn how cosmological expansion affects the back reaction of spectators.   We find that during slow roll the inflaton condensate evolves as

\be \varphi_{Isr}(t) = \varphi^{(0)}_{Isr}(t)\times \Bigg\{ \begin{array}{l}
                                                       e^{\Upsilon\,N_e(t)} ~~;~~   \Upsilon = \frac{m^2\Gamma}{9H^3}~~;~~ (\Gamma=phenomenological~ friction~term) \\
                                                       e^{\Upsilon N^2_e(t)}~~;~~ \Upsilon =  -\frac{\lambda^2}{24\pi^2 H^2}~~(\mathrm{bosons})~;~ \frac{y^2_R}{12\pi^2}~~(\mathrm{fermion})
                                                     \end{array}
\nonumber \ee where $\varphi^{(0)}_{Isr}(t)$ is the slow roll solution in absence of coupling to spectator fields and $N_e(t)$ is the number of efolds during slow roll, indicating that the phenomenological friction term is inadequate to describe the dynamics. While these results are not ``universal'' they unambiguously indicate that the correct back reaction must be studied carefully by assessing the self-energy kernels rather than uncritically invoking a phenomenological friction term.  The optical theorem, ubiquitous in S-matrix theory is extended   to a finite time domain and generalized to include cosmological expansion and coupling to a condensate to study the  production of spectator particles as a consequence of their coupling to the inflaton, their distribution function and total number of particles. Focusing on massless bosonic spectators conformally coupled to gravity, we find that their distribution function is sharply peaked a superhorizon wavelengths in contrast with Minkowski space-time where the distribution function is  constrained by  energy conservation. We find the total number of spectators produced is given by
 \be  {N}(t)= \frac{\lambda^2 \,V_{ph}(t)}{12\pi\,H}\, \Big(\varphi^{(0)}_{Isr}(t)\Big)^2  \nonumber \ee where $V_{ph}(t)$ is the \emph{physical} volume   exhibiting a rapid growth of the number of particles produced \emph{during} slow roll inflation as a consequence of their coupling to the inflaton with strength $\lambda$.

 We introduce a non-perturbative dynamical mean field theory for the self-consistent evolution of the inflaton and bosonic spectator fields from which we study particle production via a Bogoliubov transformation to the ``free particle basis''. The exact self-consistent mode functions are shown to obey  a Lippman-Schwinger integral equation which lends itself to a solution as a Born series. We obtain the lowest order radiative corrections to the inflaton equation of motion in a Born approximation, it coincides with that obtained from Schwinger-Keldysh and linear response, thereby providing yet another alternative  derivation. In this approximation we also obtain the distribution function and total number of particles in complete agreement with the results from the optical theorem.

While our study focused on specific examples of spectator fields and their coupling to the inflaton, there are several aspects that are quite general and transcend these particular cases: \textbf{i:)} a well defined framework to obtain the equations of motion of the inflaton including radiative corrections from spectator fields. \textbf{ii:)} a perturbative solution of these equations feature secular terms that grow in time, the dynamical renormalization group provides a resummation  of these yielding the asymptotic evolution of the inflaton condensate, \textbf{iii:)} a phenomenological friction term \emph{does not} describe correctly the inflaton evolution, \textbf{iv:)} radiative corrections imply the production of spectator particles, a manifestation of radiation reaction \emph{during inflation}. An extension and generalization of the optical theorem to a finite time domain including cosmological expansion and coupling to condensates yields the distribution function and total number of particles. The distribution function is in general  strikingly different from that in Minkowski space time which is constrained by energy conservation,   which is not the case with expansion because of a lack of a global timelike Killing vector.

Taken together these results raise important questions on the role of spectator fields during inflation, such as the time evolution of their energy momentum tensor, renormalization aspects, covariant conservation and their role as possible contributions to isocurvature perturbations, these aspects remain to be studied and will be the focus of future work.

\acknowledgements
  The author  gratefully acknowledges  support from the U.S. National Science Foundation through grant     NSF 2412374.

\appendix
\section{Bosonic and Fermionic correlators}\label{app:corres}
\subsection{Bosonic correlations}\label{app:boscorr}
 Using Wick's theorem, we find
  \be  \bra{0_s} :(\chi_s(\vx,\eta))^2:\,:(\chi_s(\vx',\eta'))^2:\ket{0_s}   =   2\,  G^>_s(\vx,\vx';\eta,\eta')   \label{Gis}\ee where
  \be  G^>_s(\vx,\vx';\eta,\eta')   =    \bra{0_s}   \chi_s(\vx,\eta) \, \chi_s(\vx',\eta') \ket{0_s}\,\bra{0_s} \chi_s(\vx,\eta) \, \chi_s(\vx',\eta') \ket{0_s} \,\label{ggreat}\ee it is convenient to also introduce
  \be G^<_s(\vx,\vx';\eta,\eta')   =   \bra{0_s} \chi_s(\vx',\eta') \,\chi_s(\vx,\eta)\ket{0_s}\,\bra{0_s}  \chi_s(\vx',\eta') \,\chi_s(\vx,\eta) \ket{0_s} \,, \label{gless}\ee since $\chi_s$ is a real scalar field, it follows that
  \be G^<_s(\vx,\vx';\eta,\eta') = G^>_s(\vx',\vx;\eta',\eta)\,. \label{iden}\ee  The quantization of the bosonic spectator field (\ref{chiex}) with the mode functions (\ref{gqeta}) we find
  \be G^>_s(\vx,\vx';\eta,\eta') = \frac{1}{V^2}\sum_{\vk;\vk'} g_k(\eta)g^*_k(\eta')g_{k'}(\eta)g^*_{k'}(\eta')\,e^{i(\vk+\vk')\cdot(\vx-\vx')} \,,\label{Ggreatgis}\ee

\subsection{Fermion correlations.}\label{app:fercorr}

Defining
\bea \mathcal{G}^>_f(\vx,\vx';\eta,\eta') & \equiv &   \bra{0_F}:\overline{\psi}(\vx,\eta)\psi(\vx,\eta)::\overline{\psi}(\vx',\eta')\psi(\vx',\eta'):\ket{0_F} \,,\label{Gfgreat} \\
\mathcal{G}^<_f(\vx,\vx';\eta,\eta') & \equiv &   \bra{0_F}:\overline{\psi}(\vx',\eta')\psi(\vx',\eta')::\overline{\psi}(\vx,\eta)\psi(\vx,\eta):\ket{0_F} \,,\label{Gfless}
\eea and introducing the  projectors

\be \Lambda^+_{ab}(\vk,\eta,\eta')=\sum_{\lambda=1,2} U_{\lambda,a}(\vk,\eta)  \overline{U}_{\lambda,b}(\vk,\eta') =
 \frac{f_k(\eta)f^*_k(\eta')}{2k^2} \, \left(
                                        \begin{array}{cc}
                                         \Omega(k,\eta)\Omega^*(k,\eta')\,\mathbb{I} & -\Omega(k,\eta)\,\vec{\sigma}\cdot \vk \\
                                          \Omega^*(k,\eta')\,\vec{\sigma}\cdot \vk  & -k^2 \,\mathbb{I}\\
                                        \end{array}
                                      \right)_{ab}\,, \label{projU}
 \ee

\be \Lambda^-_{ba}(\vk,\eta',\eta)=\sum_{\lambda=1,2} V_{\lambda,b}(\vk,\eta') \overline{V}_{\lambda,a}(\vk,\eta) =
 \frac{f_k(\eta)f^*_k(\eta')}{2k^2} \, \left(
                                        \begin{array}{cc}
                                         k^2 \, \mathbb{I} & -\Omega(k,\eta)\,\vec{\sigma}\cdot \vk \\
                                          \Omega^*(k,\eta')\,\vec{\sigma}\cdot \vk  & - \Omega(k,\eta)\Omega^*(k,\eta') \,\mathbb{I}\\
                                        \end{array}
                                      \right)_{ba}\,, \label{projV}
 \ee where $a,b$ are Dirac indices, and $\mathbb{I}$ is the $2\times 2$ identity, and using the conventions  of section (\ref{subsec:ferspec}) we find the fermion correlation function
\be \mathcal{G}^>_f(\vx,\vx';\eta,\eta')  = \frac{1}{V^2} \sum_{\vk}\sum_{\vk'} \mathrm{Tr} \Big[\Lambda^+(\vk;\eta,\eta')\,\Lambda^{-}(\vk';\eta',\eta)\Big]\,e^{i(\vk+\vk')\cdot(\vx-\vx')} \,,\label{corref} \ee and
\be \mathcal{G}^<_f(\vx,\vx';\eta,\eta') = \mathcal{G}^>_f(\vx',\vx;\eta',\eta)\,.\label{Gfeqs}\ee The trace in (\ref{corref}) is given by

\bea  &&  \mathrm{Tr} \Big[\Lambda^+(\vk;\eta,\eta')\,\Lambda^{-}(\vk';\eta',\eta)\Big]    =   \frac{1}{2k^2k^{'\,2}}\, \Big[f^*_k(\eta_2)f^*_{k'}(\eta_2)f_k(\eta_1)f_{k'}(\eta_1) \Big]\times \nonumber \\ && \Bigg[k^{'\,2}\,\Omega_k(\eta_1)\,\Omega^*_k(\eta_2)+k^2\,\Omega^*_{k'}(\eta_2)\,\Omega_{k'}(\eta_1)   -
  \vk\cdot\vk' \Big(\Omega_k(\eta_1)\,\Omega^*_{k'}(\eta_2)+ \Omega_{k'}(\eta_1)\,\Omega^*_{k}(\eta_2)\Big) \Bigg] \,,\label{traci}\eea in particular for $\vk^{\,'}= -\vk$ it simplifies to
  \be \mathrm{Tr} \Big[\Lambda^+(\vk;\eta,\eta')\,\Lambda^{-}(-\vk;\eta',\eta)\Big] = \frac{2}{k^2}\, f^2_k(\eta_1)\,f^{*2}_k(\eta_2) \,\Omega_k(\eta_1)\,\Omega^*_k(\eta_2)\,.\label{tracikk}\ee

\subsection{Optical theorem with condensates in real time.}\label{subsec:OT}
In this appendix we analyze the optical theorem in a finite time interval in Minkowski space-time with a straightforward generalization to the cosmological case.

The time evolution operator in the interaction picture $U_I(t,t_0)$ is unitary, namely $U_I(t,t_0)U^\dagger(t,t_0) = U_I(t,t_0)U^{-1}(t,t_0)=1$ obeying
\be i\,\frac{d}{dt}U_I(t,t_0) = H_I(t) U_I(t,t_0)~~;~~ -i\,\frac{d}{dt}U^{-1}_I(t,t_0)= U^{-1}_I(t,t_0)\,H_I(t)~~;~~ U_I(t_0,t_0) = 1 \,,\label{Uai}\ee where $H_I(t)$ is the interaction Hamiltonian in the interaction picture, it is a hermitian operator. The solutions
\be    U_I(t,t_0)   =   1-i \int^{t}_{t_0} H_I(t_1) d t_1 - \int^{t}_{t_0} dt_1 \int^{t_1}_{t_0} dt_2 H_I(t_1) H_I(t_2)  + \cdots = T \Big(e^{-i\int^{t}_{t_0}H_I(t_1) dt_1}\Big) \,,\label{timeord} \ee
 \be    U^{-1}_I(t,t_0)   =   1+i \int^{t}_{t_0} H_I(t_1) dt_1 - \int^{t}_{t_0} dt_1 \int^{t_1}_{t_0} dt_2 H_I(t_2) H_I(t_1) + \cdots = \widetilde{T}\Big(e^{-i\int^t_{t_0}H_I(t_1) dt_1}\Big)
 \,,\label{antimeord}\ee
   where $T,\widetilde{T}$ are the time and anti-time ordering operators respectively. Writing
   \be U_I(t,t_0) = 1+ i \, \mathcal{T}[t,t_0] \,,\label{defmat}\ee with $\mathcal{T}[t,t_0]$ the transition matrix,
   the optical theorem  follows from unitarity, namely
   \be -i\,\big(\mathcal{T}[t,t_0]- \mathcal{T}^\dagger[t,t_0]\big) = \mathcal{T}[t,t_0]\mathcal{T}^\dagger[t,t_0] \,,\label{otu}\ee

   In a perturbative expansion
   \be \mathcal{T}[t,t_0] = \mathcal{T}^{(1)}[t,t_0]+ \mathcal{T}^{(2)}[t,t_0]+\cdots \label{pertTmat}\ee with
   \be \mathcal{T}^{(1)}[t,t_0]   =   - \int^{t}_{t_0} H_I(t_1) d t_1 ~~;~~ \mathcal{T}^{(2)}[t,t_0] = i \int^{t}_{t_0} dt_1 \int^{t}_{t_0} dt_2 H_I(t_1) H_I(t_2)\Theta(t_1-t_2) \cdots \label{Tmatpt}\ee
   \be \mathcal{T}^{\dagger (1)}[t,t_0]   =   - \int^{t}_{t_0} H_I(t_1) d t_1 ~~;~~ \mathcal{T}^{\dagger (2)}[t,t_0] = -i \int^{t}_{t_0} dt_1 \int^{t}_{t_0} dt_2 H_I(t_2) H_I(t_1)\Theta(t_1-t_2) \cdots \label{Tmatdagpt}\ee in the second order term in (\ref{Tmatdagpt}) we can relabel the integration time variables $t_1 \leftrightarrow t_2$ yielding
   \be \mathcal{T}^{\dagger (2)}[t,t_0] = -i \int^{t}_{t_0} dt_1 \int^{t}_{t_0} dt_2 H_I(t_1) H_I(t_2)\Theta(t_2-t_1) \,.\label{relab}\ee Up to second order, the optical theorem (\ref{otu}) becomes
   \be -i\,\big(\mathcal{T}^{(2)}[t,t_0] - \mathcal{T}^{\dagger(2)}[t,t_0] \big) = \mathcal{T}^{(1)}[t,t_0]\mathcal{T}^{\dagger(1)}[t,t_0]\,, \label{otu2nd}\ee which is straightforwardly confirmed using (\ref{Tmatpt},\ref{relab}). Of course, this discussion is a generalization of well known results from S-matrix theory where $t\rightarrow \infty, t_0 \rightarrow -\infty$, to a finite time domain $t_0 \leq t_1,t_2 \cdots \leq t$. A useful relation follows from the optical theorem   by taking the expectation value of (\ref{otu}) within states $\ket{\alpha}$ and introducing the resolution of the identity in a complete set of states $\ket{\{n\}}$, which up to second order yields
   \be 2\,\mathrm{Im}\bra{\alpha} \mathcal{T}^{(2)}[t,t_0]   \ket{\alpha} =   \sum_{\{n\}}  \Big|\bra{\{n\}}\mathcal{T}^{\dagger(1)}[t,t_0]\ket{\alpha} \Big|^2   \,.\label{mtxele2nd}\ee The right hand side gives the total transition probability $\ket{\alpha} \rightarrow \ket{\{n\}}$ between $t_0$ and $t$ up to second order in the coupling.
     In Minkowski space-time and in the infinite time limit $t\rightarrow \infty,t_0 \rightarrow -\infty$, the matrix elements of the transition matrix $\mathcal{T}$ are related to the invariant scattering or decay matrix elements multiplied by overall energy-momentum conserving delta functions. The finite time counterpart is more useful in an expanding  cosmology case because   energy is not conserved as there is no global timelike Killing vector in this case, although spatial momentum is conserved in a homogeneous and isotropic cosmology.

   Let us consider Minkowski space time and a bosonic field $\chi(\vx,t)$ coupled to a \emph{homogeneous} condensate $\X(t)$  with interaction Hamiltonian in the interaction picture of
   free fields
   \be H_I(t)  = \lambda \, \int d^3 x :\chi^2(\vx,t): \,\X(t) \,,\label{appHI}\ee corresponding to the interaction Lagrangian density (\ref{lI}), with
   \be \chi(\vx,t) = \frac{1}{\sqrt{V}}\sum_{\vk} \frac{1}{\sqrt{2k}}\,\Big[ a_{\vk}\, e^{-ikt}+  a^\dagger_{-\vk}\,e^{ikt}\Big]\,e^{i\vk\cdot \vx} \,,\label{chimin}\ee   and
   consider the state $\ket{\alpha} = \ket{0}$, the vacuum state of the field $\chi$, namely $a_{\vk}\ket{0} =0$.
Using the results (\ref{Gis}-\ref{gless}) with $\eta \rightarrow t$ , we find
\bea \bra{0}\mathcal{T}^{(2)}[t,t_0]\ket{0} & = &  2\,i\, \lambda^2  \, \int^t_{t_0} dt_1 \int d^3 \vx_1 \int^{t_1}_{t_0} dt_2 \int d^3 \vx_2 \,  \,\X(t_1)\,\X(t_2)\,  G^>(\vx_1,\vx_2;t_1,t_2) \label{T2melgreat}\\
\bra{0}\mathcal{T}^{\dagger (2)}[t,t_0]\ket{0} & = &  -2\,i\, \lambda^2  \, \int^t_{t_0} dt_1 \int d^3 \vx_1 \int^{t_1}_{t_0} dt_2 \int d^3 \vx_2 \,  \,\X(t_1)\,\X(t_2)\,  G^<(\vx_1,\vx_2;t_1,t_2) \label{T2melless}\eea Since the operator $:\chi^2:$ is bilinear in the fields the first order transition matrix connects the vacuum to a two-particle state, therefore, the unitarity conditions (\ref{relab}, \ref{mtxele2nd}) become
\bea &&  2\, \lambda^2  \, \int^t_{t_0} dt_1 \int d^3 \vx_1 \int^{t_1}_{t_0} dt_2 \int d^3 \vx_2 \,  \,\X(t_1)\,\X(t_2)\,  \Big[G^>(\vx_1,\vx_2;t_1,t_2) + G^<(\vx_1,\vx_2;t_1,t_2)\Big] \nonumber \\ && = \frac{1}{2!}\,\sum_{\vk_1,\vk_2}\Big|\bra{\vk_1,\vk_2}\mathcal{T}^{\dagger(1)}[t,t_0]\ket{0}\Big|^2  = \frac{1}{2!}\,\sum_{\vk_1,\vk_2}\Big|\int d^3\vx \int^t_{t_0} \bra{\vk_1,\vk_2}:\chi^2(\vx,t_1):\ket{0}\,\X(t_1)\,dt_1\Big|^2 \,,\label{unitrelmin}\eea where the factor $1/2!$ accounts for the indistinguishability of the two particle state. This expression describes the production of particle pairs  from the vacuum as a consequence of the condensate, with total production probability
\be \mathcal{P}_{0 \rightarrow 2}[t,t_0] = \frac{1}{2!}\,\sum_{\vk_1,\vk_2}\Big|\bra{\vk_1,\vk_2}\mathcal{T}^{\dagger(1)}[t,t_0]\ket{0}\Big|^2 \,.\label{prodprob} \ee

Using (\ref{Ggreatgis}) with the Minkowski space-time mode functions $   g_k(t) = e^{-ikt}/\sqrt{2k}$  it follows that
\be \int d^3 \vx_1 \, \int d^3 \vx_2  \Big[G^>(\vx_1,\vx_2;t_1,t_2) + G^<(\vx_1,\vx_2;t_1,t_2)\Big]   =  \frac{V}{8\pi} \, \delta(t_1-t_2)\,,\label{resumg} \ee   yielding
\be 2\,\mathrm{Im} \Big[  \bra{0}\mathcal{T}^{(2)}[t,t_0]\ket{0} \Big] = \frac{\lambda^2}{8\pi}\,V\, \int^t_{t_0} \X^2(t_1)\,dt_1 \,.\label{imot}\ee Writing the condensate as
\be \X(t) = X\,e^{-imt}+ X^* \,e^{imt}\,,\label{condimin}\ee in the long time limit we find
\be 2\,\mathrm{Im} \Big[  \bra{0}\mathcal{T}^{(2)}[t,t_0]\ket{0} \Big] = \frac{\lambda^2}{8\pi}\,V\,(2\,|X|^2) \,(t-t_0) + \cdots\,,\label{ltmin}\ee where the dots stand for non-secular terms. The right hand side of (\ref{unitrelmin}) yields
\be  \frac{1}{2!}\,\sum_{\vk_1,\vk_2}\Big|\bra{\vk_1,\vk_2}\mathcal{T}^{\dagger(1)}[t,t_0]\ket{0}\Big|^2  =  {2\lambda^2\,V}\int^t_{t_0} dt_1 \int^t_{t_0} dt_2 \, \X(t_1)\,\X(t_2)\,\int \frac{d^3k}{(2\pi)^3} \frac{e^{-2ik(t_1-t_2-i\epsilon)}}{4k^2}\,,\label{rhs} \ee where we introduced the convergence factor $\epsilon \rightarrow 0^+$. Only the   contribution that is symmetric under $t_1 \leftrightarrow t_2$ from the $k-$ integral survives, finally yielding
\be  \frac{1}{2!}\,\sum_{\vk_1,\vk_2}\Big|\bra{\vk_1,\vk_2}\mathcal{T}^{\dagger(1)}[t,t_0]\ket{0}\Big|^2  =  \frac{\lambda^2}{8\pi}\,V\, \int^t_{t_0} \X^2(t_1)\,dt_1  \ee confirming explicitly the unitarity relation to leading order. Therefore, with the condensate given by (\ref{condimin}) the total probability for pair production  (\ref{prodprob}) in the long time limit is given by
\be \mathcal{P}_{0 \rightarrow 2}[t,t_0] = \frac{\lambda^2}{8\pi}\,V\,(2\,|X|^2) \,(t-t_0) + \cdots\label{totprod} \ee We can make contact with the familiar result for the probability of single particle decay into two massless particles in Minkowski space time by recognizing that in the quantum field description of the decaying scalar field, the single particle amplitude for a zero momentum state is $1/\sqrt{2Vm}$ with $m$ being the mass of the decaying field, therefore replacing $X$ by this single particle amplitude in the above expression yields for the total rate of decay of a single zero momentum particle into two massless particles as
\be \frac{\mathcal{P}_{1 \rightarrow 2}[t,t_0]}{(t-t_0)} = \Gamma = \frac{\lambda^2}{8 \pi m}\,,  \ee which is the well known decay rate of a single particle at rest into two massless particles.

Although the analysis above was carried out in Minkowski space-time, it can be simply extrapolated to an expanding cosmology in conformal time coordinates by replacing $t,t_0 \rightarrow \eta,\eta_0$ and including the appropriate scale factors $C(\eta)$ in the interaction vertices. The interaction Hamiltonian of spectator fields with the inflaton condensate $\X(\eta)$ in the interaction picture of free fields is $H_I(\eta)$
  given by equation (\ref{eomUI}) with ${\mathcal{O}^{(0)}}(\vx,\eta_0)$ given by the composite operator  in the interaction picture    (\ref{Ointpic}), namely the operator (\ref{opeO})  for free fields. Up to second order, the generalization of the optical theorem to the cosmological space-time, relates the  pair production probability of spectator particles
  \be \mathcal{P}_{0\rightarrow 2}(\eta,\eta_0) = \mathcal{S}\,\sum_{\vk_1,\vk_2}\Big|\bra{\vk_1,\vk_2}\mathcal{T}^{\dagger(1)}[\eta,\eta_0]\ket{0}\Big|^2  \,,\label{pairprod}\ee with $\mathcal{S}=1/2!,(1) $ for indistinguishable (distinguishable) particles,  to the correlation function of the composite operator ${\mathcal{O}^{(0)}}(\vx,\eta_0)$ in interaction picture,
  \bea \mathcal{P}_{0\rightarrow 2}(\eta,\eta_0) & = &   \int^\eta_{\eta_0} d\eta_1 \int d^3 \vx_1 \int^{\eta_1}_{\eta_0} d\eta_2 \int d^3 \vx_2 \,  \,\X(\eta_1)\,\X(\eta_2)\, \nonumber \\ &&   \bra{0}\Big[ {\mathcal{O}^{(0)}}(\vx_1,\eta_1){\mathcal{O}^{(0)}}(\vx_2,\eta_2) +{\mathcal{O}^{(0)}}(\vx_2,\eta_2){\mathcal{O}^{(0)}}(\vx_1,\eta_1)\Big]\ket{0}\,,\label{desitprod} \eea where $\ket{0}$ is the Bunch-Davies vacuum for all spectator fields.

  This expression can be simplified, let us write the time integrals as
  \be \int^\eta_{\eta_0} d\eta_1 \int^{\eta_1}_{\eta_0} d\eta_2 \Big\{ \cdots \Big\} = \int^\eta_{\eta_0} d\eta_1 \int^{\eta}_{\eta_0} d\eta_2 \Big\{ \cdots \Big\} \,\Theta(\eta_1-\eta_2) \,,\label{inteta}\ee and in the second term in (\ref{desitprod}) now relabel $\vx_1 \Leftrightarrow \vx_2; \eta_1 \Leftrightarrow \eta_2$ and use $\Theta(\eta_1-\eta_2)+\Theta(\eta_2-\eta_1) =1$ to find
   \be \mathcal{P}_{0\rightarrow 2}(\eta,\eta_0)  =    \int^\eta_{\eta_0} d\eta_1 \int d^3 \vx_1 \int^{\eta}_{\eta_0} d\eta_2 \int d^3 \vx_2 \,  \,\X(\eta_1)\,\X(\eta_2)\,    \bra{0} {\mathcal{O}^{(0)}}(\vx_1,\eta_1){\mathcal{O}^{(0)}}(\vx_2,\eta_2) \ket{0}\,.\label{desitprodfin} \ee

This is the main result of the optical theorem up to second order for spectator fields coupled to the inflaton condensate during a finite time domain in a cosmological space time.

\end{document}